%% file: manuscript.tex
\documentclass[sigconf,nonacm]{acmart}

\usepackage{hyperref}
\usepackage{url}
\usepackage{graphicx}
\usepackage{cleveref}
\usepackage{subcaption}
\usepackage{booktabs}
\usepackage[absolute]{textpos}

\AtBeginDocument{%
  }

\begin{document}

\begin{textblock*}{\textwidth}(0.75in, 0.5in)
  \noindent A shortened version of this paper appears in the \textit{Proceedings of the 5th ACM Symposium on Computer Science and Law (CS\&LAW ‘26)}, March 2026. This is the extended version.
\end{textblock*}

\title{A Common Pool of Privacy Problems: Legal and Technical Lessons from a Large-Scale Web-Scraped Machine Learning Dataset}


\author{Rachel Hong$^1$, Jevan Hutson$^2$, William Agnew$^3$, Imaad Huda$^2$,\\Tadayoshi Kohno$^{4}$, Jamie Morgenstern$^{1\>5}$}
\thanks{1. University of Washington Paul G. Allen School of Computer Science \& Engineering}
\thanks{2. University of Washington School of Law}
\thanks{3. Carnegie Mellon University Carnegie Bosch Institute}
\thanks{4. Georgetown University Department of Computer Science}
\thanks{5. Amazon AWS AI/ML}

\renewcommand{\shortauthors}{Hong et al.}

\begin{abstract}
  \input{sections/00-abstract}
\end{abstract}




\received{2025}

\settopmatter{printfolios=true}
\maketitle

\input{sections/01-introduction}

\input{sections/02-background}

\input{sections/03-methodology}

\input{sections/04-audit-results}

\input{sections/05-legal-analysis}

\input{sections/06-discussion}

\input{sections/07-conclusion}

\input{sections/acks}

\bibliographystyle{acm_template/ACM-Reference-Format}
\bibliography{ref}

\input{sections/appendix/appendix}

\end{document}

%% file: sections/00-abstract.tex
We investigate the contents of web-scraped data for training AI systems, at sizes where human dataset curators and compilers no longer manually annotate every sample. Building off of prior privacy concerns in machine learning models, we ask: What are the legal privacy implications of web-scraped machine learning datasets? In an empirical study of a popular training dataset, we find significant presence of personally identifiable information despite sanitization efforts. Our audit provides concrete evidence to support the concern that any large-scale web-scraped dataset may contain legally defined personal data. We use these findings of a real-world dataset to inform our legal analysis with respect to existing privacy and data protection laws. We surface various legal risks of current data curation practices that may propagate personal information to train downstream models. Based on our empirical and legal analyses, we argue for reorientation of current frameworks of ``publicly available'' information to meaningfully limit the development of AI built upon indiscriminate scraping of the internet.

%% file: sections/01-introduction.tex
\section{Introduction}
\label{intro}

With the recent popularity in foundation models like ChatGPT and Midjourney \citep{chatgpt, midjourney}, machine learning practitioners often rely on data scraped from the web to train large language or vision models \citep{solove2024great, brown2020language, rombach2022high}. DataComp CommonPool, for instance, is one of the largest publicly available image-text dataset scraped from the web with 12.8 billion samples \citep{gadre2024datacomp}. This dataset has been downloaded over 2.2 million times at the time of writing \citep{datacomp_downloads}, and its precursor LAION-5B \citep{schuhmann2022laion} was used to train well-known image generation models like Midjourney, Stable Diffusion, and Google’s Imagen \citep{midjourney, stable_diffusion, imagen}. Since machine learning models are a function of their training data, downstream models trained on the same dataset may share similar behavior \citep{bommasani2022picking}. Moreover, prior work has shown that adversaries can extract portions of training data that ChatGPT has memorized, including instances of personally identifiable information (PII) \citep{nasr2023scalable}. While presence of PII in the training data does not guarantee its extraction or leakage from a model, at web-scale it is difficult to understand the levels of potential PII in these datasets to begin with. For a widely-used dataset like CommonPool, we argue that engaging in data-centric AI governance \citep{gupta2024data} may be more effective than addressing the harms of every model one-by-one – in other words, tackling the ``root'' rather than the ``leaves'' as shown in \Cref{fig:actor_tree_small}.

In our work, we use DataComp CommonPool as a case study of web-scraping and conduct an investigation into data privacy concerns, considering both ethical and legal notions of privacy. We perform a \textit{legally-grounded audit}, one of the first to our knowledge, in which our audit findings inform our legal analysis on web-scraping, and vice versa, where privacy scholarship motivates our audit inquiries \citep{solove2024great, solove2005taxonomy}. Specifically, our audit asks: \textit{What kinds of personally identifiable information are present in DataComp? How do current data cleaning practices address privacy issues?} To do so, we draw upon prior frameworks on privacy \citep{mireshghallah2024trust}, representation \citep{diaz2024sound}, and data filtering \citep{hong2024s}.

Our legal analysis considers how use of DataComp CommonPool for AI development might trigger application of and compliance obligations under existing privacy laws for developers and downstream deployers, including US state comprehensive privacy laws and international data protection laws. We consider and problematize current interpretations of ``publicly available data'' under existing privacy laws. We also consider how privacy risks and compliance obligations triggered by the production of DataComp CommonPool propagate to downstream models trained on this dataset. Lastly, we consider ongoing privacy risks that are currently not being addressed sufficiently by data filtering and other responsible data curation and hygiene practices, which informs recommendations on how policymakers might address these risks.

In summary, we make the following contributions:

\begin{enumerate}
    \item We find instances of personal information present in DataComp CommonPool, revealing various privacy concerns in web-scraped image-text datasets. We uncover examples of personal information including credit card numbers and passport numbers, and we estimate at least 136,000 images depict resumes of individuals.
    \item We argue that ongoing cleaning methods are not sufficient to tackle privacy concerns and that these methods must be scrutinized. Specifically, we estimate that DataComp's face blurring tool fails to catch at least 100 million images of real human faces, demonstrating the importance of privacy tool assessments.
    \item We map these audit results to legal concerns to provide a critique of current data curation practices according to existing privacy laws; we also demonstrate shortcomings of existing privacy frameworks, such as the implications of exemptions for publicly available information.
\end{enumerate}

We first present the context for web-scraped machine learning dataset development in \Cref{sec:bkgd}, the stakeholders and artifacts associated in each step of the curation pipeline in \Cref{sec:stakeholders}, and related computer science and legal literature in \Cref{sec:related}. We then present our audit methodology in \Cref{sec:audit_methods} and the empirical results in \Cref{sec:audit}. We use these findings to inform our legal analysis to determine the application of various data protection laws in \Cref{sec:legal_analysis}. Finally in \Cref{sec:discussion}, we integrate the concerns revealed by our audit together with the shortcomings of existing privacy frameworks to discuss normative arguments for both policymakers and machine learning practitioners.

%% file: sections/02-background.tex
\section{DataComp CommonPool}
\label{sec:bkgd}

In April of 2023, \citet{gadre2024datacomp} released DataComp CommonPool, a publicly available image-text dataset of 12.8 billion samples collected from the web, as part of the DataComp testbed for assembling datasets to train more effective large image-text models like CLIP \citep{radford2021learning}. In this section we describe the curation process for the DataComp CommonPool dataset, describe the measures the curators took to protect privacy of individuals in the dataset, highlight the dataset's usage after its release, and situate CommonPool in relation to its predecessor LAION-5B.

\subsection{Curation process}

The steps to build CommonPool are as follows \citep{gadre2024datacomp}:

\begin{enumerate}
    \item \textbf{Gather}: The curators first gather web snapshots from 2014 to 2022, relying on Common Crawl as the data source, which is a nonprofit organization that crawls the entire web to form unformatted web dump archives \citep{commoncrawl}.
    \item \textbf{Extract}: The image URLs and accompanying alt-text (alternative text attached to the image for accessibility purposes \citep{caldwell2008web}) are extracted from the web snapshots and deduplicated. The alt-text is referred to as ``captions'' for the associated images.
    \item \textbf{Download}: The images are then downloaded from the URLs, resulting in 16.8 billion successfully downloaded samples at the time of curation.
    \item \textbf{Filter}: Next, several toxicity filters are applied in order to discard NSFW-detected images or text. In addition, a face detection algorithm is applied to annotate bounding boxes of faces in the images, as detailed in \Cref{sec:face_obfuscation}.
    \item \textbf{Deduplicate}: Finally, the images are inspected and deduplicated from evaluation sets, resulting in 12.8 billion image-text pairs that comprise DataComp CommonPool.
    \item \textbf{Release}: Rather than releasing the image content, the curators release CommonPool as a table where each sample consists of an image URL, the associated text, and additional metadata (image size, image hash, etc). To acquire the dataset, the release is accompanied with a code repository for users to run a script which instantiates a crawler that automatically downloads each image from its URL \citep{datacomp_github}. \textit{This means any dataset user must scrape the web to download CommonPool.}
\end{enumerate}

\subsection{Privacy mitigations}
\label{sec:bkgd_privacy}

In the CommonPool datasheet \citep{gebru2021datasheets}, the dataset curators disclose that due to its scale and internet sources ``it is highly likely that there is sensitive data in the dataset'' including identifying information. Therefore, they engage in several mitigations to ``prevent making sensitive content more accessible'' \citep{gadre2024datacomp}.

\subsubsection{Face obfuscation}
\label{sec:face_obfuscation}

To address privacy concerns, CommonPool is released with face detection annotations, so that the dataset download script by default hides any detected face in the image via a Gaussian blurring method \citep{yang2022study}. To create these annotations, the curators apply the SCRFD algorithm \citep{guo2021sample} to obtain bounding boxes for detected faces in each image. It is plausible that SCRFD is chosen due to its efficiency and lack of cost --- the CommonPool curators compare SCRFD to the commercial system Amazon Rekognition and find that SCRFD has worse precision and recall ($75.87\%$ \& $90.53\%$) than Rekognition ($86.09\%$ \& $93.75\%$). The curators evaluate image-text CLIP embedding models \citep{radford2021learning} trained on CommonPool with and without blurred faces and demonstrate similar model performance on their evaluation benchmarks \citep{gadre2024datacomp}. In the released artifact, the face bounding boxes thus accompany each URL-caption sample as metadata, which also inadvertently allows any user to extract faces in the dataset.

The face blurring is optional, however, as a dataset user downloading CommonPool can easily turn off face blurring through specific parameters \citep{datacomp_github}. In addition, models trained on CommonPool with blurred faces are able to zero-shot classify race and gender at rates significantly better than random chance \citep{gadre2024datacomp}. The CommonPool curators speculate that models are still absorbing sociodemographic information outside of faces, or that face blurring does not capture all human faces which we evaluate in our work.

\subsubsection{Opt-out mechanisms}
\label{sec:datacomp_opt_out}

Hugging Face, the dataset distribution platform that hosts the CommonPool URL-caption pairs, integrates with Spawning AI, a tool that allows users to search and remove their personal information from the dataset \citep{spawning}. However, as described in \Cref{sec:legal_lit:opt_out}, these opt-out policies are often not considered meaningful consent since users must first know of the presence of personal information in the first place and bear the burden to find and remove it \citep{solove2024murky}. Another mechanism which provides some attempt at privacy protection is Robots Exclusion Protocol, where a website attach a robots.txt file to instruct web crawlers what assets they can access \citep{sun2007large}. However, this protocol is not legally enforceable, and AI crawlers have recently been accused of not respecting robots.txt \citep{ai_robots_txt}. Since 2023, site hosts have modified their robots.txt files to restrict scraping for AI development, signaling some intention of site content to be kept private from training models (in addition to intellectual property concerns) \citep{longpre2024consent}. While Common Crawl does respect robots.txt \citep{commoncrawl_robots}, CommonPool relies on snapshots from 2014 to 2022 \citep{gadre2024datacomp} to aggregate URLs, before many sites began these restrictions. According to a developer of the code package used to download CommonPool, the downloading step at the time of writing does not respect up-to-date robots.txt website-level protocols due to practicality reasons, although they do respect image-level robots tags to not crawl \citep{img2dataset}, which we elaborate further in \Cref{sec:audit_unavailable}. Certain site hosts may have changed their preferences to prevent crawling across the site (and thus see no reason to update image-by-image tags), yet these preferences are not followed at the site-level.

\subsection{Usage}

According to Hugging Face, the DataComp datasets (including CommonPool and smaller subsets) at the time of writing have been downloaded 2.2 million times, with half a million downloads in the month of October 2024 alone \citep{datacomp_downloads}. CommonPool's curators explicitly state that the dataset is intended for academic research and do not condone the dataset being used to train deployed models \citep{gadre2024datacomp}. However, the URL-text pairs that comprise CommonPool are released with a CC-BY-4.0 license, which does not prohibit anyone from using the dataset for commercial purposes \citep{ccby}. The license is also specific to the URL table, rather than the images assets themselves.

Because companies that develop models often do not disclose their training data \citep{hardinges2024we}, it is unclear if any popular commercially deployed models have trained on CommonPool. However, we highlight in \Cref{sec:bkgd_laion} that the curators themselves acknowledge that CommonPool has substantial overlap \citep{datacomp_laion_overlap} with the LAION-5B dataset, which has been used to train models like Stable Diffusion, Midjourney, and Imagen \citep{midjourney, stable_diffusion, imagen}. We also observe that in several issues published to the dataset Github repository, the posters mention downloading the dataset on behalf of a company \citep{datacomp_company} --- which is unclear if the dataset is being used for research or commercial deployment. DataComp's Github repository has about 700 stars (number of users who have marked it as a favorite) \citep{datacomp_github}, while the original paper has about 400 citations. These numbers, however, do not cover the sheer number of overall dataset downloads.

\subsection{LAION-5B}
\label{sec:bkgd_laion}

DataComp CommonPool was intended as a follow-up to the LAION-5B dataset released the prior year \citep{schuhmann2022laion}, and follows many of the same data curation steps, originating from Common Crawl and releasing a similar collection of URLs rather than the image assets themselves. LAION-5B is a collection of 5 billion image-text pairs that DataComp authors state have substantial overlap with CommonPool due to reliance on the same data source \citep{datacomp_laion_overlap}. In December 2023, the LAION-5B dataset was taken down after \citet{thiel2023identifying} found presence of Child Sexual Abuse Material (CSAM) in the dataset \citep{cole2023largest}. In August 2024, the dataset developers subsequently removed links to the detected CSAM with the release of Re-LAION 5B \citep{relaion}.

\subsubsection{Privacy mitigations}
\label{sec:laion_opt_out}

In their paper, the LAION-5B authors highlight issues of PII present in the dataset. They argue their accompanying tool \texttt{CLIP retrieval}, which enables text search of LAION-5B images, grants users the ability to find their own personal content potentially present, in order to initiate takedown procedures from the dataset or website hosting provider. While the public availability of this search tool may raise awareness about the content of web-crawled data, the tool can also be used by adversaries to gather personal information; we expand on these concerns in our discussion on profiling \Cref{sec:legal:profiling}. Currently the CLIP retrieval website is no longer accessible, but code is available to run the tool locally \cite{clip_retrieval}).

Moreover, the LAION-5B authors place the responsibility on the individuals to find and remove their personal information, yet these opt-out policies again are not very meaningful (\Cref{sec:legal_lit:opt_out}). For instance, when someone found their medical records leaked on LAION-5B and wished to take them down, a LAION author responded that the hosting website was responsible since the dataset curators ``are not hosting any of these images'' \citep{laion_medical}. As we discuss in \Cref{sec:discussion_scale_issues}, content on the internet propagates and becomes incredibly hard to regulate past the time of upload.

\subsubsection{Usage}

Originally intended for research, the LAION-5B dataset also has had substantial usage to train commercially deployed models, despite authors explicitly advising ``against any applications in deployed systems without carefully investigating behavior and possible biases of models'' \citep{schuhmann2022laion}. Midjourney, Stable Diffusion, and Google's Imagen all disclose training on subsets of LAION-5B, and these models have over millions of users \citep{midjourney, stable_diffusion, imagen}. Recent work has shown that Stable Diffusion and Imagen can be subject to training reconstruction attacks, in which supplying particular captions from the training dataset into these models can generate images almost identical to training examples \citep{carlini2023extracting}. In addition, researchers have been able to fine-tune Stable Diffusion models to reconstruct training images without access to the captions \citep{li2024shake}. The wide impact of models trained on LAION-5B subsequently demonstrates that publicly-available datasets can be used without regarding the dataset curator's original intent.

\section{Stakeholders}
\label{sec:stakeholders}

\begin{figure}
    \centering
    \includegraphics[width=\linewidth]{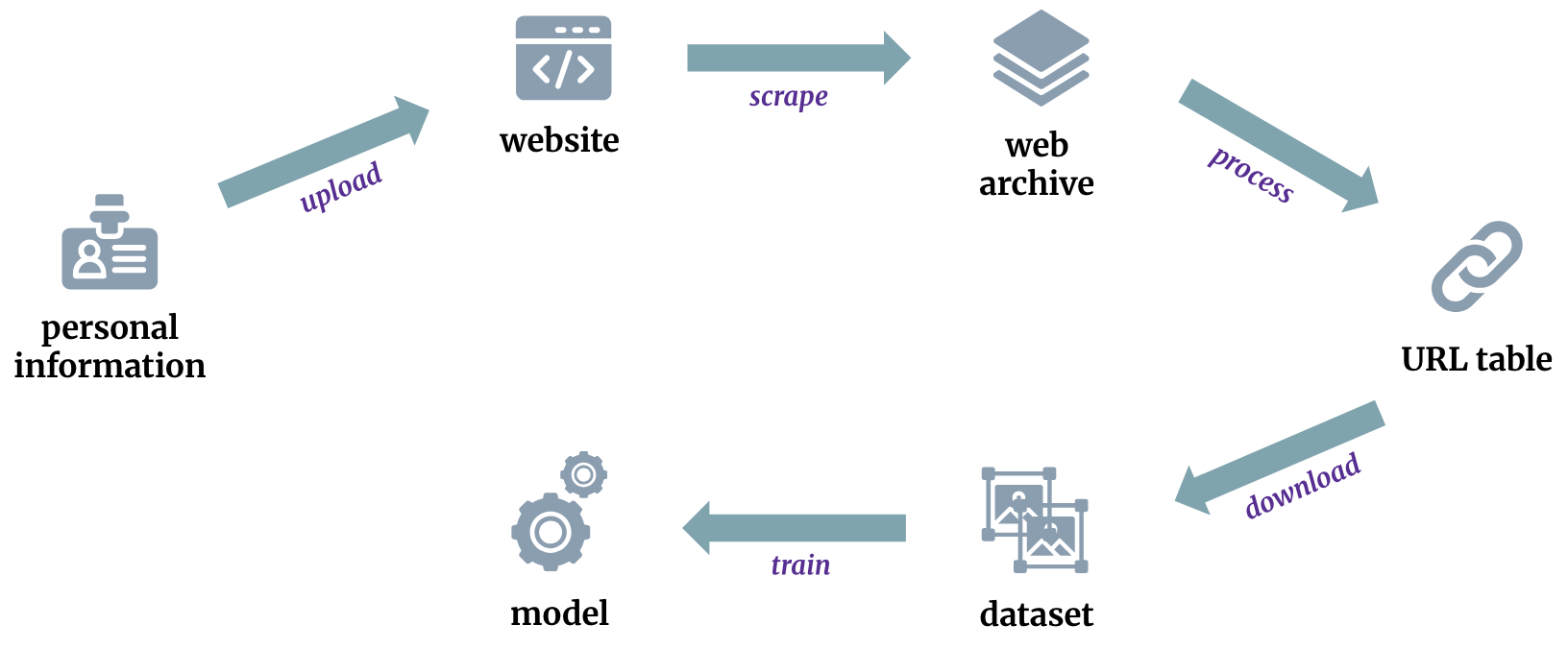}
    \caption{Data lifecycle of how personal information appears in various artifacts in the machine learning pipeline. Icons made by Freepik from www.flaticon.com \citep{flaticon}.}
    \Description{Flow chart cycle with arrows from personal information, to website, to web archive, to URL table, to dataset, to model, and back to personal information.}
    \label{fig:data_lifecycle}
\end{figure}

We now situate web-scraped training datasets like CommonPool within the broader machine learning pipeline and provide frameworks to define the various processes and actors that influence or are influenced by these datasets. Within the pipeline, personal information may be encoded in different forms, from a \textit{website} to a downstream \textit{model} trained on that data, as shown in \Cref{fig:data_lifecycle}. First, \textit{personal information} is uploaded to a \textit{website}, which then is scraped and aggregated into a \textit{web archive}, a snapshot of the entire internet obtained through web crawling. Then, the web archive is cleaned and processed into a \textit{URL table} consisting of text and a link to an image. The table is then downloaded to a \textit{dataset} which then is used to train a machine learning \textit{model} that may be subsequently deployed.

With the data lifecycle in mind, we now define the various stakeholders that are involved with the content of machine learning datasets, and in our case, focusing specifically on image-text data from the web. We follow the roles from \citet{khan2022subjects} but also add additional actors in relation to DataComp CommonPool, of whom we argue are important players that have their own incentives and consequences. As displayed in \Cref{tab:stakeholders}, the stakeholders are separated into three stages: the Internet, the Dataset, and the Usage. None of these stakeholders are mutually exclusive from each other, as there may be substantial overlap.

\begin{table}
\begin{tabular}{@{}lll@{}}
\toprule
\textbf{Internet} & \textbf{Dataset}    & \textbf{Usage} \\ \midrule
Data subject      & Web archiver        & Dataset user   \\
Data owner      & Dataset curator     & Model user     \\
Data uploader     & Dataset annotator   & Model subject  \\
Site host         & Dataset distributor &                \\ \bottomrule \\
\end{tabular}
\caption{Overview of stakeholders by stage of data lifecycle. We demonstrate how stakeholders interact with each other in \Cref{fig:actor_tree}.}
\label{tab:stakeholders}
\end{table}

\subsection{Internet}
In this stage, the actors interact with data on the internet, completely divorced from any expectation the data will be used for downstream applications.

\subsubsection{Data subject}
The data subject is the individual who the data is about. For instance, this may be the person whose face is in the image, or the person whose address is in the caption of the photo. Privacy laws often center privacy in relation to the data subject due to the presence of their personal information.

\subsubsection{Data owner (copyright holder)}
The data owner is the person who typically creates the image (and thus likely owns the copyright to the image \citep{khan2022subjects}) -- this might be a photographer who takes a photo of the data subject. Data ownership may be transferable, so the current owner may not have created the data in the first place.

\subsubsection{Data uploader}
The data uploader is the person who uploads the data to the website, which, for instance, could be the photographer's company who uploads a picture online for marketing purposes. There can be a distinction here between data uploader and data subject: the data subject may have no knowledge nor given consent to their personal information uploaded onto the web by the data uploader.

\subsubsection{Site host}
The site host is the maintainer of the website that hosts the data once uploaded. They determine how the data is depicted and accessed, as well as setting the terms of service for how the data can be automatically scraped.

\subsection{Dataset}

This stage includes actors that are a part of the dataset creation process, to be released as a public artifact explicitly intended to train machine learning models.

\subsubsection{Web archiver}

The web archiver is the entity that crawls the entire internet to aggregate data in one place as a web archive. In CommonPool's case, this is Common Crawl, the nonprofit organization that provides the data source of the same name. These web archives are typically unformatted and not usable as a training dataset.

\subsubsection{Dataset curator}

The dataset curator is the actor that compiles the dataset based on the web dump from the web archiver. The curator must establish a process to select and format data from the web archive, in order to output a dataset suitable for training models.

\subsubsection{Dataset annotator}

The dataset annotator is the person who processes or adds relevant metadata to the data. The annotator may tag information manually, or rely on automated methods built by others. For CommonPool, the data annotator is equivalent to the dataset curator --- for instance, they rely on a face detection algorithm to mark the presence of faces. This is not always the case, as dataset compilers may outsource annotation instead \citep{le2023problem}.

\subsubsection{Dataset distributor}

The dataset distributor is the entity that is in charge of hosting and distributing the dataset as an artifact for others to download. CommonPool is hosted by Hugging Face, a popular dataset distribution platform.

\subsection{Usage}

This stage describes the players involved in the downstream usage of the dataset once it is released.

\subsubsection{Dataset user (model developer)}

The dataset user is someone who downloads the machine learning dataset with the intention of processing that data. This might be to develop a model, but could also be for other purposes, such as reformatting to produce another dataset, or searching through the dataset.

\subsubsection{Model user}

If a machine learning model is developed and released by the dataset user, then the model user is the person who interacts with the model. Given the general-purpose design of foundational models like text-to-image generators, there are many potential use cases for a given model.

\subsubsection{Model subject}

The model subject, as defined in \citet{khan2022subjects}, is the person who the model makes decisions about, which may have consequences on the person's life. The model subject may also be equivalent to the data subject of an image in the model's training dataset, but not necessarily. For example, the model subject may be a job applicant who submits an application screened by a machine learning model. 

\subsection{Stakeholder network}

\begin{figure}[ht]
    \centering
    \includegraphics[width=0.8\linewidth]{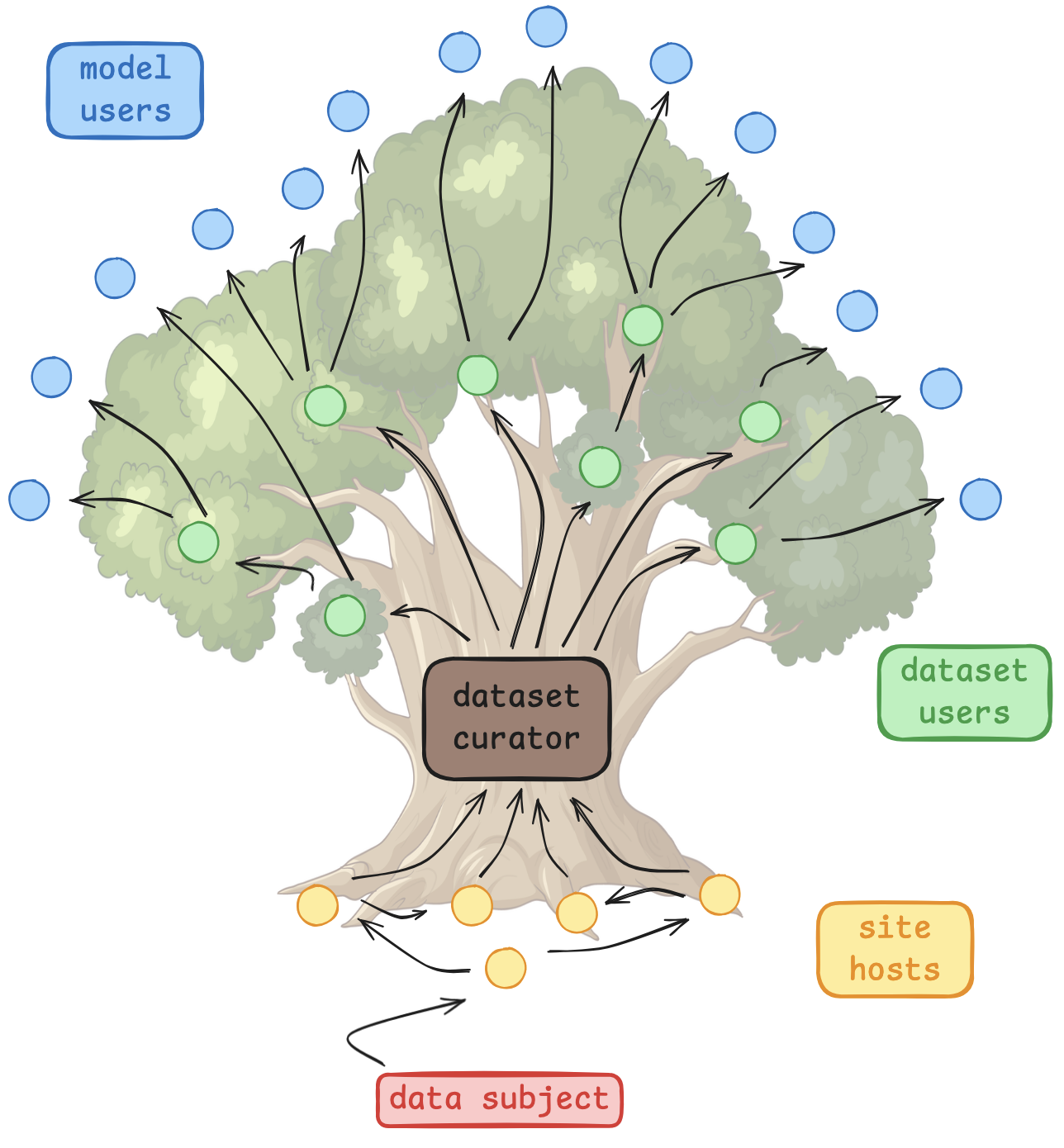}
    \caption{A high-level depiction of how personal information is aggregated by the \textit{dataset curator} and then dispersed to various \textit{dataset users} and \textit{model users}. See \Cref{fig:actor_tree} in the Appendix for a more detailed diagram. Background image source: Freepik \citep{freepik}.}
    \Description{Flow chart of nodes overlaid on top of a cartoon tree. Arrows start from the data subject to various site host nodes as the tree roots, which all point to the dataset curator at the tree base. The arrows then split from the dataset curator to various data users as the leaves of the tree, which then split off to the model user nodes.}
    \label{fig:actor_tree_small}
\end{figure}

We map the landscape of web-scraping for large-scale machine learning datasets in \Cref{fig:actor_tree_small}. We illustrate how stakeholders interact with each other to pass personal information from the \textit{data subject} to the \textit{model user} (and model subject). The diagram first depicts the various paths between \textit{site hosts} who may scrape and reupload personal information after its original upload by the \textit{data subject} (or data uploader). The \textit{dataset curator} (along with other actors in the Dataset stage) then aggregates data from the site hosts into a funnel, which then is dispersed to \textit{dataset users}, and further dispersed to \textit{model users}. The number of connections between the \textit{dataset curator} and downstream players is enormous, hence demonstrating the ``leaves'' reliant on the centralized source.

\section{Related Work}
\label{sec:related}

In this section, we present literature from both computer science and legal disciplines relevant to our work.

\subsection{Data collection}

We draw upon many prior audits of datasets to inform our approach. \citet{birhane2024sok} provide a comprehensive overview of the AI audit ecosystem and analyze the priorities of existing data audits. On the web-scraping side, \citet{dodge2021documenting} use keyword techniques and URL analysis to understand the text and websites within Common Crawl. Recent work has also audited the license and website terms of service restrictions of web-scraped machine learning datasets \citep{longpre2024consent, longpre2024large}. Several works examine the LAION-5B and DataComp data collection processes, mostly revolving around sexually explicit content, toxicity, and bias \citep{birhane2021multimodal, birhane2024into, thiel2023identifying, hong2024s}. \citet{diaz2024sound} provide a general framework to comprehensively assess representation in unstructured data like images; our work further explores the first component of their framework by examining the presence of people in unstructured datasets.

There has also been extensive work examining data curation practices from a sociotechnical lens. \citet{paullada2021data} provide a broad survey of dataset collection in machine learning research, while \citet{desai2024archival} incorporate archival studies to examine datasets --- both of which argue the need to analyze the contents of datasets and choices in assembling them. To do so, recent works have traced the values, assumptions, politics, and histories of machine learning datasets \citep{birhane2022values, scheuerman2021datasets, denton2021genealogy}. We are inspired by these critiques in our work to better understand and define web-scraping practices.

\subsection{Privacy concerns}

Researchers have demonstrated that machine learning models are susceptible to leak specific information due to memorization during training 
\citep{brown2021memorization}. Various attacks, for instance, can extract potentially personal information from large language models \citep{nasr2023scalable, lukas2023analyzing}, while on the vision side, \citet{carlini2023extracting} generate over a thousand training examples at test time from various diffusion models. Fine-tuning diffusion models can also amplify the leakage of memorized training samples \citep{li2024shake}.

To understand the privacy risks associated with a model that can output its training data, it is necessary to determine the private information present in a dataset in the first place. \citet{dou2023reducing} develop a taxonomy of various online self-disclosures, where users communicate personal information in text, and \citet{mireshghallah2024trust} use this taxonomy to find presence of personal information in user interactions with ChatGPT. For vision, most work has focused around image classification tasks like face detection or license plate detection to automatically blur regions to preserve privacy \citep{yang2022study, zhou2012principal}. Other work has focused on manually annotating more nuanced privacy concerns present in images to train models to identify privacy risks \citep{orekondy2017towards, orekondy2018connecting}. We incorporate several of these personal information detection techniques to understand the risks of DataComp CommonPool.

\subsection{Legal literature review}
\label{sec:legal_lit}

Large-scale web-scraped datasets raise long-standing privacy issues in new forms. Scholars have noted that many privacy risks posed by artificial intelligence (AI) and big data are not fundamentally novel, but rather ``remixes'' or amplifications of existing problems \citep{solove2024artificial}. 

\subsubsection{Individual control}
\label{sec:legal_lit:opt_out}
A key concern is the inadequacy of the prevailing individual control model of privacy, often termed ``privacy self-management.'' \citep{solove2013privacy} Under this model, individuals are expected to read notices and consent to data practices (or opt out), supposedly empowering them to manage their own privacy, like the mechanisms detailed in \Cref{sec:datacomp_opt_out} and \Cref{sec:laion_opt_out}. In practice, this notice-and-consent regime has been widely critiqued as ineffective and even ``farcical'' \citep{ tschider2020meaningful, richards2018pathologies, hartzog2018case, waldman2014privacy, hirsch79individual}. Solove argues that people cannot meaningfully control personal data in an AI-driven and data-intensive environment: the scale and opacity of data collection and algorithmic processing far exceed individuals' ability to understand or consent \citep{solove2024artificial}. Privacy self-management, as Solove put it, is beset by a ``consent dilemma,'' individuals face too many notices and hidden inferences to practically make informed choices \citep{solove2013privacy}. This critique, echoed by experts, underscores that reliance on individual consent is an inadequate safeguard in the age of web-scraped AI datasets \citep{ solove2024artificial, solove2024great, solove2024murky, solove2024kafka}.

\subsubsection{Inference}
Compounding the consent problem is the issue of data inference and generation. AI systems not only collect vast amounts of data as input, but also generate new data about individuals via inference \citep{solow2022information, viljoen2021relational, wachter2019right}. Machine learning algorithms can predict sensitive facts about people that they never directly revealed, blurring the line between data ``provided'' and data ``produced.'' Solove observes that inference allows companies to end-run traditional privacy protections: laws typically regulate the collection of personal data from individuals, but if algorithms can create personal data (e.g., a prediction of someone’s pregnancy or political views) from other information, those laws offer little direct control \citep{ solove2024artificial}. A famous example is Target’s analytics identifying a teen’s pregnancy from mundane purchasing data. Such inferences ``upend the traditional picture'' of privacy management, because people cannot anticipate or prevent the creation of sensitive data about them \citep{solove2024artificial}.  Scholars like \citet{ solow2022information} have termed this the “inference economy,” noting that individuals have no practical way to opt out or correct the myriad inferences drawn about them. Even inaccurate inferences can be harmful, yet privacy frameworks offer little recourse for data generated without one’s knowledge. This literature suggests that privacy law must expand its focus beyond the moment of data collection to address downstream inferencing and profiling \citep{solow2022information, viljoen2021relational, wachter2019right}.

\subsubsection{Data minimization and purpose limitation}
Another body of relevant work concerns core data protection principles, such as data minimization and purpose limitation, and how they clash with the big-data practices behind massive datasets. Data protection laws traditionally require that organizations collect only the personal data that is necessary for a specified purpose, and not repurpose it for incompatible uses. Large-scale AI datasets assembled via indiscriminate web scraping flout these principles dramatically \citep{solove2024artificial, solove2024great}. Solove and Hartzog argue that scraping ``ignores'' virtually all fair information practice principles: data is collected broadly without notice or consent, without a specified purpose, retained without regard for necessity, and used for new purposes (AI training) never contemplated by the original context \citep{ solove2024artificial, solove2024great}. This critique builds on previous scholarship recognizing that “Big Data” often demands maximal collection and open-ended future use, directly at odds with minimization and purpose limitation. For example, \citet{tene2012big} observed that organizations were beginning to effectively ``collect everything, just in case'' to exploit the potential of data, making it difficult to honor purpose restrictions or deletion obligations. DataComp CommonPool’s design --- billions of items gathered ``just because'' they may improve machine learning, exemplifies what Solove and others identify as a profound tension between big-data analytics and the foundational privacy principle of collecting the least data needed. In effect, large web-scraped datasets treat personal data as an unlimited raw resource, whereas privacy scholarship insists on data frugality and contextual integrity.

\subsubsection{Publicly available data}
Crucially, researchers have challenged the assumption that ``publicly available'' personal data are free of privacy interests. Many web-scraping efforts defend themselves by noting that the data was already public on the Internet. However, interdisciplinary scholarship has long rejected a simplistic ``secrecy paradigm'' which equates privacy solely with complete secrecy \citep{nissenbaum2020protecting, hartzog2019public}. Helen Nissenbaum’s theory of contextual integrity, for instance, posits that privacy is defined by appropriate information flow within context-specific norms, not by whether information is public or private in an absolute sense \citep{nissenbaum2011contextual}. It is well established that people maintain privacy expectations even in public arenas. They may share information on a personal blog or forum for a specific audience or purpose, yet still reasonably object to that data being mined en masse for unrelated uses. As Solove explains, individuals often disclose personal data in limited settings; privacy encompasses the ability to limit the audience and purpose of that disclosure \citep{solove2024artificial}.

The notion of privacy in public is supported by legal scholars as essential for freedom and democracy \citep{reidenberg2014privacy, nissenbaum2020protecting, hartzog2019public}. One key concept is “practical obscurity”: even if data is technically accessible, it may be difficult to find, scattered, or fleeting, which gives individuals a measure of obscurity that protects their privacy \citep{hartzog2015surveillance}. When scrapers aggregate and centralize such data, they destroy this practical obscurity – effectively a privacy loss even though the data was public before \citep{solove2024artificial}. \citet{hartzog2019public} has termed the naive belief that publicly available data is harmless the ``public information fallacy.'' Indeed, privacy law itself historically recognizes interests in public information (for example, the tort of appropriation protects against misuse of one’s public name or likeness). The emerging consensus in scholarship is that the context and method of data use matter: personal data scraped from the open web is not per se exempt from privacy concerns. \citet{solove2024artificial} and \citet{ hartzog2019public} both argue that privacy frameworks must ``safeguard obscurity'' and place limits on the unfettered harvesting of personal data from the internet.

\subsubsection{Web scraping}
Finally, there is growing interdisciplinary work examining web scraping and privacy harm. \citet{solove2024great} characterize web-scraping as a direct clash with privacy norms. They document how companies like Clearview AI have scraped social media photos to build facial recognition databases, actions that regulators around the world deemed unlawful and harmful to privacy. The Clearview incident, resulting in multimillion-dollar fines for violating data protection laws, is frequently cited as a cautionary example of treating ``public'' personal data as fair game. In the AI context, massive text and image datasets have been assembled by scraping platforms like Twitter, Reddit, Flickr, and personal websites without consent. This practice has been denounced in legal and ethics literature for sidelining individual autonomy and data subject rights. Scholars emphasize that scraping undermines nearly every element of the modern privacy toolkit: individuals typically do not receive any notice their data is taken, do not consent, cannot opt out, and often cannot even exercise rights like deletion or correction because the scraper may remain unknown to them \citep{ solove2024great}. In summary, the relevant literature paints a stark picture: large-scale dataset compilers are operating in a legal and ethical gray zone, relying on outdated notions of public data and consent. Foundational principles (data minimization, purpose limitation) are being overridden by the imperatives of ``more data at any cost.'' Commentators call for a reconceptualization of privacy law to address these challenges, shifting away from exclusive reliance on individual consent, and imposing accountability on data collectors and AI developers to respect privacy constraints even when dealing with public or inferred data.

\subsection{Privacy law review}
\label{sec:privacy_law}

\subsubsection{EU General Data Protection Regulation (GDPR)}
The GDPR provides a comprehensive privacy framework that is highly relevant to web-scraped datasets. The GDPR applies to any “controller” or “processor” who processes personal data in the context of an EU establishment, or who processes data of individuals in the EU for purposes of offering them goods/services or monitoring their behavior (Article 3). This broad jurisdictional reach means that even non-EU entities can be subject to GDPR if they scrape or use personal data from EU residents in a way that qualifies as monitoring or offering services. In practical terms, if the CommonPool dataset includes information about EU persons (highly likely given its web-scale), any entity using that data in a manner targeting the EU or involving EU operations would need to comply with GDPR requirements. There is no monetary or size threshold for GDPR coverage – it applies irrespective of company size, so long as the activity falls within its scope. 

The GDPR defines ``personal data'' very broadly as ``any information relating to an identified or identifiable natural person.'' This definition easily encompasses CommonPool’s contents: for example, an image of a person’s face or a snippet of their resume ``relates to'' an identifiable individual (even if names are not explicitly included, identifiability can be inferred from context or by combining data). Notably, unlike some U.S. laws, the GDPR does not exempt publicly accessible personal information from its scope, if the data relates to an individual, it is protected, regardless of source.  Even then, processing must occur on a proper legal basis. The GDPR also recognizes certain categories of ``special'' (sensitive) personal data that merit heightened protection. Article 9 enumerates sensitive data such as racial or ethnic origin, health information, biometric data processed for identification purposes, sexual orientation, political or religious beliefs, and information about children. Processing these special categories is generally prohibited unless a specific condition is met (such as explicit consent), including the condition of ``personal data which are manifestly made public by the data subject,'' but this is a narrow carve-out applicable mainly to the data subject’s own deliberate public disclosures. This condition does not cover all special category data in the public domain. It only covers personal data that the individual themselves has made public. In the context of DataComp, any photos revealing race or health traits, biometric identifiers (faces used for recognition), or data about children would fall under these special categories, requiring rigorous justification. 

Another core concept in GDPR is purpose specification and limitation: personal data must be collected for ``specified, explicit and legitimate'' purposes and not further processed in incompatible ways (Article 5(1)(b)). Similarly, the principle of data minimization mandates that only data which is “adequate, relevant and limited to what is necessary” for the stated purpose should be collected (Article 5(1)(c)). These principles directly speak to the DataComp scenario: repurposing people’s information from the web for AI training (a new purpose) would typically require a fresh legal basis, and collecting 12.8 billion data points ``just in case'' would seem to violate the necessity limitation. However, the GDPR does include some contextual exceptions: for example, purely personal or household use of data is exempt (Article 2), and there are allowances for scientific or statistical research that might relax certain obligations (with strict conditions and safeguards). 

Overall, GDPR sets a high bar: it prescribes legal grounds for processing (consent, contractual necessity, legal obligation, vital interests, public interest, or legitimate interests – Article 6), requires transparency to data subjects (Articles 13–14), grants individuals robust rights (access, deletion, objection, etc.), and mandates security and breach notification (Articles 32–34). If a web-scraped dataset contains personal data, a GDPR-regulated entity handling it must navigate all these obligations, unless the data can truly be anonymized such that no individual is identifiable (a standard the law and EU regulators interpret very strictly).

\subsubsection{California Consumer Privacy Act (CCPA, as amended by the CPRA)}
California’s privacy law is the first comprehensive state privacy regime in the U.S. The CCPA, as amended by the CPRA, imposes obligations on covered ``businesses.'' A business is defined generally as a for-profit entity doing business in California that meets certain thresholds: (a) annual gross revenues over 25 million; or (b) buys/sells or shares personal information of 100,000 or more California consumers or households; or (c) derives 50 percent or more of annual revenue from selling or sharing personal information. These criteria limit the law’s application to larger data handlers. For example, an academic or non-profit entity that compiled CommonPool might not be a ``business'' under CCPA, but a large tech company downloading and using it likely would be. The CCPA grants California residents rights over their personal information held by businesses, including the right to know what data is collected, to delete data, to opt out of its sale or sharing, and to non-discrimination for exercising rights. The CPRA added a right to correct inaccurate data and to limit use of ``sensitive personal information.''

The CCPA defines “personal information” in expansive terms but explicitly carves out ``publicly available'' data. Under the CCPA (2018) and its amendment via the California Privacy Rights Act (effective 2023), personal information means any information that ``identifies, relates to, or could reasonably be linked with'' a particular consumer or household. This would include typical identifiers (names, emails), internet activity data, biometric information, geolocation, and even inferences drawn about preferences or characteristics. Crucially, however, the CCPA’s coverage excludes ``publicly available information.'' Initially, ``publicly available'' was narrowly defined to mean data lawfully made available from government records. The CPRA expansion broadened this definition: now it also includes information a business has a reasonable basis to believe was lawfully made public by the individual or through widely distributed media. In other words, personal data that the consumer themselves made public , or which is in public news or media, may fall outside CCPA’s definition of regulated personal information. This could potentially exempt large swathes of DataComp content; for instance, images and text that people posted on public forums or social media without restrictive privacy settings might be considered ``publicly available'' under California law. Notably, the law excludes data that the consumer has restricted to a specific audience, and it excludes usage that is not aligned with the data’s purpose of publication. Under the law, “publicly available” also does not mean biometric information collected by a business about a consumer without the consumer’s knowledge.  These nuances mean the exemption is not absolute.   

Sensitive personal information under CPRA is a subset of personal data including items like Social Security or driver’s license numbers, financial account details, precise geolocation, racial or ethnic origin, union membership, contents of private communications, genetic data, and biometric identifiers, among others. For such sensitive data, businesses must disclose if they collect it and honor consumers’ requests to limit its use to what is necessary to provide the requested services. In the DataComp context, any scraped data like full credit card numbers, government IDs, or precise location coordinates would qualify as sensitive; California consumers could demand that businesses cease using those items beyond core functions. It’s important to note that the CCPA’s publicly-available exemption also applies to sensitive data: e.g., a person’s publicly posted phone number might not be protected as ``personal information,'' and likewise their race or religion if obviously made public by them could be deemed ``public.'' However, the presence of children’s data triggers additional rules: the CCPA requires opt-in consent (through a parent for under 13, or the minor’s consent if 13–15) before selling personal information of minors. While DataComp researchers are not ``selling'' data, any downstream commercial use of children’s personal data could implicate these protections. 

Enforcement of CCPA/CPRA is primarily by the California Privacy Protection Agency and state Attorney General (individuals have a limited private right of action for data breaches). Businesses in possession of DataComp-derived personal information would need to provide notice in their privacy policy about categories of personal info collected (potentially listing data obtained from third-party sources like web scraping), and honor deletion or opt-out requests if a California resident somehow identified their data in the dataset. In practice, exercising rights on scraped data is challenging, but the legal framework puts the onus on the business to comply where possible.

\subsubsection{Oregon Consumer Privacy Act (OCPA)}
Enacted in 2023 and effective July 1, 2024, the OCPA is part of the new wave of U.S. state privacy laws. The OCPA applies to ``controllers'' and ``processors'' meeting threshold criteria. Uniquely, Oregon’s law has no revenue threshold for applicability. It applies to any entity that conducts business in Oregon or targets products/services to Oregonians, provided that it controls or processes the personal data of at least 100,000 Oregon consumers in a year (excluding purely payment data), or of at least 25,000 consumers if deriving over 25 percent of revenue from selling personal data. This thresholds test means the law mainly catches mid-size and large data handlers. Notably, the OCPA does not exempt non-profits, making it broader in coverage than CCPA. By July 2025, many non-profit organizations will also be subject to Oregon’s requirements. 

The OCPA defines ``personal data'' as information that is linked or reasonably linkable to an identified or identifiable individual (a ``consumer'' who is an Oregon resident). Importantly, the definition excludes de-identified data and ``publicly available information.'' The statute regards data as publicly available if it is lawfully made available from government records or widely distributed media, or if the individual made the information public (in line with the CPRA’s broader approach). Thus, similar to California, Oregon’s law might exempt certain categories of CommonPool data from regulation on the premise that they were publicly accessible online. That said, OCPA’s exact definition hews closely to the individual’s intent and the nature of distribution; not everything on the internet would automatically count as ``public'' under the law’s terms. Assuming CommonPool contains typical web content, much of it could be argued to be publicly available (e.g. images from public websites), and thus outside OCPA’s scope of ``personal data.'' OCPA’s definition of sensitive data includes personal data revealing racial or ethnic origin, religious beliefs, sexual orientation, status as transgender or non-binary, immigration status, health information, genetic or biometric data, precise geolocation (within a 1,750-foot radius), and any personal data of a known child (under 13).

Key consumer rights under OCPA include the right to confirm if a controller is processing one’s data, to access a copy, to correct inaccuracies, to delete personal data, and to opt out of targeted advertising, sales of data, or certain profiling decisions. There is also a requirement to honor browser opt-out signals for selling or targeted ads. The law imposes several obligations on controllers that align with GDPR-like principles: data minimization (collect only what is ``adequate, relevant, reasonably necessary, and proportionate'' to the purposes disclosed), purpose specification (process data only for purposes that are disclosed and reasonable), and reasonable security measures. Notably, if processing ``sensitive data,'' the controller must obtain the consumer’s opt-in consent.  This means if DataComp CommonPool contains, say, images of children or data about minors, or biometric identifiers like facial scans, an Oregon-covered controller would legally need parental consent (for under 13) or the individual’s consent (for other sensitive data) before processing that data. 

In addition, OCPA mandates transparent privacy notices detailing categories of data collected, purposes of processing, categories of data shared and with whom, and how consumers can exercise their rights. It also requires controllers to conduct and document Data Protection Assessments for certain high-risk processing, such as processing sensitive data or any processing for targeted advertising, sale, or profiling that presents a significant risk of harm. 

The OCPA, enforced by the state Attorney General, thus creates a compliance regime similar to other state laws but with its own nuances (like no exemption for nonprofits and a consent requirement for all sensitive data use). For a company leveraging DataComp CommonPool, if that company has a user base or market in Oregon (or otherwise falls under OCPA), it would need to treat any personal data in the dataset in accordance with these rules – unless it can argue the data is outside the law’s scope (e.g. truly de-identified or public information).

In summary, all three frameworks (GDPR, CCPA, OCPA) share a broad view that personal data covers any identifiable information about individuals, which certainly includes much of DataComp CommonPool. The GDPR is the most encompassing, applying to essentially all personal data and imposing strict principles and rights. CCPA and OCPA similarly cover a wide range of personal information but carve out publicly available data and apply only to entities meeting certain thresholds. Each has special provisions for sensitive categories of data (especially data about children, biometric identifiers like faces, and financial or health information) and expects data handlers to practice data minimization, purpose limitation, and data security. As \Cref{sec:legal_analysis} will analyze, the presence of personal and sensitive information in CommonPool triggers these legal frameworks --- raising questions about whether those compiling or using such datasets can meet the legal obligations, and whether current exceptions (like ``public data'' loopholes) undermine privacy in practice.

%% file: sections/03-methodology.tex
\section{Audit methodology}
\label{sec:audit_methods}

We now describe the methods we used for our privacy audit, which is inspired by similar audits of web-scraped datasets that inspect images \citep{birhane2024into}, text \citep{mireshghallah2024trust}, and the websites that host the samples \citep{dodge2021documenting, hong2024s}.
Our audit is motivated by various legal definitions of personal and sensitive information \citep{sensitive_data, quinn2021difficulty} under various state and federal privacy laws, like the California Consumer Privacy Act (CCPA) \citep{ccpa} and the General Data Protection Regulation (GDPR) \citep{gdpr}. In addition, we ground our audit in ethical frameworks of the act of information collection and processing \citep{solove2005taxonomy} and user privacy perceptions of using web data based on context and consent \citep{fiesler2018participant, zimmer2018addressing}, which can include widely disseminated media like celebrity images or news articles.

To distinguish from the legal term of ``personal data,'' our audit examines personal information that is \textit{identifying} --- in the case of image and text, samples where a face or name is present. This data may not necessarily be explicitly private nor considered personal data in the legal sense (i.e. celebrity names), but we choose to focus on identifying information to (i) uncover who is represented in this dataset and how they are represented, (ii) determine if such data may be under the ambit of privacy law even if accessible on the web (as expanded upon in \Cref{sec:legal_analysis:limited_audience}), and (iii) show that indiscriminate scraping cannot respect contextual integrity \citep{nissenbaum2011contextual} (as discussed in  \Cref{sec:discussion:tracing}). Our audit explicitly refers to ``personal information'' rather than the term ``personal data,'' as that term is a legal definition that depends on the relevant privacy law, where personal information that is widely accessible may not be considered ``personal data'' in some cases.

To conduct our audit, we downloaded DataComp CommonPool in the month of April 2025, following their code package with parameters set to default in the way it is intended to be downloaded \citep{datacomp_github}. Due to space constraints, we download the \texttt{small} scale version of CommonPool, which consists of 12.8 million randomly selected samples. Because this subset is only $0.1\%$ of the entire dataset (and even at this scale still challenging to examine every sample), we use our observations to estimate 
quantities of information present in all 12.8 billion samples of CommonPool, accompanied by
confidence intervals which quantify the probable estimation error due to sampling. We do not aim to capture \textit{all} possible privacy concerns in our audit, but rather establish a lower bound based on our various approaches to inform our legal analysis. As we describe further in \Cref{sec:legal_analysis}, the determination of whether privacy laws are triggered is not always based on scale; the presence of sensitive content alone can be enough to inform our legal analysis. Thus, to demonstrate this presence, throughout our audit we surface individual images as case studies for legal implications.

\subsection{Audit techniques}

We use a variety of tools to understand the privacy concerns of web-scraped datasets. The techniques that are specific to search categories, such as sociodemographic information or children's information, we define in \Cref{sec:audit}. In this section, we highlight the study-wide techniques incorporated throughout our analysis of the contents of CommonPool.

\subsubsection{Optical character recognition}
\label{sec:ocr}

To examine the contents of scraped images, such as documents or screenshots, we use optical character recognition, or OCR, to extract the text that is contained in every image. Prior OCR comparisons \citep{vedhaviyassh2022comparative} are difficult to extend to this dataset, as images on the web may be of lower quality or depict non-document text and therefore likely follow a different distribution than these benchmarks. As a result, we perform an evaluation of various popular open-source OCR methods on a random subset of CommonPool samples in \Cref{sec:appx_ocr_eval} and determine that PaddleOCR \citep{ppocr} is most effective. We defer to \Cref{sec:appx_text_viz} for an overview of the OCR-extracted text and captions.

\subsubsection{Entity extraction}

To surface examples with personal information, we apply Microsoft Presidio's PII detection tool (version 2.2.357) to the captions and OCR-extracted text of CommonPool \citep{presidio}. Presidio's recognizers incorporate regular expression matching and named entity recognition \citep{nadeau2007survey} to find sensitive data like credit card numbers, social security numbers, and individual names, as shown in \Cref{tab:presidio_pii}. As in prior work \citep{mireshghallah2024trust} we find errors in detected PII entities, so we therefore use Presidio to flag content as \textit{possibly} containing PII. Dependent on our search category (for example, sociodemographic dimensions in \Cref{sec:audit_text} or identity documents in \Cref{sec:audit_img}), we then manually inspect the flagged content and only the samples that satisfy our criteria are included in our counts. We additionally flag content using basic keyword searches from prior works \citep{birhane2021multimodal, dodge2021documenting} and also manually inspect matches for inclusion in our counts.

\begin{table}
\begin{tabular}{@{}lll@{}}
\toprule
\textbf{PII Entity}    & \textbf{Caption} & \textbf{OCR} \\ \midrule
\textbf{Name}          & 1.3M             & 3.3M         \\
\textbf{Address}       & 370K             & 1.4M         \\
\textbf{Date Time}     & 500K             & 880K         \\
\textbf{Demographics}  & 86K              & 240K         \\
\textbf{Email}         & 16K              & 3.0K         \\
\textbf{Medical}       & 8.9K             & 8.2K         \\
\textbf{URL}           & 5.2K             & 2.4K         \\
\textbf{Government ID} & 4.1K             & 2.8K         \\
\textbf{Business ID}   & 3.0K             & 2.3K         \\
\textbf{Financial ID}  & 1.8K             & 1.2K         \\
\textbf{IP Address}    & 494              & 22           \\
\textbf{Vehicle ID}    & 171              & 133          \\ \bottomrule \\
\end{tabular}
\caption{Sample counts of Presidio-detected PII entities in captions and OCR-extracted text of \texttt{small} scale dataset (12.8 million samples, or 0.1\% of CommonPool). Upon manual inspection, many of these detections are false positives.}
\label{tab:presidio_pii}
\end{table}

\subsubsection{URL analysis}
Because CommonPool is released as an index of URL-caption pairs, we evaluate
the URLs storing the images as well as the 
image and text content of the samples. We follow similar URL analysis from recent audits \citep{dodge2021documenting, hong2024s} to assess the website category and earliest recorded timestamp of the URL. Source analysis helps us understand how certain kinds of PII may have been uploaded onto the internet.

\subsection{Limitations}
\label{sec:limitations}

In conducting an audit on a dataset of this size, our methodology may have certain limitations, some of which are inherited from tools we use. These limitations also extend to addressing privacy concerns via automated cleaning methods in the curation and usage of web-scraped training datasets more broadly, which we highlight in \Cref{sec:discussion}.

\paragraph{False Positives and Mitigation} Certain algorithms like OCR or URL categorization may make incorrect predictions, so these predictions cannot be treated as ground-truth. We also observe that Presidio's PII detection tool flags random sequences of numbers or letters as identification numbers or financial accounts. Due to these errors, we manually inspect the samples flagged by either PII entity recognition or OCR-based keyword search and confirm which samples contain personal information.

\paragraph{False Negatives} We additionally recognize that our audit will miss certain kinds of information. PII detection tools do not capture all personal information, especially nuanced content or text that does not match regular expressions \citep{mireshghallah2024trust, xin2025false}. For annotation of text, our analysis is focused on the English language, and expanding privacy audits to include non-English languages is an important avenue for future work. As a consequence, reliance on PII tools and the use of manual inspection constrains the scale of our audit to the millions rather than billions. As we discuss in \Cref{sec:discussion:ml}, our work speaks to the challenges of building web-scraped datasets more broadly at scales in which every sample can no longer be individually examined.

\paragraph{Selection bias.} Given that our audit examines the \texttt{small} scale of the dataset ($0.1\%$ of CommonPool), we acknowledge concerns with selection bias of this random subset. For instance, we investigate in \Cref{sec:audit_unavailable} that certain images are no longer available, and duplicate images in the \texttt{small} scale may not be representative of all of CommonPool, especially the tails. While our findings do not allow us to extrapolate exact estimates to the entire dataset, auditing 12.8 million samples is large enough to give non-vacuous confidence intervals even with hundreds of samples. We focus on the \textit{presence} of personal information and how these measurements may translate to larger numbers for a dataset in the billions.

\subsection{Ethical considerations}
\label{sec:ethics}

Our institution's IRB did not consider this study to involve human subjects research due to the dataset being collected from the web, including the online presence validation approach described in \Cref{sec:audit_resume}. Nevertheless, as IRB approval alone is not sufficient to guarantee that a study is ethical \citep{abbott2011systematic, heimer2010bureaucratic}, we carefully considered ethics throughout our study, beginning with study conception. Given the sensitive nature of our study that can reveal personal information, we store images and indirect identifiers to these images on a secure server. In our results in \Cref{sec:audit}, we aggregate all measurements, carefully anonymize examples to preserve privacy, and ensure that searching our redacted text on the web or the dataset does not return the actual samples.

The ethical tensions of studying public data that may violate individual privacy have long been discussed in social computing and computer security research \citep{bruckman2002studying, zimmer2018addressing, kohno2023ethical}.
We follow best practices from prior works on Internet user perceptions of the use of their data for research \citep{fiesler2018participant, dym2020ethical}. To do so, in presenting our work we obfuscate personal information and rewrite verbatim text in individual case studies, such that the image and caption cannot be directly retrieved. We also follow the level of heavy disguise from \citet{bruckman2002studying} and deliberately introduce false details so that privacy concerns are demonstrated in spirit without allowing the data subject to be recognized.

We also address various potential ethical implications of our work. (1) Our methods may be easily replicated, yet this set of 12.8 billion URLs has already been crawled over two million times \citep{datacomp_downloads}. (2) Our privacy audit may indirectly encourage future machine learning systems to become even less transparent for fear of legal risk, but prior work has already observed the lack of transparency of model training \citep{hardinges2024we}. (3) There may also be second-order effects, as removing personal information may alleviate legal risk, although recent work shows that PII removal may lead to model memorization of leftover PII \citep{borkar2025privacy}. 

We investigate the privacy implications of DataComp CommonPool to raise awareness to the degree to which privacy concerns and legal risks may arise in web-scraped data in general. We acknowledge that the datasheet for CommonPool references the presence of sensitive data and clarifies intention as a research artifact \citep{gadre2024datacomp}, yet CommonPool's licensing does not restrict the commercial deployment of models trained on this dataset --- speaking to the difficulty of regulating the use of web-scraped data in general. In \Cref{sec:discussion}, we expand upon these ethical considerations for future dataset use cases as well as the tension between publicly available data and human subjects research.

%% file: sections/04-audit-results.tex

\section{Audit Results}
\label{sec:audit}

\begin{table*}
\resizebox{\linewidth}{!}{%
\begin{tabular}{@{}lll@{}}
\toprule
\textbf{Modality} & \textbf{Search category} & \textbf{Results} \\ \midrule
\textbf{Text (\ref{sec:audit_text})} &
  \begin{tabular}[c]{@{}l@{}}1. Sociodemographics\\ 2. Celebrities \end{tabular} &
  \begin{tabular}[c]{@{}l@{}} We find captions that disclose the full name along with sexual orientation, religion, race, or ethnicity.\\ Individual names mentioned in samples refer to celebrities mostly from the U.S. and U.K. \end{tabular} \\ \midrule
\textbf{Image (\ref{sec:audit_img})} &
  \begin{tabular}[c]{@{}l@{}}1. Identity documents\\ 2. Resumes\end{tabular} &
  \begin{tabular}[c]{@{}l@{}}We find credit cards, drivers licenses, social security  numbers, passports, and birth certificates.\\ We estimate at least $136,000$ images depict resumes of individuals with public online presence.\end{tabular} \\ \midrule
\textbf{URL (\ref{sec:audit_url})} &
  \begin{tabular}[c]{@{}l@{}}1. Children’s information\\ 2. Unavailable images\end{tabular} &
  \begin{tabular}[c]{@{}l@{}}We find children's names, faces, and birth certificates, passports, and health status. \\ Of the $21.4\%$ of links that fail to download, $19.0\%$ of those links fail due to lack of access permissions. \end{tabular} \\ \midrule
\textbf{Metadata (\ref{sec:audit_metadata})} &
  \begin{tabular}[c]{@{}l@{}}1. Image Exif tags \\ 2. Face bounding boxes\\ \end{tabular} &
  \begin{tabular}[c]{@{}l@{}}We find that Exif tags attached to images reveal full names and precise geolocations.\\We estimate at least 100 million images of human faces are not covered by bounding boxes.\end{tabular} \\ \bottomrule \\
\end{tabular}%
}
\caption{Summary of audit findings by data modality and search category for the April 2025 download of DataComp CommonPool.}
\label{tab:results_summary}
\end{table*}

When downloaded, each sample in the CommonPool URL-caption artifact contains various components: the caption which contains \textbf{text}, the \textbf{URL} which upon downloading gives the \textbf{image} (and may contain additional text extracted through OCR), and accompanying \textbf{metadata} relating to the image. To organize our results, we structure our audit into four sections based on the data modality of our search; each section is narrowed down by search category as motivated by existing privacy laws:
\begin{enumerate}
    \item \Cref{sec:audit_text} highlights sociodemographic information and presence of celebrities found in \textbf{text} (both captions and OCR-extracted text).
    \item \Cref{sec:audit_img} covers identification documents and resume documents visually presented in the downloaded \textbf{images}.
    \item \Cref{sec:audit_url} surfaces platforms relating to children's information as well as samples that are no longer available, based on the \textbf{URLs}.
    \item \Cref{sec:audit_metadata} demonstrates issues relating to the image Exif tags and face bounding boxes in the attached \textbf{metadata}.
\end{enumerate}

In our analysis, there is substantial overlap in modality, as we obtain OCR-extracted text from the images, investigate the URLs of the sociodemographic information, or use text keyword search to find visual documents. However, we ascribe each search category to its main data component (for instance, documents are verified upon visual inspection, not via text) and emphasize this groupings is for the purposes of organization rather than a contribution in and of itself. \Cref{tab:results_summary} gives an overview of our findings split by data modality.

\subsection{Text}
\label{sec:audit_text}

We search for query keywords in the captions or OCR-extracted \textbf{text} of samples to find matches that may contain PII. As described in \Cref{sec:audit_methods}, we surface both measurements and individual case studies to inform legal analysis in \Cref{sec:legal_analysis}. In this section, we describe findings that mention sociodemographic information related to individuals as well as the presence of celebrity names. \Cref{sec:audit_img} later covers personal information that may appear in particular types of documents. 

\subsubsection{Sociodemographic information}
\label{sec:audit_sociodemographics}

\begin{figure}
    \centering
    \includegraphics[width=\linewidth]{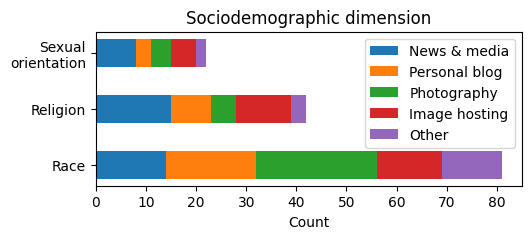}
    \caption{Number of annotated samples that link a name with sociodemographic information. Each bar represents the sociodemographic query described in \Cref{sec:audit_sociodemographics} broken down by website type, which is categorized manually or verified via Cloudflare \citep{cloudflare} for the \texttt{Image hosting} category.}
    \Description{Graph of three bars for different sociodemographic dimensions of sexual orientation, religion, and race, split by website type. The bar for sexual orientation has 22 samples, religion has 42 samples, and race has 81 samples, with news and media websites being a very common source.}
    \label{fig:sociodemographic_pii}
\end{figure}

\begin{figure*}
    \centering
    \includegraphics[width=\linewidth]{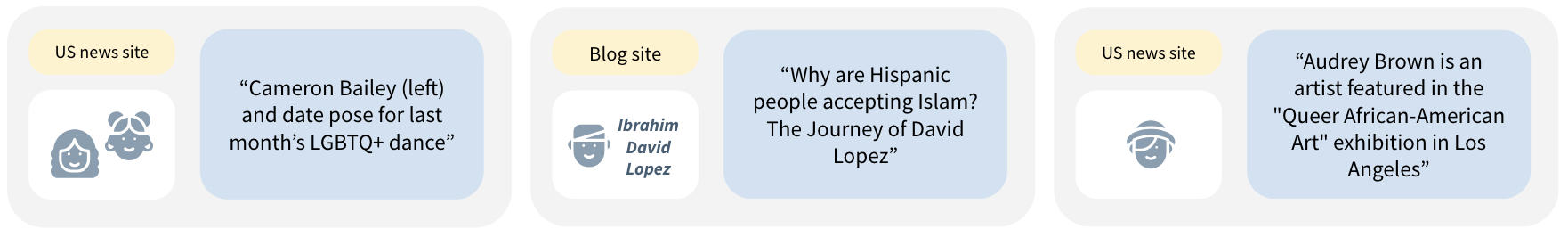}
    \caption{Examples of identifying sociodemographic information found in CommonPool's small scale dataset. For each sample, the type of URL site is shown at the top left, the image in the bottom left, and the caption in quotes on the right. All personal information has been replaced, and text has been paraphrased to avoid direct quotations. Images have been redacted to show the presence of faces without identifying the individuals.}
    \Description{Three example instances of image, website source, and caption. On the left: an image of two faces from a U.S. news site with caption stating "Cameron Bailey (left) and date pose for last month's LGBTQ+ dance." In the middle, an image of a face with the name "Ibrahim David Lopez" from a blog site with caption stating "Why are Hispanic people accepting Islam? The Journey of David Lopez." On the right, an image of a face from a U.S. news site with caption stating "Audrey Brown is an artist featured in the 'Queer African-American Art' exhibition in Los Angeles."}
    \label{fig:sociodem_examples}
\end{figure*}

\paragraph{Approach} Presidio's named entity recognition \citep{presidio} first flags samples with names mentioned in the OCR-extracted text or caption. To narrow down samples, we manually discard names that do not consist of two words, as well as names of cartoons like ``Peter Pan'' or historical figures like ``George Washington.'' Among this set, we query the captions and OCR-extracted text for keywords matching regular expressions related to \textbf{religion} (following the most popular world religions and religious sects \citep{religion_list}), \textbf{race and ethnicity} (such as \texttt{African}, \texttt{Asian}, \texttt{Caucasian}, \texttt{Hispanic}, \texttt{Latinx}, \texttt{Indian} \citep{dodge2021documenting}), and \textbf{sexual orientation} (such as \texttt{queer}, \texttt{lgbtq}, \texttt{homosexual}, \texttt{gay}, \texttt{lesbian}, \texttt{bisexual}). These sociodemographic dimensions are all considered sensitive data under the CCPA and GDPR \citep{ccpa, gdpr}. We manually examine these queried samples as an initial exploration and highlight various instances of sociodemographic information linked to names. Of these instances, in \Cref{fig:sociodemographic_pii} we manually categorize the websites (relying on Cloudflare \citep{cloudflare} to determine image-hosting sites), while \Cref{fig:sociodem_examples} depicts individual case studies.

\paragraph{We find captions that disclose full names paired with sexual orientation, several of which originate from news sites} Keyword search and manual examination surfaces 22 examples depicting the names of certain individuals who identify as LGBTQ+, with some images including the person's face. As depicted in \Cref{fig:sociodemographic_pii}, eight ($36.4\%$) of these samples come from news sites. For instance, \Cref{fig:sociodem_examples} shows a picture of a couple with the caption describing the name of a high school student attending a queer event --- this image is part of an article in which the student was interviewed and submitted the photo. In this case, the individual likely disclosed their name and information for the purposes of the news article, rather than consent to use their sociodemographic information to train a model, which we discuss in \Cref{sec:legal:purpose}.

\paragraph{We observe samples that reveal religion, race, or ethnicity terms paired with full names, originating from image hosting or blog sites and some news sites} We flag 42 instances that disclose the religion and full name of an individual, as well as 81 instances that disclose the race or ethnicity of an individual. Keywords like \texttt{African} or \texttt{Indian} may describe geographic regions or origins, so we only count examples that describe the individual, such as the phrase ``Asian artist.'' A significant portion of samples that disclose religion or race comes from news articles ($17.3\%$ for race and $35.7\%$ for religion) similar to our findings on sexual orientation. As an example in \Cref{fig:sociodem_examples}, it may also be likely that the individual disclosed this information for the purpose of the article. Many examples describe celebrities in which race may be inferred or common knowledge, such as ``first African American president,'' or referring to a religious leader, such as ``rabbi.'' We expand on the prevalence of self-disclosed religion, race, and national origin at a document level in \Cref{sec:audit_resume}.

\paragraph{Of the 142 unique samples that mention full names and sociodemographic keywords relating to sexual orientation, race, or religion, all but three samples depict human faces.} We examine the images of these samples and find that 139 samples contain images of human faces. However, only 119 of of the 139 include bounding box annotations that would blur the faces by default at the time of download, which motivates the evaluation of DataComp's face detection algorithm in \Cref{sec:face_obfuscation}.

\subsubsection{Celebrity names}
\label{sec:audit_celebrity}

\paragraph{Approach} Presidio's named entity recognition tool \citep{presidio} extracts the detected names in the captions and OCR text of the samples. We clean any false positives and manually discard names that represent clothing brands, fictional characters, and deceased figures, in order to determine which individuals are described. As an alternative approach, we search the captions and OCR text for celebrity names from Pantheon 2020 \citep{yu2016pantheon}, a dataset of 48 thousand well-known individuals (alive at the time at the time of collection), based on the criteria that their Wikipedia biographies have been translated into at least 15 languages. We again exclude names shared with designer brands and names that may be used in text outside of describing an individual (such as ``50 Cent'').

\paragraph{We find about 113 thousand CommonPool samples mention names of celebrities, with Donald Trump being mentioned significantly more than any other name.} The Pantheon celebrity search returns mentions of 45,829 unique names (most of the original Pantheon dataset), corresponding to 113 thousand CommonPool samples. \Cref{fig:pantheon_celebrity_search} plots the sample frequency of the top 50 most common Pantheon celebrity names. We observe that Donald Trump is mentioned more than twice as many times compared to any other celebrity, followed by various other United States politicians, athletes, musicians, and authors. When breaking down by occupation (across all celebrity mentions), we observe in \Cref{fig:pantheon_occupation} that actors, athletes, musicians, and politicians are the most common. In terms of country of origin, \Cref{fig:pantheon_country} shows that a majority of the samples mention celebrities originate from the United States and United Kingdom. We find similar results with Presidio's named entity recognition in \Cref{fig:presidio_celebrity_search}.

\subsection{Image}
\label{sec:audit_img}

In addition to personal information present in text, we search explicitly for specific types of identification or resume documents that may raise privacy concerns. We first incorporate keyword search and the Presidio PII detection tool to surface samples with matching text descriptions of certain documents. Then, we manually verify and discard samples that do not depict documents through visual inspection of the \textbf{image} component.

\subsubsection{Identity documents}
\label{sec:audit_id_doc}

\paragraph{Approach} We use both simple keyword search (relating to drivers license, credit card, etc.) as well as the Presidio PII detection tool to surface examples relating to identity numbers. We then examine images manually to find documents with government identifiers.

\begin{figure*}
    \centering
    \includegraphics[width=0.8\linewidth]{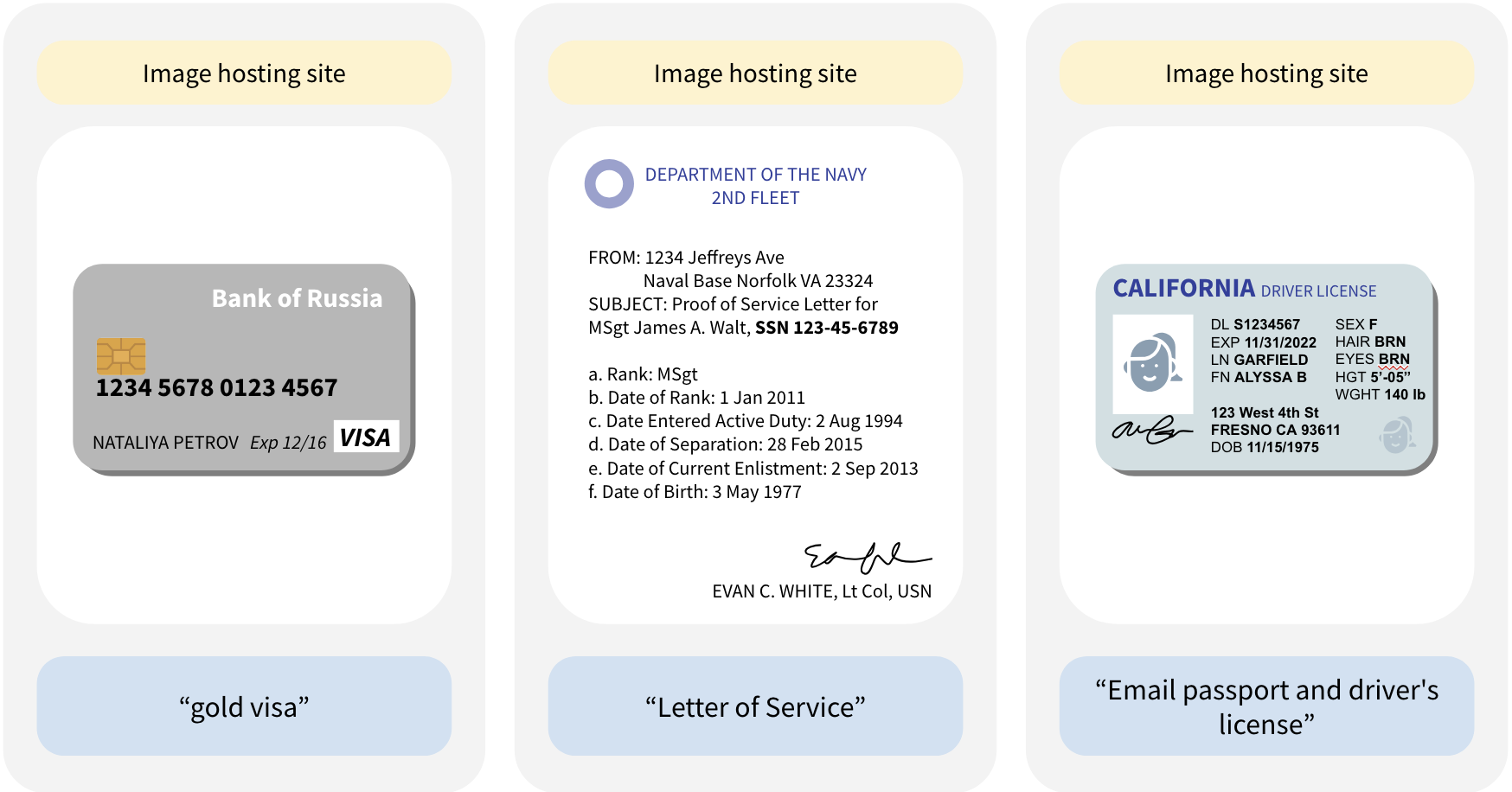}
    \caption{Examples of identity-related documents found in CommonPool's small scale dataset, showing a credit card, social security number, and a driver's license. For each sample, the type of URL site is shown at the top, the image in the middle, and the caption in quotes below. All personal information has been replaced, and text has been paraphrased to avoid direct quotations. Images have been redacted to show the presence of faces without identifying the individuals.}
    \Description{Three example instances of image, website source, and caption. On the left, an image of a Bank of Russia credit card including numbers, name, and credit card from an image hosting site with caption stating "gold visa." On the middle, an image of a proof of service letter for the Navy disclosing an individual's social security number from an image hosting site with caption stating "Letter of Service." On the right, an image of a California driver's license revealing name, date of birth, driver's license number, and address from an image hosting site with caption "Email passport and driver's license."}
    \label{fig:identity_doc_examples}
\end{figure*}

\paragraph{We find images that depict credit cards, drivers licenses, social security numbers, passports, and birth certificates} We find pictures or screenshots of credit card numbers with full names and security codes. We also find documents or pictures of U.S. drivers licenses, social security numbers, as well as passports from various countries. We identify birth certificates, mainly of celebrities, although U.S. states often make these documents publicly accessible. A few redacted examples appear in \Cref{fig:identity_doc_examples}.

\paragraph{These identity documents appear on image hosting sites, even as some seem to be uploaded by the data subject themselves} Many of these identity documents are uploaded to various image hosting sites, which makes it difficult to determine if the document comes from a data breach or is uploaded by the data subject themselves. In one specific case, where a U.S. social security number is included in a military document depicted in \Cref{fig:identity_doc_examples}, we trace the document to having been uploaded by a social media account sharing the individual's name. This same image appears on another image hosting site which is crawled by CommonPool, so even if the social media user had taken down the document, the image would still exist elsewhere. We also find a few samples with captions that describe how to generate or purchase credit card numbers and social security numbers, in which the images show examples of fake identity documents.

\subsubsection{Resumes}
\label{sec:audit_resume}

\paragraph{Approach} The sociodemographic keywords from \Cref{sec:audit_sociodemographics} indicated certain kinds of sensitive data present in professional resumes from job seekers. As such, we perform a measurement study of the types of the types of PII disclosed in job application materials and investigate the origins of these resumes. We search for CommonPool images that contain OCR-detected text relating to \texttt{resume}, \texttt{curriculum vitae}, \texttt{cv}, or \texttt{cover letter} and exclude samples with filler text in the image or caption, like \texttt{lorem ipsum} or \texttt{sample text}. For the sake of readability, we refer to ``resumes'' to describe all resume, cover letter, or curriculum vitae documents.

This initial query surfaces 3,770 samples (out of 12.8M), of which 3,634 images are successfully downloaded. We then engage in several rounds of annotating: (1) \textbf{Clean}: We first remove any sample that does not depict a legible resume document or clearly represents a fake individual, resulting in 805 samples. (2) \textbf{Validate}: We confirm which resumes and letters describe individuals with online presences. We find public LinkedIn profiles or news media mentioning the same name that have at least three \textit{points of equivalence} to the resume --- meaning that both sources share at least three identical attributes like middle name, job title, city, graduation year, or educational institution. (3) \textbf{Annotate}: We then manually tag the 168 validated resumes and cover letters for the types of personally identifiable information present. (4) \textbf{Automate}: Finally, we automatically analyze the URLs of the validated documents via Cloudflare's URL categorization \citep{cloudflare} and the Wayback Machine \citep{wayback} to understand the origins of these images. 

\begin{figure*}
    \centering
    \includegraphics[width=0.8\linewidth]{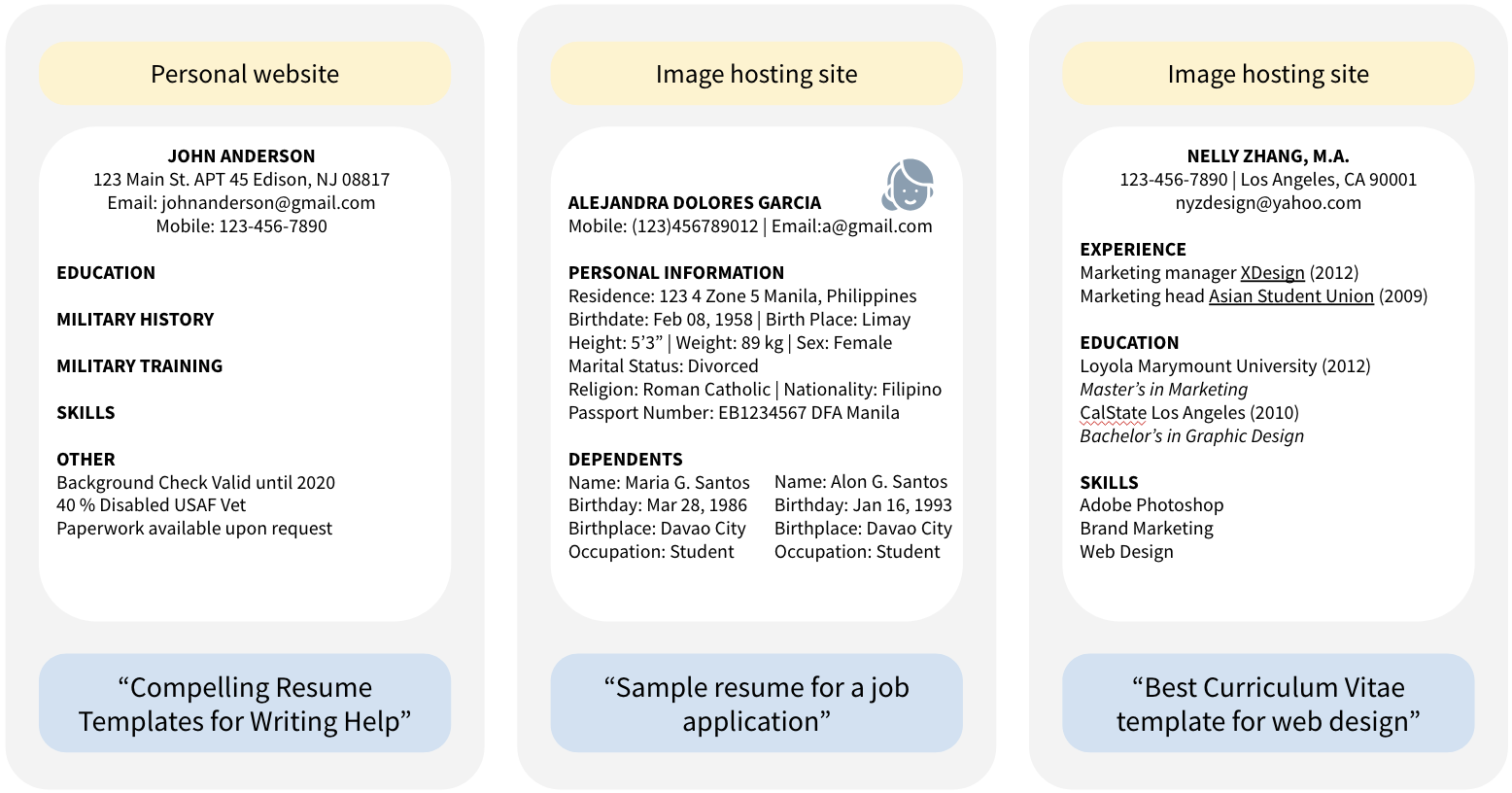}
    \caption{Examples of resume documents and personal disclosures found in CommonPool's small scale dataset. For each sample, the type of URL site is shown at the top, the image in the middle, and the caption in quotes below. All personal information has been replaced, and text has been paraphrased to avoid direct quotations. Images have been redacted to show the presence of faces without identifying the individuals.}
    \Description{Three example instances of image, website source, and caption. On the left, an image of a person's resume disclosing background check and disability status from a personal website with caption stating "Compelling Resume templates for Writing Help." In the middle, an image of a person's resume disclosing a face and dependents' date of birth, birthplace, and occupation from an image hosting site with caption stating "Sample resume for a job application." On the right, an image of a person's resume disclosing race membership from an image hosting site with caption stating "Best Curriculum Vitae template for web design."}
    \label{fig:resume_examples}
\end{figure*}

\paragraph{We find specific examples of resumes that disclose background check, disability status, the birth dates and places of dependents, and race} Searching keywords within the captions surfaces additional samples (around 14 thousand) but cannot all be annotated due to scale constraints. To complement our measurement, in \Cref{fig:resume_examples}, we surface several individual examples with captions that contain resume-related words and Presidio-detected names, with additional linking to online profiles.

\paragraph{Overall, we estimate at least $136,000$ images in all 12.8 billion samples of CommonPool depict resume documents linnked to users with public online presence} As shown in \Cref{tab:resume_overview}, out of the 3,634 downloaded images, 805 samples depict resume documents that are not visually fake. Of those, we confirm the public online presence of 168 resume documents, mostly through LinkedIn profiles but some Facebook or news articles sites as well. Given the search is within a random 0.1\% subset of CommonPool, at a 95\% confidence interval (adjusted by Bonferroni correction), we estimate between 136 thousand and 200 thousand images depict resume documents of individuals with public online presence. This number again is a lower bound, as the keyword search does not uncover all resumes; moreover, during the validation step, some resumes may depict individuals but their profiles may be private or non-existent.

\paragraph{Of the validated resumes, we observe careers relating to technology and academia, and many resumes are duplicated on image hosting sites.} Of the most recent jobs listed in each sample, there is a high presence of careers relating to information technology, engineering, graphic design, and marketing. We also find six samples of PhD student resumes and five samples of professor resumes. Due to the external nature of certain types of jobs, it may be reasonable to expect that professional experiences are publicly available, especially as some resumes are uploaded by LinkedIn profiles with the same name. At the same time, 20 of these resumes are duplicated across image hosting sites.

\begin{figure*}
    \centering
    \includegraphics[width=0.9\linewidth]{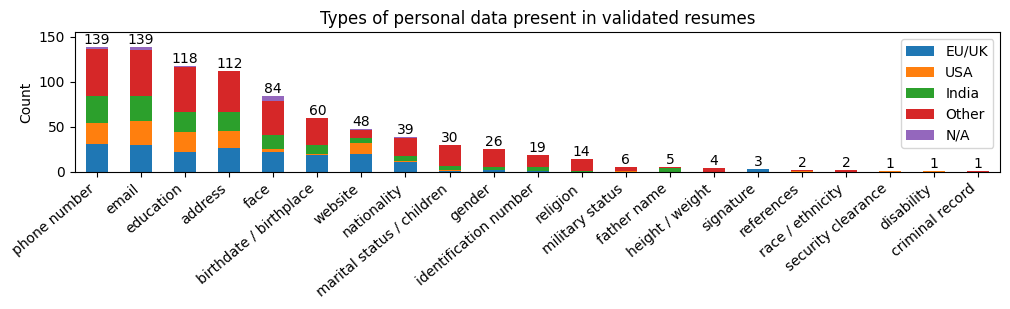}
    \caption{Sample counts of annotated personal information present in the 168 resume documents with validated online presence, broken down by region (if disclosed in resume). We highlight the United States and countries from European Union due the focus of our legal analysis (grouping the United Kingdom with the EU due to their current application of GDPR \citep{uk_gdpr}). Some resumes (10 out of 168) do not include addresses and are labeled ``N/A.''}
    \Description{Bar graph of frequency of personal information type. Phone number, email, education, address are the most common with at least 100 samples. Bar graph is also broken down by geographic region, with many from the USA and EU/UK regions.}
    \label{fig:resume_pii_types}
\end{figure*}

\paragraph{Among the resumes with online presence, we find disclosures of contact information, individual faces, government identifiers, and sociodemographic information, as well as the personal information of other individuals.} In \Cref{fig:resume_pii_types}, we manually annotate the types of personal information present in the 168 validated resumes. The majority of these images contain contact information including phone number, email, education, and physical address. Of the 112 samples with physical addresses, 11 explicitly include residential addresses, and 6 include work addresses, while the rest are not clarified. A significant number of resumes also include a photo of the individual, their date and place of birth, personal website URLs, and even government identification numbers like driver's licenses or passports. We also find presence of certain kinds of sociodemographic information like gender, marital status, number of children, religion, race or ethnicity, disability, height and weight, and criminal record. While an individual creating a resume may have disclosed personal information for job-seeking purposes, we observe information relating to other individuals, such as contact information for references, the name of the individual's father, or dates of birth of children as seen in \Cref{fig:resume_examples}. In our legal analysis in \Cref{sec:legal_analysis}, we comment on the nature of consent of sensitive data like race or religion disclosed in resumes that are later scraped to build datasets to train models.

\paragraph{We find most resumes have addresses associated with India and the United States (and states with consumer privacy laws), with some associated with the European Union.} We manually annotate the country associated with validated resumes, in order to inform the possibility of legal attachment based on data subjects in various jurisdictions. Each sample is tagged according to country of address in \Cref{fig:resume_country_address} and national origin or citizenship in \Cref{fig:resume_nat_origin}. We find that India and the United States are the most common countries associated with the 168 validated resumes. A substantial number of resumes come from countries in the European Union. Within the United States, addresses correspond to 15 unique states, notably including California, Texas, Colorado, New Jersey, Massachusetts, Indiana, Oregon, and Illinois.

\begin{table}
    \begin{subtable}{.5\linewidth}
      \centering
        \begin{tabular}{ll}
            \toprule
            \textbf{Resume}\\\textbf{annotation stage} & \textbf{Count} \\
            \midrule
            Overall   & 3634 \\
            $\rightarrow$ Cleaned & 805 \\
            $\qquad\rightarrow$ Validated   & 168 \\
            \bottomrule \\
        \end{tabular}
    \end{subtable}%
    \begin{subtable}{.5\linewidth}
      \centering
        \begin{tabular}{ll}
            \toprule
            \textbf{Websites}\\ (out of 168) & \textbf{Count} \\
            \midrule
            bing.net & 54 \\
            pinimg.com     & 53 \\
            slidesharecdn.com    & 27 \\
            \bottomrule \\
        \end{tabular}
    \end{subtable} 
    \caption{Sample counts of resume annotation process detailed in \Cref{sec:audit_resume}. Left: Funnel of annotation stages, resulting in $168$ samples of resumes that have a validated online public presence (e.g. LinkedIn). Right: Breakdown of most common site origins of validated resumes.}
\label{tab:resume_overview}
\end{table}

\paragraph{Most validated resume images come from image hosting or photography sites.} The final automation step examines the types of websites that serve the validated resume images. \Cref{tab:resume_overview} shows the most common websites: \texttt{bing.net}, \texttt{pinimg.net}, \texttt{slidesharecdn.net}. These findings again demonstrate the prevalence of personal information appearing on image hosting sites, potentially being propagated or not uploaded to the site by the data subject themselves. In \Cref{fig:resume_timestamp}, we plot the frequency of the earliest timestamp tracked by the Wayback Machine \citep{wayback} of the resume URLs. The Wayback Machine only found records for 70 of the 168 resumes, noting potential inaccuracies of the earliest recorded timestamps, which also signifies the challenges of tracing the origins of content on the web. Of the sites that were recorded, most images existed before 2022, which aligns with the fact that CommonPool is sourced from Common Crawl snapshots from 2014 to 2022. As a result, most of these resume documents may have been uploaded before the existence of popular generative AI systems \citep{cao2023comprehensive}, yet are now being downloaded over a million times to train models.

\subsection{URL}
\label{sec:audit_url}

This section presents results on privacy concerns relating to the \textbf{URL} component of CommonPool. \Cref{sec:audit_children} describes searching for children-related websites to find children's information present in samples, and \Cref{sec:audit_unavailable} investigates URLs that fail to download due to DataComp's web crawler.

\subsubsection{Children's information}
\label{sec:audit_children}

While children's information does not fall under the definition of sensitive data in the CCPA and GDPR (but does for the Oregon Consumer Privacy Act \citep{ocpa}), both laws consider special provisions for children's information. Moreover, in the United States, the Children's Online Privacy Protection Rule (COPPA) protects the use of personal information from children under 13 years of age \citep{coppa}. While COPPA explicitly covers the collection of personal information \textit{from} children, we cannot determine whether this information is collected \textit{from} versus \textit{about} children, just as we argue in \Cref{sec:discussion:tracing} that web-scraping cannot determine either.

\paragraph{Approach} We search for personal information relating to children, primarily focusing on online services directed towards children (due to the scope of COPPA requirements). To do so, we initially perform a manual keyword search for samples mentioning \texttt{child} and related words in the caption, in order to find individual case studies. We next identify samples that come from children's related websites. We rely on Cloudflare website categorizations \citep{cloudflare} from prior work \citep{hong2024s} on a random subset of 100,000 domains from CommonPool and isolate samples belonging to sites in the \texttt{Safe for Kids} category. While Cloudflare categorizations have been shown to be accurate \citep{ruth2022toppling}, samples that come from these websites may not necessarily be considered children's information. We thus follow up with an alternative approach to examine samples from sites participating in COPPA safe harbor programs. The safe harbor provision enables industry groups to self-regulate its members to follow COPPA's guidelines \citep{coppa_safe_harbor}. Because members that join these safe harbor programs intend to be in compliance with COPPA, we search for sites from iKeepSafe \citep{ikeepsafe}, kidSAFE \citep{kidsafe}, and PRIVO \citep{privo} certification programs, which display their members publicly. \Cref{tab:children_overview} gives an overview of the sample counts and number of sites by each approach. We examine samples associated with these sites for any personal information.

\begin{figure*}
    \centering
    \includegraphics[width=0.8\linewidth]{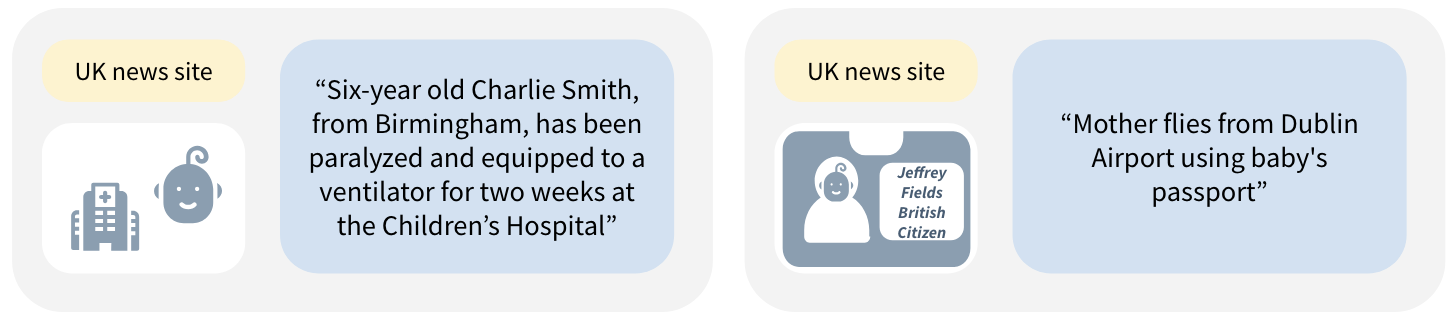}
    \caption{Real examples of children's information found in CommonPool's small scale dataset. For each sample, the type of URL site is shown at the top left, the image in the bottom left, and the caption in quotes on the right. All personal information has been replaced, and text has been paraphrased to avoid direct quotations. Images have been redacted to show the presence of faces without identifying the individuals.}
    \Description{Two example instances of image, website source, and caption. On the left, an image of a child at a hospital from a U.K. news site with caption stating "Six-year old Charlie Smith, from Birmingham, has been paralyzed and equipped to a ventilator for two weeks at the Children’s Hospital." On the right, an image of a child's passport disclosing their name and face from a U.K. news site with caption stating "Mother flies from Dublin Airport using baby's passport."}
    \label{fig:children_examples}
\end{figure*}

\paragraph{We locate samples that depict children's birth certificates, passports, and health status, originating from news articles or blogs} With keyword search, we discover various images of children's birth certificates and passport information. As shown in \Cref{fig:children_examples}, we also find an image of a child's passport and also an image depicting a child unconscious on a hospital bed, with the caption including their full name and health condition. These samples often come from news articles or online blogs, in which it may be plausible that the use of these photographs and full names may have obtained parental consent specifically for inclusion in the article.

\begin{table}
\centering
    \begin{tabular}{lll}
        \toprule
        \textbf{Site source} & \textbf{Unique sites} & \textbf{Count} \\
        \midrule
        Cloudflare Safe for Kids  & 493 & 12698 \\
        COPPA Safe Harbor & 52 & 315 \\
        \bottomrule \\
    \end{tabular}
    \\
    \begin{tabular}{ll}
        \toprule
        \textbf{COPPA safe harbor}\\\textbf{PII presence} & \textbf{Count}\\(out of 315) \\
        \midrule
        Adult's face & 20 \\
        Child's face     & 14 \\
        Name    & 14 \\
        \bottomrule \\
    \end{tabular}
    \caption{Sample counts relating to children's information detailed in \Cref{sec:audit_children}. Top: Number of unique sites and sample count corresponding to each approach, Cloudflare's categorization \citep{cloudflare} or list of COPPA safe harbor programs. Bottom: Breakdown of most common site origins of validated resumes.}
\label{tab:children_overview}
\end{table}

\paragraph{Sites categorized as ``safe for kids'' include child-targeted companies like Hasbro or Disney, as well as platforms in Japan and the United Kingdom, mainly depicting toys or cartoons rather than personal information.} We find about 500 websites categorized as \texttt{Safe for Kids} by Cloudflare, corresponding to about 13,000 samples, of which 3,000 samples have faces detected by CommonPool annotations. \Cref{fig:cloudflare_kids} provides a breakdown of the most frequent sites that Cloudflare categorizes as \texttt{Safe for Kids}, including various commercial platforms targeted towards children, such as Hasbro or Disney. Some popular sites also have country domain names associated with Japan, Russia, South Africa, and the United Kingdom. We also observe that 34\% of samples come from \texttt{rcgroups.net}, which is an image hosting site for RCGroups, a radio control forum (not necessarily related to children). Upon further examination of these samples, we find pictures of toys or cartoons rather than personal content, so presence of samples from Cloudflare-categorized sites do not reveal much evidence of private information.

\paragraph{We find child's names and faces revealed in samples from sites that are members of COPPA Safe Harbor platforms} There are 315 samples from COPPA Safe Harbor sites, with most frequent sites are shown in \Cref{fig:coppa_sites}. We manually annotate all of these samples for presence of PII and find that some examples show adults' or children's faces, as well as full or first names (in \Cref{tab:children_overview}). While these numbers are small, not all platforms under COPPA safe harbor programs are captured, and this search only covers a random 1/1000th subset of all of CommonPool. With a 95\% confidence interval (after Bonferroni adjustment), we estimate between 272 thousand and 358 thousand samples are from our set of sites that intend to be COPPA-compliant.

\subsubsection{Unavailable images}
\label{sec:audit_unavailable}

\paragraph{Approach} At the time we download CommonPool in April 2025, $21.4\%$ or 2.7 million image URLs fail to download, yet during the time of CommonPool's creation in March of 2023, all the images could be successfully scraped from the web, otherwise these samples would have been removed \citep{gadre2024datacomp}. Even if there are unavailable images in our current version, their corresponding captions, site URLs, and accompanying image annotations still exist with the URL-caption artifact. Because websites frequently change or are no longer maintained, and as some download errors may be a result of our server or connection issues, here we investigate the types of HTTP download errors.

\paragraph{We find that the most common download error is due to broken or dead links ($35.4\%$), while the next most common is due to a lack of permission to access the link ($19.0\%$).} In \Cref{fig:http_error_freq}, we plot the most common HTTP download errors (after manually merging and renaming similar errors) and find \texttt{Not Found}, \texttt{Forbidden}, and \texttt{Service Not Known} as the most common errors. While many errors relate to a failure to reach the image URL, which perhaps indicates the website or image asset has moved, we find a substantial number errors relating to permissioning with confirmation that the image asset exists. For instance, we observe that some samples with \texttt{Forbidden} errors do in fact render manually, which means that the web server may have recognized the download script \citep{datacomp_github} as a web crawler and subsequently blocked access. The \texttt{Forbidden} error is distinct from \texttt{Unauthorized}, as it indicates that the web server recognizes the DataComp crawler and verifies a lack of permission. In other words, there is an explicit rejection of consent for users of this dataset to automatically scrape site content, but the images have been scraped in the past, and the captions, URL, and metadata are still available.

\paragraph{We observe that the tool to download DataComp CommonPool by default respects image robots tags when crawling, but not site-level robots.txt protocols.} \Cref{fig:http_error_freq} shows that \texttt{Robots Disallowed} tag is also a common error for 51 thousand image URLs that fail to download. Upon investigation, we find that the crawler for DataComp by default respects X-robots-tag (unless explicitly modified by the user) \citep{img2dataset_robots_tag}. The X-robots-tag is specified by the site host in the image URL HTTP header when the link is crawled \citep{robots_tag}, and is distinct from the robots.txt protocol which is surfaced at the main webpage (and not each individual image URL). While the DataComp crawler respects site host consent preferences at the image-level, the crawler in its current form requests the \textit{entire} URL content. These requests may increase load on the web server and therefore increase costs for the site host, especially if fetched over two million times! If a site host wishes to prevent crawling at the image-level for server performance or cost reasons, the load increase effectively defeats this purpose. While the tool maintainers are aware of this issue, at the time of writing it has not been resolved \citep{img2dataset_robots_tag}. Moreover, if wishing to prevent web-scraping for other purposes, a site host that disallows crawling on its robots.txt file would have to continually attach X-robots-tags to every image URL on the site just to avoid the image content being scraped. The DataComp crawler's current setup to ignore robots.txt runs counter to best practices from the World Wide Web Consortium \citep{web_crawler_practice}.

\begin{figure*}
    \centering
    \includegraphics[width=0.8\linewidth]{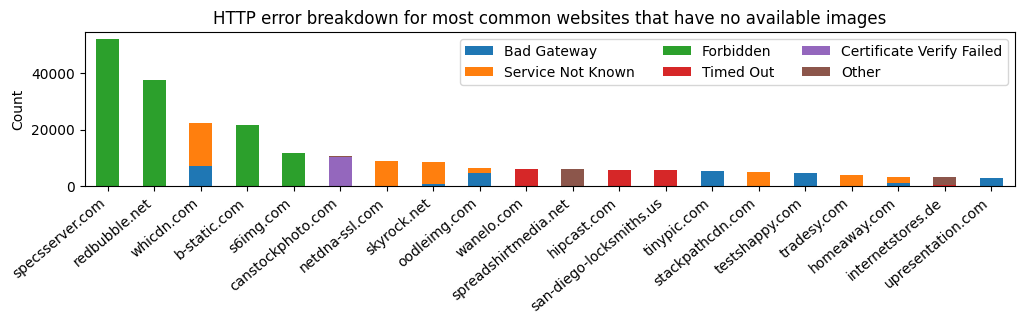}
    \caption{Error breakdown of most common websites of which all samples failed to download for the \texttt{small} scale, but at the time of CommonPool curation and for early download versions image URLs on these sites were successfully downloaded.}
    \Description{Frequency bar graph of common websites that have no available images with "specserver.com," redbubble.net," "whicdn.com," and "b-static.com" as the most common with above 20 thousand samples. Bars are split by HTTP error where the "Forbidden" error is the only error for four of the top five websites.}
    \label{fig:paywall_failed_sites}
\end{figure*}

\paragraph{Several websites' entire set of samples fail to  download in our evaluation set, of which most of these websites no longer load, and one website has a login screen to access these images.} Of the 2.7 million image URLs that failed to download in the \texttt{small} scale of CommonPool, we find that 1.2 million of these are from website domains that have successfully-downloaded samples, while the leftover 1.5 million image URLs are from websites that are no longer available. As it is plausible that these websites may have available images URLs on larger scales of CommonPool, we examine the error breakdown of these ``failed'' sites in \Cref{fig:paywall_failed_sites}. The most popular failed sites have unavailable image URLs for a variety of reasons, but we note that of the top five, the most common reason is due to a \texttt{Forbidden} error. Of the sites listed in \Cref{fig:paywall_failed_sites}, we try to load the main site page and find the majority of these sites fail to load. However, we find that the most common website \texttt{specsserver.com}, which composes $0.4\%$ of CommonPool, renders a login screen. If the login screen of the website existed at the time of CommonPool creation, the image URLs that were once available to download may be considered not legally public as described in \Cref{sec:legal:public_avail}. If the login screen and authentication to the image URLs were added after CommonPool's release, then the site host may have blocked access to the image assets, although they would have been downloaded in earlier versions.

\paragraph{Of the random subset of failed-to-download images with Wayback Machine records, most of the image URLs had earliest timestamps before 2022.} We track the earliest timestamp recorded by the Wayback Machine \citep{wayback} for a random subset of $1000$ image URLs that fail to download. We find records for $21.3\%$ of these URLs in \Cref{fig:failed_url_timestamp} of which most of these existed before 2022 (and now are no longer available). We examine whether image URLs that fail to download have earlier records than image URLs that successfully download --- perhaps due to older sites lacking maintenance --- and observe that the distribution of existing are roughly similar (shown in \Cref{fig:success_url_timestamp}). To compare the distributions of successfully-downloaded and failed-to-download groups, we randomly select 1000 samples per group to measure statistical differences in the sample means as shown in \Cref{tab:failed_t_tests}. Compared to a random subset of successfully-downloaded samples, failed-to-download images on average are larger, have more detected faces, and have higher CLIP-similarity scores (DataComp's measure for image quality), although differences are slight.

\paragraph{We find that captions of samples that mention invoices, social security numbers, and credit cards are associated with higher-than-average download error rates, but by a small amount.} Because the captions are still available of image URLs that fail to download, we examine the association between the download error rate and the presence of PII in the caption. We query samples for regular expression matches with personal information, like driver's license, passport, resumes, etc. In \Cref{fig:pii_mention_error_rate}, we see that captions that mention invoices are associated with higher-than-average error rates, in addition to credit cards and social security numbers, while captions that mention resumes have substantially lower download error rates.

\subsection{Metadata}
\label{sec:audit_metadata}

We now focus on the \textbf{metadata} component associated with each URL-caption pair. \Cref{sec:audit_exif_tags} describes the image Exif tags that are extracted when downloading the image assets of CommonPool. Investigation of these tags reveals precise geolocation data accompanied with full names. We then examine the face detection metadata in \Cref{sec:audit_indv_faces}: as described in \Cref{sec:face_obfuscation}, the released CommonPool artifact comes with bounding box annotations from a face detection algorithm so that when downloaded, the detected faces can blurred in the dataset (unless overridden by the user). We search through the face annotations to evaluate whether this face obfuscation technique effectively anonymizes the presence of faces.

\subsubsection{Exif tags}
\label{sec:audit_exif_tags}

\paragraph{Approach} Each web image is embedded with Exif tags, which can be added manually or automatically by cameras at the time of image creation. However, \citet{henne2014awareness} show that users often are not aware of metadata, which can disclose personal information, being shared when an image is uploaded to the web. DataComp's download tool explicitly extracts the image tags according to the Exif standard for every sample, which means that additional information is also being stored at the time of download. In this section, we investigate these Exif annotations and search for presence of individuals.

\begin{figure}
    \centering
    \includegraphics[width=0.7\linewidth]{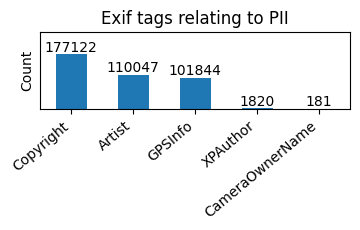}
    \caption{Sample counts of non-empty Exif tags relating to PII for all successfully-downloaded images in small-scale of CommonPool. Upon inspection, many name-related tags, like \texttt{CameraOwnerName} or \texttt{Artist} contain full names. We re-extract a random set of non-empty \texttt{GPSInfo} tags and find that $28.6\%$ contain precise geolocation.}
    \Description{Bar graph of Exif tags relating to PII including Copyright (177,122 samples), Artist (110,047 samples), GPSInfo (101,844 samples), XPAuthor (1,820 samples), and CameraOwnerName (181 samples).}
    \label{fig:img_exif_tags}
\end{figure}

\paragraph{We find non-empty Exif tags relating to timestamps, geolocation, and individuals, which upon inspection many of which disclose full names.} \Cref{fig:img_exif_tags} plots the frequency of non-empty Exif tags that may disclose personal information, where there are hundreds of thousands of samples that are embedded with metadata detailing timestamps, geolocation information, and individual presence. We investigate various Exif tags relating to individuals and find that while some reveal companies or photography studios, a significant amount of metadata text for the \texttt{CameraOwnerName} and \texttt{Artist} tags include full names.

\paragraph{We re-extract the Exif tags for GPS information and find that $28.6\%$ of \texttt{GPSInfo} tags point to precise geolocation, of which $6.1\%$ of those also come with full names.} Of the $102$ thousand images with \texttt{GPSInfo} Exif tags, we observe that the tag information is malformed at the time of download, so we replicate the Exif tag extraction process for a random subsample of $5$ thousand geolocation-tagged images. We determine that $28.6\%$ of extracted \texttt{GPSInfo} tags contain precise latitude and longitude locations attached to the images, which cameras may include by default \citep{henne2014awareness}. Of those images with precise geolocations, about $5.9\%$ of those images also have Exif tags contain full names.

\subsubsection{Face biometric data}
\label{sec:audit_indv_faces}

\paragraph{Approach} As described in \Cref{sec:face_obfuscation}, the CommonPool artifact includes annotations of bounding boxes automatically-detected via the SCRFD algorithm. The script to download the images will by default blur those detected face regions, although the dataset user can override the obfuscation step (and can even use these annotations to extract images of detected faces). To evaluate DataComp's method to detect and obfuscate faces, we apply Amazon Rekognition's face detection algorithm applied to 100 thousand randomly selected samples to surface samples uncaught by SCRFD. Among the manually confirmed images of faces, we annotate the presence of a name, location, and whether the face unambiguously depicts a child under 13 years of age. We also investigate the site origins of the images with human faces, as well as any differences in the distributions of human faces caught and uncaught by SCRFD. While prior work has demonstrated that individuals with faces blurred can still be identified \citep{mcpherson2016defeating, oh2016faceless}, we leave this investigation to future work, especially as models trained on CommonPool with faces blurred seem to identify signals to predict gender and race \citep{gadre2024datacomp}. 

\paragraph{We estimate at least 100 million images of human faces are not obfuscated when downloading CommonPool with default parameters, although a majority of these facial images are small} Out of the 100 thousand random samples, SCRFD detects 25 thousand samples containing at least one face, while Rekognition detects about $19,000$ samples containing at least one face. For about three thousand samples, Rekognition detects more faces than SCRFD, of which $1,445$ images SCRFD does not detect a single face. In \Cref{tab:face_detection}, we manually annotate those $1,445$ samples and find that $59.1\%$ of these samples depict human faces missed by SCRFD, $16.2\%$ are false positives, and $24.7\%$ are depictions of drawings or cartoons. At this miss rate, for all 12.8 billion samples of CommonPool, we estimate at a $95\%$ confidence interval (after Bonferroni adjustment) of 100-119 million samples contain human faces uncaught by DataComp's current face obfuscation mechanism (a lower bound based on the samples surfaced by Rekognition). We observe that a majority of these images contain human faces with bounding boxes of less than 400 square pixels, which means that the depicted faces are low quality --- and while this finding implies that identification may be difficult based on the facial image alone, we still find many high-quality images of people's faces.

\begin{table}
    \begin{subtable}{.5\linewidth}
      \centering
        \begin{tabular}{ll}
            \toprule
            \textbf{Annotation} \\ (out of 1.4K) & \textbf{Count} \\
            \midrule
            Human   & 854 \\
            Drawing & 357 \\
            False   & 234 \\
            \bottomrule \\
        \end{tabular}
    \end{subtable}%
    \begin{subtable}{.5\linewidth}
      \centering
        \begin{tabular}{ll}
            \toprule
            \textbf{PII presence}\\ (out of 854) & \textbf{Count} \\
            \midrule
            Location & 155 \\
            Name     & 52 \\
            Child    & 48 \\
            \bottomrule \\
        \end{tabular}
    \end{subtable} 
    \caption{Manual annotation counts of samples without facial bounding boxes. Left: Annotation of $1,445$ samples  with Rekognition positive classifications (at least one face detected) and SCRFD negative classifications (no faces detected). Right: Annotated PII presence of samples with manually verified human faces that were not covered by DataComp's face blurring.}
\label{tab:face_detection}
\end{table}

\paragraph{We find presence of personal information relating to name, location, and depiction of children among the non-obfuscated human faces} We then manually annotate for the types of personal information present among the 854 samples with human faces uncaught by SCRFD. In \Cref{tab:face_detection}, we find mentions of location and name in the caption or image, as well as images that depict children with faces non-obfuscated. We observe various examples of screenshots with un-blurred profile pictures with full names present.

\paragraph{Compared to the images of faces obfuscated by DataComp, the images of faces not obfuscated are on average smaller and have lower pixel brightness}
Among the manually confirmed samples of Rekognition predictions that contain a single human face, we randomly draw 400 images each from the subset with faces detected by SCRFD and the subset with faces undetected by SCRFD. We examine differences in the means of these two sampled groups with respect to various image-related variables: Rekognition's bounding box area, average pixel brightness, proportion predicted as \texttt{Female} by Rekognition, and the age predicted by Rekognition. Age and gender may be unreliable measures due to biases encoded by Rekognition \citep{schwemmer2020diagnosing}, which make it difficult to make valid inferences. In \Cref{tab:face_t_tests}, we find a statistically significant difference (at a 99\% confidence level) between the image-related variables of facial images detected and undetected by SCRFD. We accept the alternative hypothesis that unblurred facial images on average have smaller bounding boxes, less bright in average pixel value, and younger Rekognition-imputed ages than facial images that would have been blurred.

\paragraph{Observed differences may be situated against ongoing work of sociodemographic biases in face detection} While these image-related variables represent noisy imputed signals (i.e. Rekognition has its own classification error and biases), these statistical differences in pixel brightness may demonstrate certain instances in which the task to detect faces is not as accurate. These results may possibly relate to other demographic biases, as prior work has established well-known biases in face or person detection along the lines of skin tone \citep{wilson2019predictive, menezes2021bias}.

\paragraph{Many images in the set of human faces not covered by bounding boxes originate from image-hosting sites, blogs, and media platforms with earliest records before 2022.} Finally, we examine the website URLs of the various images of human faces undetected by SCRFD. In \Cref{sec:appx_face_url}, we find many images originate from image-hosting sites, blogging sites like Wordpress, and media sites. We also confirm that earliest timestamps recorded by the Wayback Machine for most samples with records are before 2022, before the existence of popular generative AI systems \citep{cao2023comprehensive}.

%% file: sections/05-legal-analysis.tex
\section{Legal Analysis}
\label{sec:legal_analysis}

Our audit results now inform legal analysis of the treatment of personal data in CommonPool, determination of legal attachment and obligations, and the sufficiency of sanitization attempts.

\subsection{Does DataComp CommonPool contain ``personal data,'' and if so, how do privacy laws treat it? }
Given the empirical findings of our audit, it is clear that CommonPool contains extensive personal information, triggering the definition of ``personal data'' under GDPR and the equivalent terms under U.S. laws. As highlighted in \Cref{tab:results_summary}, the dataset includes identifiable human faces, full names and contact details on resumes, government ID numbers, financial information (such as credit card numbers with security codes), and even content involving children (e.g. birth certificates). Under the EU’s GDPR, virtually any information relating to an identifiable person is personal data. A photograph of a face, for example, is personal data because a person can be identified from it (either directly by recognition, or indirectly via facial recognition technology or matching with other data). Likewise, a resume image clearly ``relates to'' an identifiable person (the individual named on the resume). Thus, almost all the examples uncovered, faces, resumes, names, emails, credit card details, qualify as personal data under GDPR. The fact that this data was scraped from publicly accessible websites does not remove it from GDPR’s ambit.  In practice, a photo posted on a personal blog or an image on Flickr is protected personal data in Europe despite being publicly viewable.

Under California’s CCPA/CPRA, the scraped information also largely counts as ``personal information.'' The law explicitly includes inferences about a person within the definition of personal information, meaning even any tags or labels inferred in the dataset (for instance, if the dataset or subsequent model infers someone’s age or occupation from an image) are considered personal information about the individual. California regulators have affirmed that internally generated profiles or inferences are covered just like collected data. However, California’s law has a notable exclusion for ``publicly available'' information. The DataComp curators might argue that because they scraped data from public internet sources, the data is ``publicly available'' and thus not subject to CCPA. This argument has some force only if the data squarely fits the statutory definition of publicly available, it was lawfully made available through government records or widely distributed media, or was broadly made public by the individual. Some subset of CommonPool likely does come from widely distributed media (for example, news websites), and some comes from individuals’ public postings on social platforms. To that extent, a business could claim those specific portions are exempt. However, it is not a blanket escape hatch. The CPRA version of ``publicly available'' still requires a case-by-case look at how the data was made public and by whom. For instance, a leaked database posted on a forum would not count as ``lawfully made available.'' A personal photo taken from behind a login-only site (i.e. the example found in \Cref{sec:audit_unavailable} if the login screen existed at time of crawling) would not be ``publicly available.'' Even data that was public may lose the exemption if used in a manner different from the purpose for which it was published (the law implies that the context of publication matters). Moreover, any derivative information in the dataset (such as embeddings or metadata added) wouldn’t be “publicly available” in origin. In short, while California’s law could deem parts of DataComp CommonPool outside its scope, a large portion, especially the more sensitive bits like driver’s license images, personal communications, or anything not obviously from a public-facing source, would still be considered personal information subject to the CCPA. And crucially, being ``public'' does not strip individuals of all protections: if a California resident finds out a business is using their personal photo or essay from the web, they could still exercise rights (like deletion), since the CCPA’s public-data exception mainly affects whether the law applies at all, not what happens once data is in play. Publicly available sensitive data can also potentially be regulated if used in certain ways, for example, a business could still be restricted from using a publicly posted race or health detail for targeted advertising without offering an opt-out, because that would be profiling on sensitive grounds.

Under the Oregon Consumer Privacy Act, the definition of personal data likewise covers CommonPool’s contents with an exclusion for publicly available data. Oregon’s definition of ``publicly available'' information is similar to California’s broadened definition (encompassing information a person has made public or that is available through public sources). If the CommonPool entries are determined to be publicly available under OCPA, they would not be considered ``personal data'' under that law. For example, an image scraped from a publicly viewable Instagram profile might be deemed public information. 

The OCPA mandates opt-in consent for processing sensitive data, which includes “any personal data of a child” and biometric identifiers. One cannot simply scrape a child’s personal details from a public website and evade Oregon’s consent requirement, because the statute treats all children’s data as sensitive regardless of source. Thus, personal data about children in DataComp CommonPool is a particularly problematic category under all frameworks: GDPR accords it special protection (requiring parental consent for young children in many cases), CCPA/CPRA imposes opt-in consent for selling minors’ data and heightened duties to protect children, and OCPA flatly requires consent to process kids’ data at all. Our audit unearthed content like birth certificates and photos of children in the dataset– these likely pertain to minors, meaning any entity subject to OCPA or even general consumer protection could face legal risk in using that data.

In sum, the personal data in CommonPool does not cease to be personal data simply because it was scraped from the web. GDPR treats it as fully regulated personal data. CCPA/CPRA and OCPA carve out some public data, but not in a way that would categorically exempt the majority of a massive, wholesale-scraped collection. At minimum, identified or identifiable individuals are present throughout the dataset, and thus privacy laws recognize their personhood in the data. The inclusion of inferred data (e.g. algorithmically generated labels about individuals) also falls under these definitions. The GDPR is clear that profiling data or any information ``relating to'' an individual is in scope, and California explicitly lists inferences as personal information. Therefore, any notion that CommonPool’s billions of samples are completely anonymized or not ``personal'' cannot be sustained. Despite some sanitization, our findings (such as the images of real, unblurred human faces remaining after filtering in \Cref{sec:audit_indv_faces}) indicate that identifiable data is abundant. Each such face image is biometric data linked to a person; each resume is a dossier of someone’s identity. Privacy law is concerned with exactly these kinds of data.

\subsection{When and how do these privacy laws ``attach'' to DataComp CommonPool or its use? }
The applicability of GDPR, CCPA, and OCPA depends on the circumstances of the entity processing the data. DataComp CommonPool itself is an artifact, a collection of files, and not a legal entity. Thus, the laws apply to the controllers or processors who handle that personal data. Different scenarios illustrate when obligations would kick in:

\subsubsection{The dataset creators/distributors} Suppose the team that compiled CommonPool (Gadre et al., per the DataComp paper) is based in the U.S. and released the dataset publicly for research. If they have no business operations in California or Oregon and are an academic/non-profit entity, CCPA and OCPA likely did not apply to their act of compilation (CCPA covers only for-profit businesses, and OCPA only from 2024 with inclusion of non-profits). GDPR might apply if, for example, EU personal data was scraped (such as resumes with disclosed addresses from EU countries in \Cref{fig:resume_pii_types}) and the act of scraping is considered monitoring behavior of EU residents (web crawling could be seen as a form of monitoring). Clearview AI’s scraping of EU citizens’ photos led EU regulators to assert GDPR’s jurisdiction, even though Clearview was a U.S. company. Under GDPR Article 3(2)(b), monitoring individuals’ behavior in the EU (which continuous scraping and analyzing of EU websites could qualify as) brings the activity under GDPR. Additionally, if any EU-based researchers or organizations are involved in hosting or curating CommonPool, GDPR directly binds them. We see that privacy laws can attach even at the dataset creation stage if the compilers meet jurisdictional criteria. However, enforcement at that stage is murky --- for instance, if an academic merely scrapes data for research without any commercial purpose, they might invoke exceptions for research or freedom of expression (though GDPR’s research exemption still requires safeguards and doesn’t nullify data subject rights entirely).

\subsubsection{Downstream users (companies or researchers training models on DataComp CommonPool)} This is likely the more consequential point of attachment. Any organization that obtains CommonPool and processes it to train an AI model becomes a data controller (determining the purposes and means of processing personal data in the dataset) or a processor for some other controller. If that organization is in the EU, GDPR straightforwardly applies. If it is outside the EU but offering an AI system to EU residents or monitoring EU individuals’ behavior through the model, GDPR also applies extraterritorially. For example, a U.S. company training a photo recognition model on CommonPool, which might later identify EU individuals, is arguably processing EU persons’ data and could be seen as monitoring them (especially if the model can recognize EU citizens from scraped images, which is precisely what EU regulators objected to in the Clearview case). Under California law, if the user of CommonPool is a for-profit business that does business in California (which includes virtually any larger tech company or any company selling services in CA) and meets a threshold (say they have over 25M revenue or deal in large volumes of data), then any personal information in CommonPool pertaining to California residents falls under the CCPA. It may be hard to know which entries are Californians, but realistically, a significant portion likely are (given California’s large online population). The law would require that business to, at minimum, include those categories of data in its privacy disclosures and honor any consumer rights requests related to them. Oregon’s law similarly attaches if the user ``conducts business in Oregon or targets Oregon residents'' and crosses the 100k-resident data processing threshold. The threshold count (100k individuals’ data) could easily be met by a dataset of billions (even random sampling would include more than 100k Oregonians). Notably, OCPA has no revenue threshold, so even a smaller company (or a non-profit, starting in 2025) would be covered if they process data about 100k people in Oregon. In essence, any substantial deployment of DataComp CommonPool by a tech company or organization is likely to trigger one or more of these privacy regimes. The only actors who might be outside the laws’ reach are, for example, a researcher using the data in a purely non-commercial setting and not sharing the model or outputs in regulated markets. But the moment the data or any model derived from it enters commerce or is made available to individuals in regulated jurisdictions, the privacy laws become relevant.

\subsubsection{Thresholds and exemptions} It is worth noting specific threshold quirks: CCPA’s threshold of 100k consumers/households for buying/selling data might conceivably rope in the dataset distributor if, for instance, over 100k Californians’ data was exchanged (even freely). But since the dataset is openly published (not sold) and the compilers presumably don’t have a traditional business relationship with California consumers, CCPA likely wouldn’t label the compilers as a ``business.'' Conversely, a big tech company using the data definitely has annual revenue > 25M (threshold a) and will derive value from the data (even if not selling it, simply retaining it counts as processing). OCPA’s inclusion of non-profits means if, say, a non-profit research consortium in Oregon curates or uses CommonPool and it involves >100k individuals, they would have to comply as well (OCPA from 2025 covers non-profits processing large data volumes). GDPR of course has no threshold, even processing data of one EU person can invoke rights and obligations, but enforcement priorities might focus on large-scale systematic processing, which CommonPool certainly is (processing on a “large scale” triggers requirements like Data Protection Impact Assessments under GDPR, per Article 35).

In summary, these laws attach wherever personal data from DataComp CommonPool is processed by an entity within their reach. In practice: a company in California using CommonPool is under CCPA/CPRA; any company of significant size anywhere in the U.S. using it might fall under some state law (if not California’s, then perhaps another similar state law, since many states now have comparable statutes). Any company or researcher in Europe using it must comply with GDPR. Even a non-EU company could be subject to GDPR if EU individuals’ data in CommonPool is involved in offering a service (for instance, offering a generative image service that might recreate someone’s image or personal details). Therefore, the mere presence of regulated personal data in the dataset ``anchors'' legal obligations to anyone who takes possession of it, unless they undertake robust anonymization (which, as we discuss, was attempted in part but not fully successful).

\subsection{What obligations are triggered once these laws apply? }
If an entity is subject to GDPR, CCPA, or OCPA while using DataComp CommonPool, a suite of legal duties follow. We outline the most pertinent obligations:

\subsubsection{Lawful basis / consent} Under GDPR, every processing of personal data requires a lawful basis (Article 6). For a dataset like CommonPool, it is hard to imagine a lawful basis other than legitimate interests or consent, and consent of the individuals whose data was scraped has not been obtained in any direct way. Legitimate interests (Article 6(1)(f)) might be invoked by an AI developer, arguing that training a model is in their (and perhaps societal) legitimate interest. However, this basis requires a balancing test weighing the impact on individuals’ rights. Given the dataset includes sensitive info and people have no expectation of this use, the balance may tip against the controller’s interest. Moreover, for special categories of data (GDPR Article 9) like biometric identifiers (faces) or health data that may be present, legitimate interests cannot be used at all – a specific condition like explicit consent or ``data manifestly made public by the subject'' (Art 9(2)(e)) would be needed. It is highly doubtful that individuals depicted in these images explicitly consented to this use of their data (training an AI). Thus, a GDPR-compliant processor of CommonPool would either have to filter out all special-category data or find an Art 9 exception (scientific research could be one, Art 9(2)(j), but that requires meeting strict necessity and proportionality requirements and providing appropriate safeguards). Under CCPA and OCPA, the concept of lawful basis is less formal, consent is generally not required just to collect or use regular personal data (except for sensitive data under OCPA). However, if the data will be used for certain secondary purposes, consent or opt-outs become relevant: for instance, if a business were to sell any CommonPool personal information (selling in CCPA includes any disclosure for value), it would need to provide an opt-out mechanism. If it engages in ``profiling'' or automated decision-making that produces legal or similarly significant effects on individuals, some laws (like OCPA and forthcoming CPRA regs) may require consent or at least assessments. Oregon’s OCPA explicitly requires opt-in consent for processing sensitive data. So any CommonPool entries that fall under sensitive data (which includes biometric data, specific geolocation, or a child’s data) legally mandate obtaining consent from the individual before using. Obviously, in a scraped dataset context, obtaining individual consent post-hoc is nearly impossible (the controller often doesn’t even know the identities or have contact with data subjects). This puts the controller in a position of non-compliance by default if they proceed to use sensitive personal data without consent. The only workaround is to exclude or anonymize those pieces – which again raises the question of how effective the dataset’s filtering was.

\subsubsection{Notice and purpose specification} \label{sec:legal:purpose} Privacy laws uniformly require transparency about data practices. A company using DataComp CommonPool to train a model would need to disclose in their privacy notice/policy that they collect and use personal data, potentially from third-party sources (and describe categories such as photos, resumes, etc.). Under GDPR (Articles 13-14), if personal data is collected indirectly (not from the individual), the controller must provide the individual with a privacy notice including the source of the data and the purposes of use. Complying with GDPR’s notice obligation for a scraped dataset is logistically daunting, one would have to somehow inform millions of individuals worldwide that their publicly posted content is now being used for machine learning development, giving them details and rights information. In many cases this is practically impossible, and GDPR acknowledges this by allowing exceptions if providing notice is ``impossible or would involve disproportionate effort'' (Art 14(5)(b)), but then the controller must instead publicly post the information. This means at minimum a public-facing notice should exist. For example, LAION (a German non-profit that created a 5-billion image dataset similar to DataComp) published an online notice listing broad details of the processing and offering an opt-out email for copyright or personal data takedown requests \citep{laion_takedown}. A business in California would similarly need to include in its CCPA-required privacy notice the categories of personal information it collects (which would include those scraped categories) and the purposes (e.g. ``to train and improve AI models''). Oregon’s law also mandates clear notices with purpose statements. Furthermore, purpose limitation means the controller should process the data only in ways compatible with those purposes. If the data was originally collected by websites for other purposes, a strict reading of purpose limitation (especially under GDPR) suggests that using it for model training is a new purpose that might not be ``compatible'' with the original context, absent certain conditions or consent.

\subsubsection{Data minimization and scope of collection} All regimes encourage minimizing personal data use. GDPR’s Article 5 and OCPA explicitly require that only data which is necessary for the stated purpose be collected/used. In the context of DataComp, one must ask: is each piece of personal data needed to achieve the aim of training a useful model? Likely not; the collection is opportunistic (grab as much as possible). A privacy officer evaluating this under GDPR would be hard-pressed to justify that, say, 136,000 individuals’ resumes (found in \Cref{sec:audit_resume}) are necessary to train a general image-caption model. Similarly, OCPA’s requirement that collection be ``reasonable in relation to the purposes'' might be violated if highly sensitive or irrelevant personal info was included beyond what’s needed. This principle might force a controller to actively filter out or minimize the personal data from the dataset (e.g., perhaps hashing faces or excluding text that looks like contact info) to align with the law. DataComp’s curators did attempt some minimization by blurring faces and removing some not-safe-for-work content, but our findings in \Cref{sec:audit_indv_faces} show these measures were incomplete. We estimate at least 100 million unblurred faces remain, and of our subsample we find many alongside identifiable context like names or locations (\Cref{tab:face_detection}), which is far from the minimum data necessary for any specific training objective; it’s rather an artifact of imperfect filtering. Under privacy law principles, a controller using CommonPool would be expected to proactively weed out unnecessary personal data. Failing to do so could be seen as a breach of the duty to implement Privacy by Design (GDPR Art 25), which requires controllers to integrate data protection principles (like minimization) into the processing activities.

\subsubsection{Data subject rights and control} Once personal data is being processed, individuals have rights that the controller must be able to honor. Under GDPR, these include the right to access their data, rectify inaccuracies, erase data (the ``right to be forgotten''), restrict or object to processing, and not be subject to certain automated decisions (Art 15–22). For a company holding CommonPool data, responding to such requests is extremely challenging. How would they find one person’s data among billions of samples, especially if the person only knows, for example, ``there might be a photo of me scraped from my blog?'' It’s not impossible,  the controller could at least attempt to search if provided with identifying details (some projects use perceptual hashes or embeddings to identify specific images). But the scale is prohibitive for tracing all instances, especially as our audit in \Cref{sec:audit_id_doc} shows government identifiers like social security numbers can propagate to various image hosting sites. Still, legally, if an EU citizen made a GDPR access or deletion request specifically referencing this dataset, the controller would have to attempt compliance. Failure to comply could result in regulatory action or penalties. Under the CCPA, Californians have the right to request that a business disclose what personal info it has about them and delete it (with some exceptions). A business leveraging CommonPool data would need to have processes for such requests. They might rely on the exception for data collected from a third party that is not maintained in a manner that would be considered personal (for instance, if truly anonymized or if they cannot verify the person in the data). But regulators might not look kindly on ``we have your data but can’t delete it because we can’t find it'' as a response. OCPA and similar laws also provide rights to access and delete. Thus, the legal analysis reveals a tension: the very nature of giant scraped datasets makes individual control and data subject rights nearly impossible to operationalize. This is one reason scholars argue that the notice-and-choice or individual control model breaks down at scale. In practice, if individuals start invoking their rights against AI training datasets, controllers might opt to delete broad swathes of data or refrain from use, as compliance on a piecemeal basis could be infeasible.

\subsubsection{Special rules for sensitive data} All three regimes impose stricter conditions on sensitive personal data, which DataComp CommonPool unquestionably contains. Under GDPR, as mentioned, processing data like facial images (biometric data) for identification, health information, or data revealing race/ethnicity (which a resume or photo can do, see \Cref{fig:sociodem_examples} and \Cref{fig:resume_examples} for examples) is prohibited unless an exception applies (Art 9). One possible GDPR argument is that the data subjects ``manifestly made public'' these special-category data themselves – for instance, someone publicly posts their own photo, resume, or medical info. That exception (Art 9(2)(e)) might allow processing, but it’s a gray area; arguably they made it public for a certain audience or purpose, not for any use whatsoever. Also, if the person in the photo is not the one who posted it (e.g., a news article about someone’s health in \Cref{fig:children_examples}), the exception doesn’t apply. DataComp CommonPool has many images of people taken by others (indicated by mentions of celebrity names in \Cref{sec:audit_celebrity}), so ``manifestly public by the subject'' fails. Thus for a GDPR-compliant approach, a controller would need either explicit consent from each person (impossible at scale), or to fit under the research exception (Art 9(2)(j)) which requires that processing be necessary for research in the public interest and subject to EU or member state law providing appropriate safeguards. A commercial company training a product likely cannot claim the research exemption; an academic might, but even then must implement safeguards like de-identification. Under CCPA/CPRA, ``sensitive personal information'' such as account passwords, financial info (found in \Cref{sec:audit_id_doc}), precise geo-location (found in \Cref{sec:audit_exif_tags}), or contents of communications can be used by a business only for limited purposes (generally, what is necessary to provide the service, or as permitted with notice) if a consumer directs them to limit it. If a business were, say, using CommonPool and it contained login credentials or credit card numbers (which it does, in some images of documents shown in \Cref{fig:identity_doc_examples}), that’s sensitive info that should never be exploited beyond necessary security research. OCPA goes further to require consent for any processing of sensitive data. In context, any use of a child’s image or personal details from CommonPool without parental consent (see \Cref{sec:audit_children}) is a clear violation of OCPA. Also, any biometric data (like using faces to improve a face recognition algorithm) would technically require prior consent of the individual in Oregon. Even if enforcement is unlikely, legally the obligation is there. So a controller would need to filter out all children’s data and biometric identifiers or risk non-compliance. Another sensitive category is resumes --- these often contain contact info, education, employment history. While not ``sensitive'' by statutory definition, they are highly personal. If a resume includes something like a Social Security number or driver’s license (which some do, as shown in \Cref{fig:resume_pii_types}), that becomes sensitive (government ID number). CommonPool was found to have images of passports and driver’s licenses in \Cref{sec:audit_id_doc}, which are both sensitive and highly regulated (for example, storing driver’s license numbers triggers breach notification duties if breached, under various U.S. laws). Financial data like credit card numbers in the dataset raise data breach concerns: under all U.S. states’ laws, if a company inadvertently exposed those, they’d owe notifications. Thus, even beyond privacy-specific laws, holding such data creates liability if it leaks or is hacked.

\subsubsection{Security and breach notification} Privacy laws also require securing personal data. GDPR Article 32 mandates appropriate technical and organizational measures to protect data. CCPA requires ``reasonable security'' and provides a private right of action (lawsuit) for consumers whose sensitive data (like certain ID numbers) is breached due to lack of reasonable security. OCPA similarly obligates reasonable data security practices. In the context of DataComp, any entity storing the dataset or integrating it into systems must implement strong protections against unauthorized access. This is especially important because the dataset contains some very sensitive elements (e.g., full credit card details, identity documents). If, hypothetically, a company using CommonPool got hacked and the hackers obtained these personal entries, that company could face breach notification duties to potentially millions of individuals (though identifying and contacting them would be almost impossible, which doesn’t absolve the duty). The inability to notify affected persons (because the data was scraped without emails or phone numbers perhaps) is a nightmare scenario, it means the company simply cannot fully comply with breach laws if an incident occurred. This is a legal risk of assembling data that you cannot trace back. Regulators might view the initial decision to compile such data as negligent if it could never be properly safeguarded or managed.

\subsubsection{Automated decision-making and profiling} \label{sec:legal:profiling} One might ask if GDPR’s provisions on automated decisions (Article 22) apply. Article 22 gives individuals the right not to be subject to a decision based solely on automated processing (including profiling) that produces legal or similarly significant effects. Training an AI model on DataComp CommonPool doesn’t directly make decisions about those individuals, so Article 22 isn’t directly triggered by the training process. However, if the resulting model is used in a way that profiles or affects people, then those individuals have rights regarding how their data was used to create that model. This is a cutting-edge area: there’s debate about a person’s right to opt out of being included in training data that will be used in profiling. GDPR doesn’t yet explicitly give a right to opt out of processing for AI model training (unless it’s considered processing for a legitimate interest to which they object under Art 21). But some have argued using personal data to materially inform algorithmic decisions about people could trigger obligations of fairness or explanation. For example, if CommonPool were used to build a facial recognition system that is then used by police, EU citizens might challenge the legality of processing under GDPR’s law enforcement provisions or human rights law.

\subsubsection{Data Protection Impact Assessment (DPIA)} Under GDPR Article 35, if processing is likely to result in high risk to individuals (especially using new technologies on a large scale with sensitive data), a DPIA must be conducted. A controller planning to use DataComp CommonPool should perform a DPIA evaluating the risks to rights and freedoms of individuals whose data is in the set. It would almost certainly conclude there are significant risks (e.g., unauthorized disclosure, bias, misuse of personal images). Mitigation measures (like additional filtering, encryption, access controls, or not using certain data) should then be taken. OCPA similarly requires documented risk assessments for processing that presents a heightened risk of harm, such as processing sensitive data or profiling that could lead to unfair outcomes. Training an AI on personal images might qualify as profiling with potential disparate impact (imagine the model reinforces biases or misidentifies certain demographic groups). Thus, these laws demand a proactive, documented examination of the privacy impacts of using CommonPool, something that currently, many AI practitioners might not be doing.

\subsection{Were DataComp’s own filtering and anonymization efforts legally sufficient?}
\label{sec:legal_analysis:filtering}

The dataset creators did implement some privacy filters, notably, automated face blurring to obscure identities in images. However, our audit showed this was far from comprehensive: in \Cref{sec:audit_indv_faces} a proportion of real people’s faces went unblurred due to the tool’s failure to detect them. Legally, if one tries to anonymize personal data but the anonymization is incomplete, the data must still be treated as personal data. GDPR, for instance, considers data ``anonymous'' only if individuals are no longer identifiable taking into account all means reasonably likely to be used to identify them. A simple blur or pixelation on a face may not meet that standard, especially at scale --- advances in AI can reverse blurring \citep{mcpherson2016defeating} or at least identify individuals from unblurred parts (hair, posture) or by correlating with other images \citep{oh2016faceless}. Moreover, blurring the face in an image does nothing if the caption or surrounding text mentions the person’s name or other info. In DataComp, even where faces were blurred, in \Cref{tab:face_detection} we find instances where the accompanying alt-text still states, ``Photo of [Name] at [event].'' That remains personal data. Thus, from a GDPR perspective, the dataset as released was not effectively anonymized and should be treated as personal data. The legal expectation for anonymization is very high (truly irreversible de-identification). Short of that, one might pursue pseudonymization, replacing identifiers with codes, but here the images are inherently identifying (a face is a unique identifier). The DataComp curators also did not engage in methods to remove obvious PII strings (using tools to detect things like emails or SSNs), and while we found plenty of ID numbers, names, and contacts, prior work shows that PII detection tools are not sufficient \citep{mireshghallah2024trust} and create a ``\textit{false sense of privacy}'' \citep{xin2025false}. This underscores that ``no automated cleaning can remove all PII,'' as we demonstrated in our dataset audit. Privacy law would likely concur: if personal data remains, the controller cannot claim exemption by saying ``we tried to filter it.'' Instead, the controller must continue to handle the data under applicable law or take further steps to mitigate risk.

One could ask: does blurring faces reduce the legal risk at all? It might mitigate it somewhat. For example, a fully blurred face might no longer be ``biometric data'' because you cannot recognize the person from it. If the blur is strong enough that the person is not reasonably identifiable, that particular image might fall out of definitions of personal data. However, if at least 100 million faces were missed (\Cref{sec:audit_indv_faces}), the effort fails to appreciably lower the overall risk. Additionally, partial mitigation could demonstrate awareness of privacy issues, which regulators could use to argue the controller knew of the risk yet didn’t do enough. In the U.S., attempting to de-identify data can provide some safe harbor (like CCPA says de-identified data is not “personal information” if it meets certain criteria). But de-identified in that context means data that ``cannot reasonably be used to infer information about, or otherwise link to, a particular consumer.'' Given the residual personal info in DataComp CommonPool, it’s hard to argue it’s de-identified. For example, an image showing a credit card with the numbers visible and a person’s name (which our example findings revealed in \Cref{fig:identity_doc_examples}) is clearly identifiable to that cardholder. No amount of general dataset size changes that. Thus, legally, the filtering was not sufficient to escape privacy obligations. At best, it was an attempt at data protection by design, but an underinclusive one.

An interesting legal question is whether releasing the dataset with incomplete blurring could be seen as a form of data processing for research that is privileged. Some laws and courts recognize that publishing personal data for public interest research or journalism can be protected by freedom of expression. But here it’s not journalistic, and the personal data belongs to numerous unsuspecting individuals whose interests were not considered individually. The lack of a specific legal basis (no consent, etc.) means that incomplete anonymization doesn’t cure the issue; it just demonstrates that a risk was acknowledged. Regulators like the UK’s ICO have fined companies even when they attempted anonymization that proved inadequate (e.g., characterizing poor de-identification as essentially an unauthorized disclosure of personal data).

\subsection{Is relying on ``publicly available'' data a defensible legal strategy for AI datasets?}
\label{sec:legal:public_avail}

Relying on “publicly available” data may sound like a legal shortcut, but in the context of AI training datasets, it’s increasingly a trapdoor. As privacy laws evolve, accessibility is no longer a proxy for permissibility. Laws like the GDPR, CCPA, and OCPA make clear that just because data is online doesn’t mean it’s free for the taking. All three legal regimes reject the simplistic notion that data is “public” merely because it can be accessed online.  Without a reasonable understanding of user intent, context, and consent, sweeping up personal information from the web and calling it “public” is a legally risky and often indefensible strategy. 

The DataComp CommonPool dataset does not qualify as “publicly available” data under state consumer privacy laws like the Oregon Consumer Privacy Act (OCPA) and the California Consumer Privacy Act (CCPA), and should not be exempt from legal protections. Though some of the information in CommonPool may have been posted online, the legal definition of “publicly available” is more nuanced than mere accessibility. Both the OCPA and CCPA impose specific conditions to prevent the misuse of personal data that individuals did not affirmatively and knowingly place into the public sphere for unrestricted use.

\subsubsection{\textbf{Indiscriminate scraping fails the ``reasonable basis'' standard}}

Under both the OCPA and the CCPA, information is not considered ``publicly available'' simply because it can be found on the internet. The laws require that a controller or business have a \textit{reasonable basis} to believe that the data was lawfully made available to the public \textit{by the consumer}. For example, OCPA 646A.570·(13)(b)(B) allows an exemption only where ``a controller reasonably has understood [the data] to have been lawfully made available to the public by a consumer.'' Similarly, CCPA 1798.140(v)(2)(B)(i)(II) requires that the business “has a reasonable basis to believe” that the consumer made the information publicly available. 

In the case of DataComp, this standard cannot be met. The dataset is created by automated systems that crawl and scrape data indiscriminately, without human oversight or consumer context. These scrapers cannot discern whether data was posted intentionally for public reuse or under restricted circumstances, such as within a social media profile, a comment section, a classroom forum, or a personal blog with limited viewership. As a result, they cannot reasonably determine consumer intent or consent.

What’s more, the volume and automation of this data collection preclude any individualized assessment of context. If a business or controller is scraping billions of data points with no mechanism for filtering out user-restricted or audience-limited disclosures (such as sites from \Cref{sec:audit_unavailable}), it cannot credibly claim to have a ``reasonable basis'' for believing the data was lawfully made publicly available. The law contemplates thoughtful, contextual evaluation, not mass extraction based on surface availability.

This same concern is even more pronounced under the GDPR. Article 6 requires that personal data processing be grounded in a valid legal basis, such as consent, legitimate interest, or performance of a contract. Even publicly accessible data may still require a legal basis for further use. Moreover, Recital 47 of the GDPR states that reliance on ``legitimate interest'' must be balanced against the reasonable expectations of the data subject. A data subject posting on a message board, publishing a blog post, or uploading a photo cannot reasonably expect their content to be scraped, stored in perpetuity, and used to train AI models --- especially if the image was uploaded before these technologies even existed (for instance, some resumes found in \Cref{sec:audit_resume} or facial images found in \Cref{sec:audit_indv_faces}). The absence of notice, transparency, or opportunity to object violates both the GDPR’s fairness principle (Article 5(1)(a)) and the requirement of transparency under Articles 12–14. Without a valid legal basis and fair processing, scraping and reuse of such materials is not lawful, even if the content is technically accessible online.

\subsubsection{\textbf{Widely distributed media $\neq$ automatically public under privacy law}}

Both laws provide a narrow carveout for information made available through ``widely distributed media.'' But that does not give blanket immunity to scraped web content. This provision exists to exclude traditional journalistic content and intentionally public communications, like letters to the editor or public government filings, not to exempt all content accessible via a search engine.

For instance, the CCPA makes clear that \textit{biometric data} collected without the consumer’s knowledge is \textit{not} “publicly available,” even if it was technically accessible. This reflects an underlying principle: data shared without meaningful understanding or consent is still protected.

Similarly, data scraped from discussion forums or social media platforms may be technically accessible but not ``widely distributed'' in the sense intended by the law. Many platforms have shifting or ambiguous privacy settings, and users often do not realize that their content is publicly indexed, especially if they are not sophisticated about data privacy. This ambiguity undermines any argument that consumers clearly and affirmatively made such data available to the general public.

\subsubsection{\textbf{Disclosures to a limited audience are not ``publicly available''}}
\label{sec:legal_analysis:limited_audience}

California further clarifies that even when consumers disclose information online, it is not “publicly available” if it was disclosed to a specific person or group \textit{with an expectation of audience limitation}. CCPA 1798.140(v)(2)(B)(i)(III) explicitly excludes from the “publicly available” exemption any data that the consumer shared \textit{with audience restrictions}. Again, DataComp’s scraping model does not (and cannot) distinguish between content shared globally and content disclosed to a limited group.

As such, data shared in online forums, academic or professional listservs, group chats, or social platforms with customizable privacy settings would often fall outside the CCPA’s definition of publicly available information. If the dataset includes these types of data (and preliminary audits suggest that it does, e.g. a social security number originating from LinkedIn \Cref{fig:identity_doc_examples}, resumes from Pinterest in \Cref{tab:resume_overview}, and a screenshot of an online forum in \Cref{fig:ocr_sample_images}), they are clearly out of scope for the exemption.

\subsubsection{\textbf{Public access does not equal public availability under law}}

Finally, both the CCPA and OCPA rest on the understanding that “publicly available” is a legal term of art, not a synonym for “can be found online.” Treating all internet-accessible information as “publicly available” would render the statutory exemptions meaningless and invite systemic abuse by companies that profit from mass scraping. The laws instead require a deeper inquiry into the \textit{source}, \textit{intent}, and \textit{context} of the data shared.

The GDPR, while not using the term “publicly available” as a formal exemption, still requires controllers to consider context and user expectations. The European Data Protection Board has made clear that the publication of data online does not strip individuals of their rights under the GDPR. Even data shared voluntarily does not give downstream actors carte blanche to reuse it for incompatible purposes. Any secondary use, especially for high-impact applications like AI training, requires a fresh legal basis and must be compatible with the original context of collection (per GDPR Article 6(4)). 

The act of scraping publicly \textit{accessible} content does not transform it into “publicly available” data under the law. Both the CCPA and OCPA limit this exemption to cases where there is a reasonable, contextual understanding that the data was knowingly placed into the public domain. The GDPR goes further, requiring not just accessibility, but lawful processing grounded in purpose compatibility, transparency, and data subject rights.

The DataComp CommonPool dataset, by design, ignores these safeguards. It amasses personal data without meaningful legal justification, often in direct contradiction to user expectations and platform norms. Its indiscriminate, large-scale scraping practices circumvent not just the spirit but the letter of modern privacy and data protection laws. Policymakers and regulators should be skeptical of claims that internet scraping inherently falls outside privacy regulation. In reality, web-scraped datasets like CommonPool raise urgent legal and ethical questions that warrant scrutiny, not exemption.

\subsection{Summary}

In conclusion, the legal analysis shows that using a dataset like DataComp CommonPool creates significant compliance challenges under prevailing privacy laws. The personal data in the dataset is squarely within the scope of GDPR, CCPA, OCPA, etc., meaning entities cannot simply ignore those obligations. They must consider jurisdiction (very likely at least one law will apply), then fulfill duties of transparency, lawful basis, and data subject rights – tasks that, given the dataset’s nature, are extremely burdensome if not impossible at scale. The attempts at anonymization (face blurring) were not sufficient to remove these obligations because large amounts of identifiable data remain. Indeed, those attempts, while well-intentioned, illustrate the difficulty of truly de-identifying unstructured big data. Relying on the data being ``public'' is not a silver bullet; privacy laws provide some leeway for public data but also contain important caveats and are backed by a broad consensus that privacy is not forfeited upon disclosure. Ultimately, current notice-and-consent frameworks falter in this scenario --- individuals were neither notified nor asked. Thus, any organization leveraging DataComp CommonPool should adopt a very cautious approach: aggressively filter out known personal identifiers, limit the purposes to something justifiable, conduct risk assessments, and be prepared to cease using or delete portions of the data if individuals exercise their rights. They should also monitor legal developments, as regulators are actively grappling with how to apply existing laws to AI datasets and may issue guidance or take enforcement action (for example, enforcement against Clearview AI signals that mass scraping of faces is unacceptable under data protection law). In many ways, DataComp is a test case for the tension between innovation through massive data aggregation and compliance with privacy principles. Our analysis suggests that without changes, either in how datasets are curated or in the legal approach, there is a substantial compliance risk and a broader normative concern that individuals’ privacy is being compromised at scale without the tools to effectively protect it.

%% file: sections/06-discussion.tex
\section{Discussion}
\label{sec:discussion}

We now present the recommendations and implications of our audit results and legal analysis in terms of using datasets like DataComp CommonPool to train models or for other purposes. Our discussion is not specific to the DataComp dataset, as it is likely that other web-scraped large-scale datasets contain similar risks of personal data, despite automated sanitization efforts. Our audit and legal analysis reveal deep structural tensions between large-scale dataset curation practices and the enforcement mechanisms of contemporary privacy law. As shown in \Cref{sec:legal_analysis}, existing “publicly available” exemptions cannot be coherently applied to entire web-scraped training datasets under current regulatory frameworks. Nonetheless, we argue that prevailing laws remain insufficient to address core ethical privacy principles, such as control, consent, and contextual integrity, particularly in the contemporary AI web-scraping ecosystem. As the collection, redistribution, and downstream use of personal information become increasingly automated and diffuse, the foundational assumptions underlying existing privacy frameworks are rendered ineffective, if not obsolete. We therefore advance a set of recommendations for both policymakers and machine-learning practitioners.

\subsection{Challenges of existing privacy laws at scale}
\label{sec:discussion_scale_issues}
The findings of our audit illuminate deep structural tensions between the practices of large-scale dataset curation and the enforcement mechanisms of modern privacy law. As the collection, redistribution, and downstream usage of personal information becomes increasingly automated and dispersed, the foundational assumptions of existing privacy frameworks, namely individual control, meaningful consent, and data minimization, are rendered ineffective or outright obsolete. Below, we outline four interlocking challenges that arise when applying current privacy laws to web-scale data practices like those underpinning DataComp CommonPool.
\subsubsection{The collapse of individual control}
As articulated by privacy scholars, the model of ``privacy self-management,'' which expects individuals to read privacy policies, understand potential downstream uses, and assert their rights has collapsed under the weight of modern data practices \citep{solove2024murky, solove2024great}. Our audit underscores this failure: the individuals whose resumes, government IDs, or children’s medical information appear in CommonPool could not have meaningfully understood or anticipated these downstream uses at the time of upload. Indeed, in many cases, the content was posted years before the rise of large-scale foundation models, making the notion of ``informed consent'' retroactively implausible.
Even where opt-out mechanisms exist, such as the integration of Spawning AI with Hugging Face, these tools presume a level of awareness, technical skill, and effort that is unrealistic at scale. Data subjects are not just unaware that their data has been scraped; they are unaware that it ever could be. And even if they discover their data’s inclusion in a training dataset, privacy law does not adequately address revocation of consent post hoc. For example, \Cref{sec:audit_unavailable} of our audit shows that a nontrivial portion of CommonPool (over 21 percent of sampled URLs) now fails to download, including roughly 0.4 percent of samples (an estimated 50 million) from a site that currently requires a login. The dataset nonetheless retains the metadata and text of these entries, and models trained on earlier versions may retain the visual content. There is no legal mechanism to retroactively purge these artifacts.
The focus on consent at the point of collection falters when the act of collection is hidden from the data subject. Furthermore, it fails to address the dynamic nature of web content: data that was once ``public'' may later be made private, but this change has no bearing on datasets already scraped. The law thus fails to honor either the data subject’s evolving intentions or their right to meaningful withdrawal of consent.
\subsubsection{Web-scale data is ``too big to privacy''}
\label{sec:discussion:tracing}

A second, compounding problem is that tracing data provenance at internet scale is functionally infeasible. As web pages and digital images propagate across the internet (shared, reposted, and mirrored) tracing any given image or caption back to its original context becomes an exercise in futility. This makes it nearly impossible to assess whether scraped data was originally behind a login wall, taken from a compromised database, or uploaded with restricted permissions.
Our audit provides a striking example in \Cref{fig:identity_doc_examples}: a social security number uploaded by an individual to a social media site was later scraped from an entirely different image hosting service, raising questions about provenance and the individual's original intent. This demonstrates that even if a user attempts to revoke their disclosure, downstream propagation renders such efforts ineffective.
\Cref{sec:audit_unavailable} further reveals that many previously accessible images have since become restricted. However, there is no reliable way to determine whether a login requirement was in place at the time of scraping, or if the access was revoked afterward, much less whether the takedown was initiated by the data subject themselves. In either case, the privacy harm remains: prior downloads, including by researchers and commercial entities, retain access to data that is no longer meaningfully public, with no obligation to re-evaluate its lawfulness.
This collapse in traceability not only undermines consent but exposes the limits of relying on ``public availability'' as a legal safe harbor. The practical inability to distinguish between a deliberately public blog post and an inadvertently leaked medical record suggests that more robust frameworks are needed, ones that prioritize data integrity and context, not just accessibility.
\subsubsection{Dataset monoculture and the waterfall of harm}
A third challenge is the monoculture effect inherent in web-scale datasets: once personal data is included in a widely used dataset like CommonPool, it is replicated and amplified through every model trained on it. This is not merely a matter of one model misusing personal data, it is hundreds or thousands of models, potentially deployed across commercial, academic, and governmental contexts.
Unlike previous concerns about model behavior, the risk here stems from the centralization of data sources. As described in Figure 1 and throughout Section 3, the pipeline of AI model development is premised on reusing existing datasets to build ever-larger and more generalizable systems. But the reuse of CommonPool, and its release as a URL index rather than a frozen corpus, means every download triggers a fresh crawl, potentially retrieving newly restricted or outdated content. The same image (if propagated to other image hosting sites) may be downloaded millions of times, long after the data subject has removed or placed the original behind a paywall.
Moreover, the burden on data subjects to file data subject access requests (DSARs) for each instantiation of their data is untenable. Even if a subject were to discover their image in CommonPool, they would need to issue requests to every model developer who has used the dataset, a task rendered impractical by the absence of dataset tracking or provenance tools. And even if a request were granted, deleting personal data from a model’s training corpus is a deeply unresolved technical and legal problem \citep{hutson2024america}.
The transition from a ``web of documents'' to a ``web of training data'' demands a rethinking of data governance. The internet made data widely \textit{available}; web-scraped AI pipelines have made data widely \textit{processed}. This shift changes not just the scale, but the very stakes of privacy.
\subsubsection{Incomplete anonymization and the limits of de-identification}
Finally, while some developers attempt to mitigate risk through de-identification, our audit demonstrates the limits of those efforts. As detailed in \Cref{sec:legal_analysis:filtering}, CommonPool employs face blurring and filters to remove sensitive content, but these methods are incomplete. An estimated 100 million images of real human faces are not blurred despite the default tooling, and OCR-extracted text still reveals credit card numbers, names, and even birth certificates.
The result is a system in which purported anonymization is neither comprehensive nor verifiable. Worse, the existence of such flawed de-identification gives dataset developers a false sense of compliance, while leaving data subjects exposed to downstream inference, reidentification, and profiling.
Privacy laws allow for certain processing exemptions where data has been sufficiently anonymized. However, as our audit makes clear, the reasonability of these claims collapses at scale. At a dataset of 12.8 billion samples, even a 0.1\% failure rate translates into millions of instances of potential privacy harm.
Legal frameworks must move beyond the binary of ``identified'' vs. ``anonymous,'' and instead impose robust standards for anonymization that are tied to dataset size, processing purpose, and downstream risks. Moreover, data minimization principles must be enforced irrespective of data identifiability: collecting vast troves of ``possibly anonymous'' data still imposes measurable risk.
Pending legislation like California’s SB 2013 and Colorado’s AI Act offer a glimpse of reform. These laws include requirements for dataset documentation and impact assessments. But they must go further, mandating privacy-preserving evaluations of curation practices and requiring public disclosure of filtering tools and their performance metrics.

\subsection{Recommendations for law \& policy}

\subsubsection{\textbf{State Attorneys General should enforce aggressively to preserve the integrity of ``publicly available'' exceptions}}

Attorneys general in states with comprehensive privacy laws, like California and Oregon (among others), should act decisively to prevent the hollowing out of consumer privacy protections through the misuse of ``publicly available'' exceptions. The DataComp dataset represents a paradigmatic abuse: personal data scraped indiscriminately, at scale, and without regard for user expectations or context. Permitting companies to sidestep liability merely because data was technically accessible online eviscerates the spirit of these laws.

Both laws impose a ``reasonable basis'' standard for treating data as publicly available, a deliberately higher bar than mere access. Yet controllers exploiting datasets like DataComp often bypass this safeguard entirely, relying on automation and volume to collect content without any contextual analysis. This practice not only violates the letter of the law but undermines its purpose: to restore agency to individuals over their personal data.

Enforcement authorities should use existing statutory tools to investigate companies that use datasets like DataComp without proper diligence. They should:

\begin{itemize}
    \item Challenge the assertion that scraped personal data—especially biometric, children’s, or resume-related information—was lawfully made public by the data subject.
    \item Use their investigative powers to examine whether businesses using such data conducted meaningful context assessments.
    \item Issue interpretive guidance to clarify that mass scraping fails the ``reasonable basis'' test by default unless extraordinary safeguards are in place.
    \item Pursue enforcement actions against high-profile users of DataComp as a deterrent, signaling that privacy law will not permit large-scale circumvention through technical loopholes.
\end{itemize}

Enforcement is not only legally justified; it is normatively essential. The legitimacy of privacy laws depend on the idea that individuals retain rights over their personal data, even when that data is visible online. If regulators do not defend this principle, the web will become a de facto public domain for surveillance, profiling, and commodification.

\subsubsection{\textbf{State legislatures should close the web-scraping loophole and modernize the ``publicly available'' exception}}

Legislatures should act to modernize the ``publicly available'' exception in consumer privacy statutes by drawing clear lines against the misuse of scraped data. The current definitions were crafted in an era of limited data sharing and do not account for the reality of AI today, in which indiscriminate webscraping is used to vacuum up billions of personal records, often without the knowledge or consent of the data subject.

To address this, state laws should be amended to include the following reforms:

\paragraph{1. \textbf{Express Carveout for Indiscriminately Scraped Data}}

Amend the definition of ``publicly available'' data to exclude personal data collected through automated or indiscriminate web-scraping, unless the controller can demonstrate that the data was lawfully made available to the public by the data subject with clear intent for unrestricted downstream reuse. This preserves limited, contextual reuse of truly public information (e.g., letters to the editor, government records) but shuts the door on practices like DataComp, where no intent, consent, or meaningful context is established.

\paragraph{2. \textbf{Presumptive Protection for Sensitive or Contextualized Data}}

Automatically disqualify the following from the ``publicly available'' exemption: (i) sensitive data, including biometric data and children’s data, even if accessible online; (ii) any data disclosed on platforms that allow audience restrictions, unless disclosed with a public license or tag; (iii) data disclosed in contexts where reuse is materially incompatible with the original purpose (e.g., training AI on personal essays or support forum posts). This approach aligns with reasonable expectations of privacy and recognizes that technical access does not equate to waiver of privacy interests.

\paragraph{3. \textbf{Mandate Transparency and Attribution}}

Require any controller invoking the ``publicly available'' exemption to document: (i) the source of the data; (ii) why the data was considered publicly available; (iii) how the controller confirmed the user’s intent and awareness; and (iv) whether the platform terms of service allowed for scraping and reuse. This shifts the burden of justification to the party exploiting the data, not the data subject.

Without legislative reform, the ``publicly available'' exemption becomes a backdoor for pervasive surveillance and privacy harm at scale. It was never meant to immunize AI companies from obligations simply because they used a webscraper. States that claim to lead on consumer privacy cannot permit exceptions that swallow the rule. To that end, we propose the following possible language to modernize the "publicly available" data exception to close the gap that allows AI developers to harvest massive quantities of personal data under the pretense that it is publicly available.

\textbf{Section [X]: Amendments to the Definition of ``Publicly Available'' Information}

\begin{quote}

\textbf{(A)} \textit{Revised Definition of Publicly Available Data}

 ``Publicly available information'' means any personal data that:

\begin{enumerate}
    \item Is lawfully made available from federal, state, or local government records;
    \item Is lawfully made available to the general public by the consumer with clear and affirmative intent for such information to be broadly accessible without restriction; or
    \item Is available in widely distributed media intended for general public dissemination (such as news broadcasts or publicly licensed publications).
\end{enumerate}

\textbf{(B)} \textit{Exclusions from Publicly Available Data}

 Notwithstanding subsection (A), the following categories of personal data shall not be considered ``publicly available'':

\begin{enumerate}
    \item Personal data collected or processed through automated, large-scale, or indiscriminate webscraping methods, unless the controller demonstrates that:
    \begin{itemize}
        \item (i) The data subject explicitly made the data publicly accessible with no audience restriction;
        \item (ii) The data subject had actual knowledge and intent to permit unrestricted downstream use; and
        \item (iii) The platform’s terms of service clearly authorized such scraping and reuse.
    \end{itemize}
    \item Personal data disclosed in contexts where audience limitation, expectation of privacy, or contextual sensitivity is apparent, including but not limited to:
    \begin{itemize}
        \item (i) Social media posts with non-public or limited visibility settings;
        \item (ii) Content from discussion boards, classroom platforms, or professional forums not intended for general indexing;
        \item (iii) Any disclosure where the data subject did not manifestly intend the information to be used for unrelated commercial or algorithmic training purposes.
    \end{itemize}
    \item Sensitive data, including biometric data and a child's personal data.
\end{enumerate}

\textbf{(C)} \textit{Transparency and Documentation Requirements}

 A controller or processor relying on the ``publicly available'' exception must maintain internal records demonstrating:
\begin{enumerate}
    \item The source of the data;
    \item The legal basis for concluding the data was publicly available as defined in this Section;
    \item That reasonable measures were taken to assess the data subject’s intent and the original context of disclosure; and
    \item That any scraping or automated collection complied with the originating platform’s access terms and community guidelines.
\end{enumerate}

\textbf{(D)} \textit{Purpose Limitation}

 Data obtained under the ``publicly available'' exception may only be processed for purposes that are compatible with the context in which the data was originally disclosed. The use of such data for:

\begin{itemize}
    \item (i) Training or developing algorithmic models;
    \item (ii) Profiling; or
    \item (iii) Commercial repurposing unrelated to the original context

 shall not be presumed compatible without specific consumer consent.

\end{itemize}
\end{quote}
By adopting a clarified and modernized exception to the definition of publicly available data, state legislatures can better align privacy law with the realities of contemporary data practices and the technical architectures of AI development. The proposed reform does not prohibit the use of public data outright; rather, it imposes necessary constraints on the indiscriminate scraping and repurposing of personal information in ways that disregard user context, intent, or consent. It operationalizes key privacy principles—purpose limitation, data minimization, and transparency—by ensuring that public accessibility is not conflated with unconditional legal availability. Importantly, it also harmonizes domestic privacy law with emerging international norms, particularly those reflected in the GDPR’s emphasis on contextual fairness and lawful reuse. Where AI systems increasingly rely on large-scale ingestion of personal data, these clarifications are essential to preserving the normative foundations of privacy and data protection law. Without such legislative intervention, the exception for publicly available data risks becoming a structural loophole, one that undermines individual rights at scale and erodes the practical enforceability of privacy protections in the age of AI.

\subsection{Recommendations for machine learning}
\label{sec:discussion:ml}

In this section, we discuss recommendations for \textit{machine learning practitioners and researchers} based on our audit findings and legal analysis, as well as contributions from current work.

\subsubsection{Misperceptions of publicly available data}

As stated in \Cref{sec:legal:public_avail}, legally ``publicly available'' data is not equivalent to data that is ``accessible'' via web-scraping. The ''publicly available'' exception in various data protection laws may not to apply data posted from a breach or data behind a login screen, or in the case of GDPR, data not uploaded by the data subject themselves. Moreover, the ``vacuuming'' of the internet counters the data protection principle of data minimization to gather only the data ``necessary for the stated purpose.''
The machine learning community should be aware of the distinction between data that is legally public versus data that is available online.

Based on current mechanisms for individual consent, opt-out policies should also be followed, despite the Robots Exclusion Protocol being unenforceable \citep{ai_robots_txt}. Websites that disallow web-scraping indicate a revoking of consent of their data being used. Datasets and web-scraping tools therefore should respect these protocols at the \textit{site level} (rather than image-level \cite{img2dataset_robots_tag} and \textit{at the time of downloading} (see \Cref{sec:audit_unavailable}), to align with web-crawling best practices \citep{web_crawler_practice}. Sites like spawning.ai \citep{spawning} that ask for individual opt-out may give some indication of consent, but may not be enough to remedy privacy harms, as it is implausible for individuals to know where their personal data exists on the internet.

\subsubsection{General-purpose models}

The development of foundation models that are agnostic to a purpose run counter to the data protection principle of purpose specification. Machine learning practitioners should explicitly define narrow use cases of training models \citep{bommasani2021opportunities}, especially in cases where consent is required to collect training data. Large vision models can also be considered face recognition tools, as many vision-language models have been shown to generate or classify celebrities in their training datasets \citep{chen2023text, zhao2025magicnaming}. This implies the collection of facial images as biometric data for recognition purposes, even if models are not explicitly trained to do so. Practitioners and researchers should be aware that prior notions of ``facial geometry'' may change as legal definitions and technologies evolve.

\subsubsection{Researching web-scraped data}

Specifically for researchers, we repeat the need to release datasets with constricted use licenses like the RAIL license \citep{contractor2022behavioral} (although enforcement still remains an open question) rather than restriction-free licenses when artifacts are not intended to be deployed for commercial use. While research often builds on top of other work for collaboration and advancements, these dependencies become more difficult to trace as literature proliferates --- researchers should be more critical of the practice of training on web-scraped datasets just because prior work has done so.  Even as research exemptions are included in existing privacy laws, there are still certain expectations that collected data for research aligns with data protection principles: GDPR safeguards in Article 81 \citep{gdpr}, for instance, require data minimization and anonymization to fulfill the research purpose, and that the use of personal data does not affect individuals.

Our work reveals that privacy risks still remain with studying personal data on the internet for research purposes, as we highlight settings where data may not be considered ``legally public'' even if collected from the web. University institutional review boards should re-evaluate existing web-scraped datasets that previously gained IRB approval and re-examine their exemption protocols, given the amount of personally-identifiable information of real individuals found on the web (whether publicly available or not).

\subsubsection{Alternatives to web-scraping}

Given the wide usage of datasets like DataComp CommonPool, we ask, is it too late? Where do we go from here? We list several alternatives and concrete remediations, although we acknowledge that these approaches may not completely remove all personal information or downstream privacy harms.

\begin{enumerate}
    \item While automated sanitization methods may never guarantee complete removal of personal data \citep{hong2024s, xin2025false}, proactive evaluation of data cleaning and justification of the use of existing techniques will alleviate privacy risk more than post-hoc audits.
    \item We encourage empirical research on implementing data minimization when training models through techniques like anonymization \citep{goldsteen2022data}, early stopping \citep{shanmugam2022learning}, and data pruning \citep{ganesh2024data}.
    \item Dataset curators should ideally ask for explicit assent for inclusion in datasets or model training \citep{andrews2023ethical}. For example, Mozilla Common Voice \citep{ardila2019common} is a community-driven open speech corpus where contributors volunteer to record. On the text side, the Common Corpus is an open dataset of two trillion tokens that are either uncopyrighted or under permissible licenses \citep{langlais2025common}.
    \item To address the propagation of images across websites, there have been several mechanisms to maintain attribution on the web, such as Adobe's Content Credentials \citep{adobe_credentials}. Not only do Content Credentials label generated media \citep{gamage2025labeling}, they are also attached to content when shared and can indicate creator's usage preferences, although the extent to which these standards are adopted remains to be seen.
    \item For large models that have already been trained, \citet{lee2025beyond} argues that algorithmic disgorgement may not be necessary. Technical interventions such as machine unlearning \citep{nguyen2022survey}, data attribution \citep{li2023survey} to avoid output attributed to personal data, and fine-tuning \citep{singh2024whispered} may prevent privacy leakages, although we caution that these techniques require careful intervention, as prior work has found that fine-tuning can may increase the rate of training data extraction \citep{li2024shake}.
\end{enumerate}

%% file: sections/07-conclusion.tex
\section{Conclusion}
\label{sec:concl}

In this work we present an empirical analysis of a popular machine learning dataset and demonstrate that, even with intentions to remove PII, personal data can remain. These results can trigger existing privacy laws for downstream uses of web-scraped data.

%% file: sections/acks.tex
\begin{acks}

We are incredibly grateful for the invaluable comments from the CS\&LAW conference shepherd and reviewers. We also are thankful for the helpful feedback from the 2025 Privacy Law Scholars Conference and especially comments from Inyoung Cheong. The first author is supported by the National Science Foundation Graduate Research Fellowship Program. 
This work was supported in part by U.S.\ National Science Foundation awards CCF-2045402 and CNS-2205171, the Carnegie Bosch Postdoctoral Fellowship, the University of Washington Tech Policy Lab, the  McDevitt Chair in Computer Science, Ethics, and Society, and a grant from the Simons Foundation.

\end{acks}

%% file: sections/appendix/appendix.tex
\appendix
\onecolumn

\section{Stakeholder network}
\label{sec:stakeholder_network}

\begin{figure}[ht]
    \centering
    \includegraphics[width=0.8\linewidth]{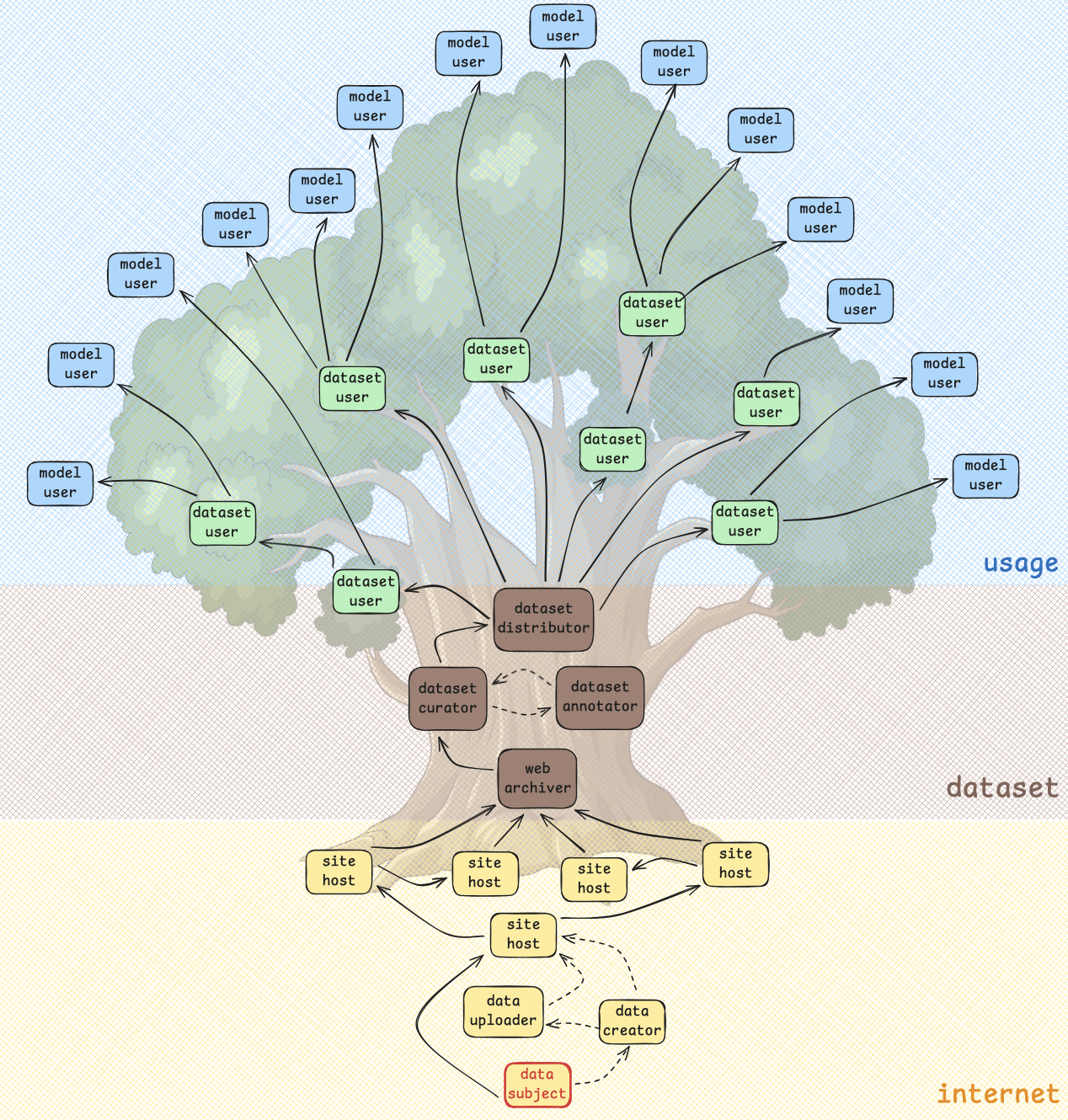}
    \caption{The stakeholder network demonstrates the potential flow of personal information between actors in the Internet (yellow), Dataset (brown), and Usage (blue) stages. Personal information starts from the data subject and is uploaded to the site host, potentially due to the data creator or data uploader without the data subject's knowledge (in dashed lines). Data may pass between site hosts through crawling and uploading, but is all aggregated by the web archiver. The dataset curator gathers from the web archiver and may optionally rely on an eternal dataset annotator (in dashed lines). The curated URL table is given to the dataset distributor who in turn passes data to multiple dataset users, who may create other datasets to pass to other dataset users, or deploy a model for various model users. This diagram demonstrates the trunk as the centralized source before wide dissemination of a popular training dataset like CommonPool with over two million downloads.}
    \Description{Flow chart of nodes overlaid on top of a cartoon tree. Arrows start from the internet stage with the data subject pointing to various site host nodes as the tree roots, which all point to the stakeholders from the dataset stage (web archiver, dataset curator, dataset annotator, and dataset distributor) at the tree base. The arrows then split from the dataset curator to various data users as the leaves of the tree, which then split off to the model user nodes.}
    \label{fig:actor_tree}
\end{figure}

\Cref{fig:actor_tree} provides an overview of the data flow between stakeholders defined in \Cref{sec:stakeholders}.

\section{Methodology details}
\label{sec:appx_methods}

In this section, we provide additional details on our audit methodology for OCR method selection.

\subsection{Optical character recognition evaluation}
\label{sec:appx_ocr_eval}

\begin{figure}
   \centering
   \begin{subfigure}[b]{0.5\textwidth}
       \includegraphics[width=\textwidth]{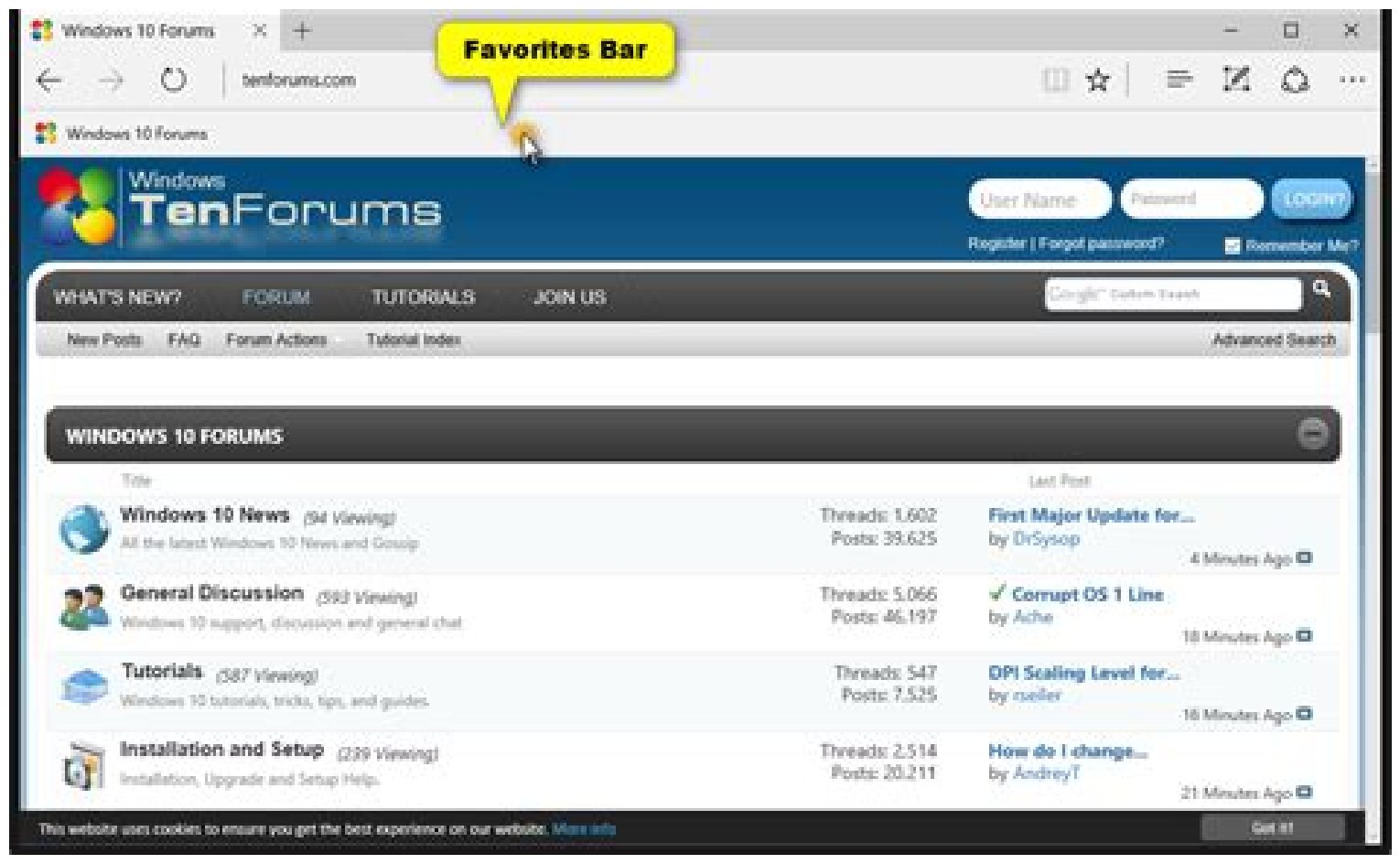}
       \caption{Screenshot of webpage}
   \end{subfigure}
   \begin{subfigure}[b]{0.25\textwidth}
       \includegraphics[width=\textwidth]{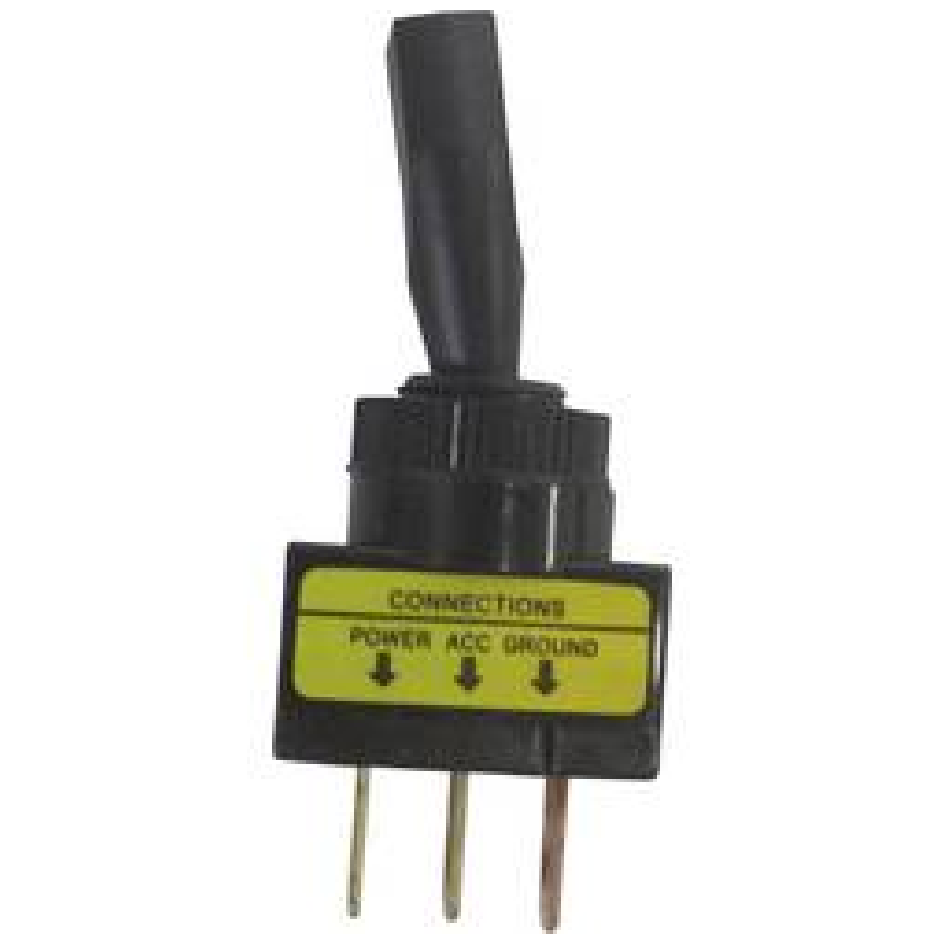}
       \caption{Picture of product}
    \end{subfigure}
    \caption{Example CommonPool images that contain text}
    \Description{On the left, example image of a screenshot of the "tenforums.com" forum including posts of users. On the right, a picture of a black toggle switch.}
    \label{fig:ocr_sample_images}
\end{figure}

We consider several open-source state-of-the-art optical character recognition (OCR) methods: EasyOCR \citep{easyocr}, PaddleOCR \citep{ppocr}, Tesseract \citep{tesseract}, and TROCR \citep{trocr}. Because most OCR methods are intended for handwriting detection or document extraction \citep{memon2020handwritten, vedhaviyassh2022comparative}, we perform our own evaluation for web-scraped images in CommonPool. We subsequently manually annotate the visible, legible text contained in 100 randomly-selected images and treat these annotations as ground truth. Images in our evaluation set often include screenshots or products of varying image quality (see \Cref{fig:ocr_sample_images} for examples), which present text differently from typical OCR use cases.

\textbf{Metrics:} As shown in prior work \citep{neudecker2021survey}, OCR evaluation metrics are not consistent, as the particular choice of metric may change which OCR tool is considered more accurate. In particular, character error rate (CER) based on Levenshtein distance \citep{levenshtein1966binary} is often used as a performance metric, which relies on a particular ordering of words in documents. In our analysis, however, ordering does not matter for screenshots or products and not as necessary for keyword queries. We also desire word accuracy rather than close characters, since the words are later searched or fed as input into Presidio's named entity recognition model. As such, we rely on the bag-of-words model for our OCR evaluation, which tracks the number of words that are accurately recognized \citep{neudecker2019ocr}.

\begin{table}
    \centering
    \begin{tabular}{llll} 
     \toprule
    \textbf{OCR method} & \textbf{Accuracy} & \textbf{Precision} & \textbf{Recall} \\ [0.5ex] 
     \midrule
     \textbf{EasyOCR} \citep{easyocr} & 0.426 & 0.442 & 0.469 \\ 
     \textbf{PaddleOCR} \citep{ppocr} & \textbf{0.579} & \textbf{0.567} & \textbf{0.684 }\\
     \textbf{Tesseract} \citep{tesseract} & 0.168 & 0.184 & 0.177 \\
     \textbf{TROCR} \citep{trocr} & 0.054 & 0.049 & 0.078 \\
     \bottomrule
    \end{tabular}
    \caption{Evaluation of OCR tools according to bag-of-words accuracy, precision, and recall.}
    \label{tab:ocr_eval}
\end{table}

\textbf{Results: }\Cref{tab:ocr_eval} shows that on our evaluation set, PaddleOCR outperforms the other methods on every bag-of-word metric. Specifically, PaddleOCR has higher recall than the other methods, and thus encompasses the most recognizable words, which we then use to flag samples and subsequently manually verify due to the error in OCR.

\section{Additional audit results}

In this section we provide additional audit results.

\subsection{Text visualizations}
\label{sec:appx_text_viz}

\begin{figure}
   \centering
   \begin{subfigure}[b]{0.4\textwidth}
       \includegraphics[width=\textwidth]{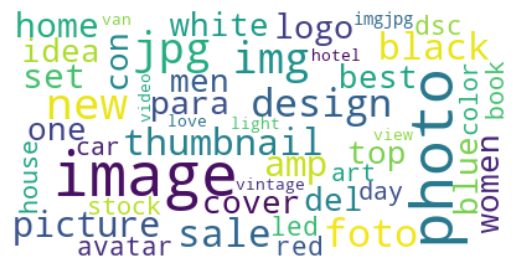}
       \Description{A word cloud with common words like image, thumbnail, photo, design, picture.}
       \caption{Word cloud where size of the word corresponds to relative size.}
   \end{subfigure}
   \begin{subfigure}[b]{0.4\textwidth}
       \includegraphics[width=\textwidth]{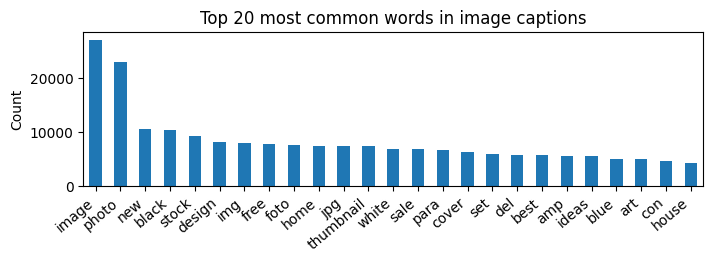}
       \Description{Bar graph with most common words including image, photo, new, black each occurring at least ten thousand times.}
       \caption{Top 20 most common words.}
    \end{subfigure}
    \caption{Word visualizations of captions (without stop words) of a 1 million random subsample of CommonPool.}
    \label{fig:caption_word_viz}
\end{figure}

\begin{figure}
    \centering
    \begin{subfigure}[b]{0.4\textwidth}
       \includegraphics[width=\textwidth]{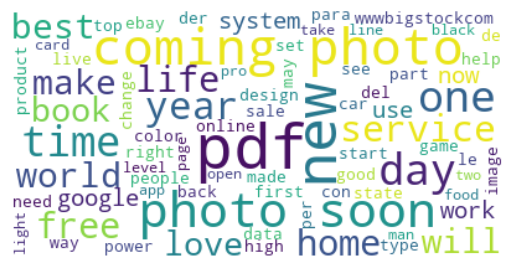}
       \Description{A word cloud with common words like pdf, photo, coming, soon, service, new.}
       \caption{Word cloud where size of the word corresponds to relative size.}
   \end{subfigure}
   \begin{subfigure}[b]{0.4\textwidth}
       \includegraphics[width=\textwidth]{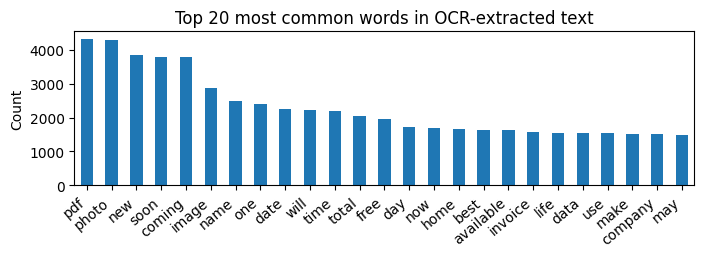}
       \Description{Bar graph with most common words including pdf, photo, new, soon, coming each occurring at least three thousand times.}
       \caption{Top 20 most common words.}
    \end{subfigure}
    \caption{Word visualizations of OCR-extracted text (without stop words) of a 1 million random subsample of CommonPool.}
    \label{fig:ppocr_word_viz}
\end{figure}

In this section we show text visualizations as a cursory analysis of the types of content in CommonPool. \Cref{fig:caption_word_viz} shows that image captions are related to stock photos or describing images, which aligns with their usage as alt-text \citep{caldwell2008web}. \Cref{fig:ppocr_word_viz} also shows that detected text in images describes stock photos and images, but in addition we observe that \texttt{invoice} is one of the most common words identified in the OCR-extracted text. We find several images of invoices that display business and customer names and addresses, as well as payments issued between them. While these invoices are not considered personal information, they may reveal corporate information that may not be intended to be public.

\begin{figure}
    \centering
    \begin{subfigure}[b]{0.45\textwidth}
       \includegraphics[width=\textwidth]{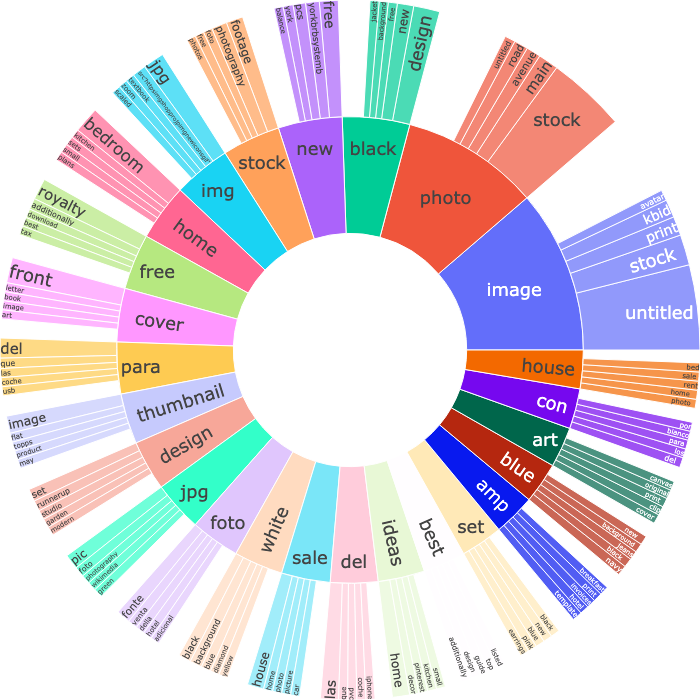}
       \caption{Caption.}
   \end{subfigure}
   \hfill
   \begin{subfigure}[b]{0.45\textwidth}
       \includegraphics[width=\textwidth]{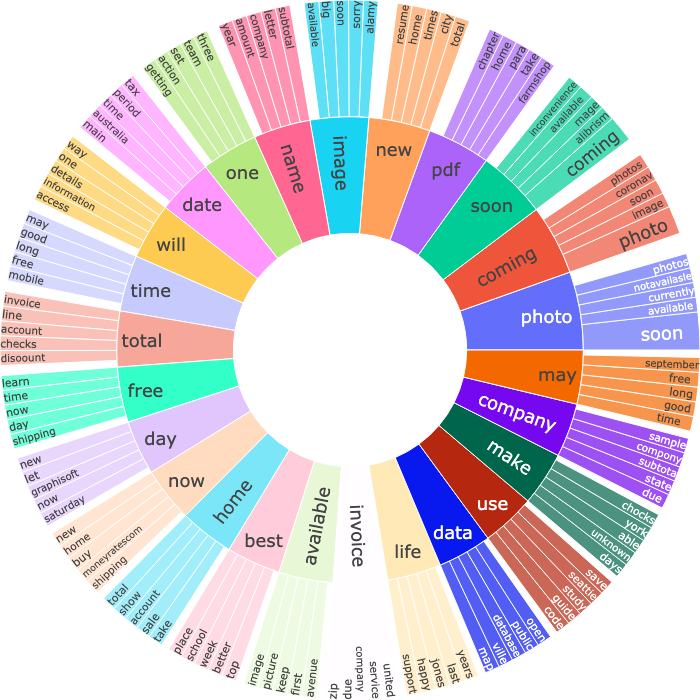}
       \caption{OCR-extracted text.}
    \end{subfigure}
    \Description{Disk visualizations of common words like image, photo, black, new, stock for the captions on the left, and photo, coming, soon, pdf, new for OCR-extracted text on the right.}
    \caption{Bigram disk visualizations in the caption and OCR-extracted text of 1 million random subsamples of CommonPool. The words in the first ring have sizes relative to their frequency, but isolated to the top 25 non-stop words for visibility. The outer ring depicts frequent words that appear directly after the corresponding inner word. Within a single segment, the outer words have sizes relative to their frequency, but isolated to the top 5 words within its inner segment and bounded for readability.}
    \label{fig:disk_viz}
\end{figure}

\subsection{Celebrity name search}
\label{sec:appx_celebrity}

\begin{figure*}
    \centering
    \includegraphics[width=\linewidth]{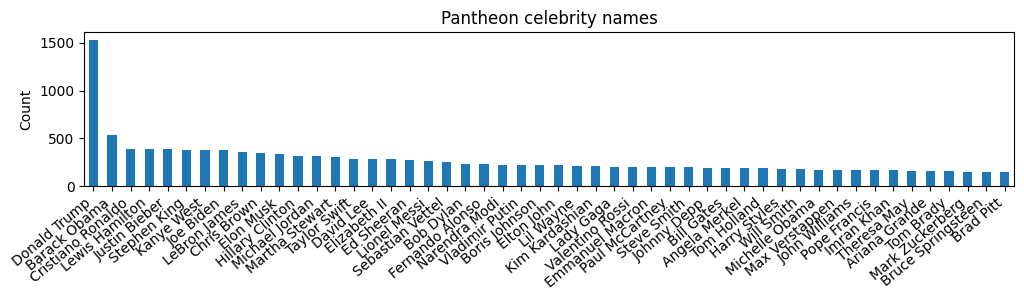}
    \caption{Top 50 most common celebrities from Pantheon 2020 dataset \citep{yu2016pantheon} mentioned in CommonPool captions and OCR-extracted text.}
    \Description{Bar graph with most celebrity mentions including Donald Trump occuring at least 1500 times, and celebrities like Barack Obama, Cristiano Ronaldo, and Lewis Hamilton following.}
    \label{fig:pantheon_celebrity_search}
\end{figure*}

\begin{figure}[ht]
    \centering
    \begin{subfigure}[b]{0.58\textwidth}
       \includegraphics[width=\textwidth]{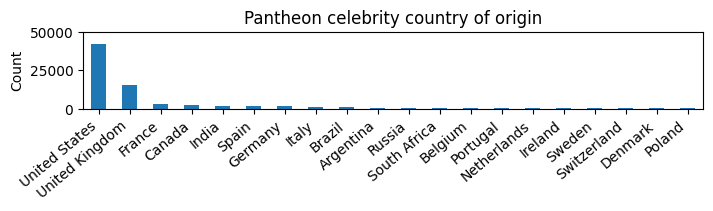}
       \caption{Top 20 most common countries of origin.}
       \Description{Bar graph with most celebrity country of origin with the United States occurring at least 35 thousand times.}
       \label{fig:pantheon_country}
   \end{subfigure}
   \hfill
   \begin{subfigure}[b]{0.38\textwidth}
       \includegraphics[width=\textwidth]{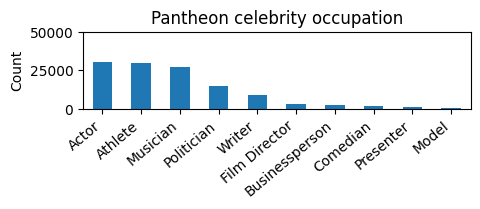}
       \Description{Bar graph with most common occupations including actor, athlete, and musician each occurring at least 25 thousand times.}
       \caption{Top 20 most common occupations.}
       \label{fig:pantheon_occupation}
    \end{subfigure}
    \caption{Additional bar graphs from searching Pantheon 2020 celebrity names \citep{yu2016pantheon} from CommonPool captions and OCR-extracted text.}
    \label{fig:pantheon_breakdown}
\end{figure}

\begin{figure}[ht]
    \centering
    \includegraphics[width=\linewidth]{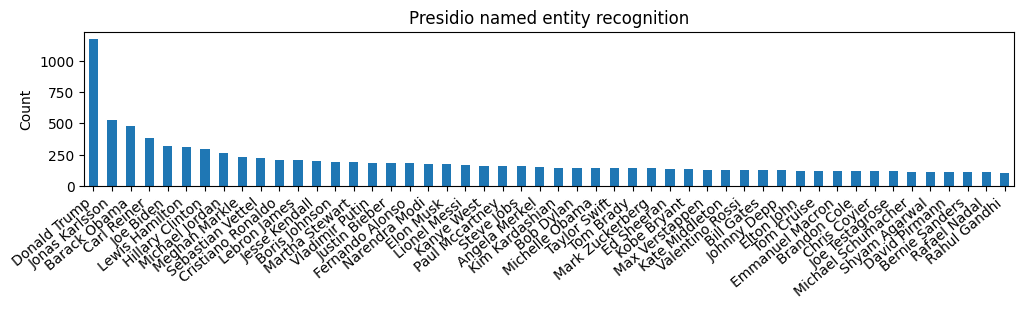}
    \caption{Top 50 most common Presidio-detected names from CommonPool captions and OCR-extracted text.}
    \Description{Bar graph with most celebrity mentions including Donald Trump occuring at least 1000 times, and celebrities like Jonas Karlsson, Barack Obama, Carl Reiner, and Joe Biden following.}
    \label{fig:presidio_celebrity_search}
\end{figure}

This section provides results from searching for celebrity names from the Pantheon dataset (\Cref{fig:pantheon_celebrity_search} and \Cref{fig:pantheon_breakdown}). \Cref{fig:presidio_celebrity_search} also shows the most common names using Presidio outside of brands.

\subsection{Resume documents}

\begin{figure}[ht]
    \centering
    \includegraphics[width=0.8\linewidth]{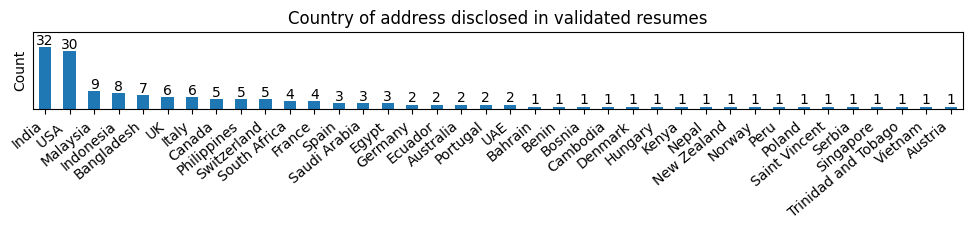}
    \caption{Sample count breakdown of country of address disclosed by validated resumes.}
    \label{fig:resume_country_address}
    \Description{Bar graph with most common disclosed country of address with India and USA having at least 30 samples.}
\end{figure}

\begin{figure}[ht]
    \centering
    \includegraphics[width=0.8\linewidth]{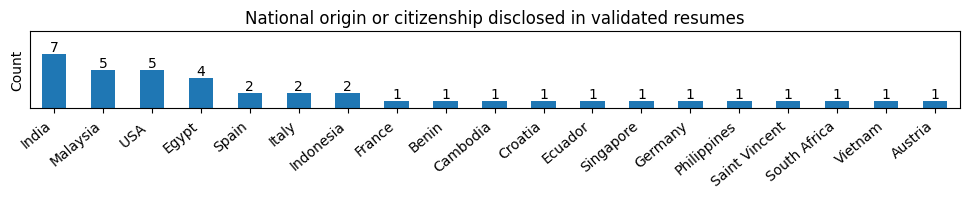}
    \caption{Sample count breakdown of national origin or citizenship disclosed by validated resumes.}
    \Description{Bar graph with most common disclosed national origin with India, Malaysia, and USA having at least 5 samples.}
    \label{fig:resume_nat_origin}
\end{figure}

In this section, we show additional results of the associated geographic locations of the resume documents associated with online presence of individuals. \Cref{fig:resume_country_address} refers to the disclosed address, while \Cref{fig:resume_nat_origin} refers to the disclosed national origin or citizenship.

\begin{figure}
    \centering
    \includegraphics[width=0.3\linewidth]{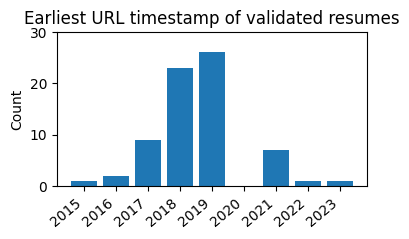}
    \caption{Earliest timestamp of URLs of validated resumes according to Internet Archive's Wayback Machine \citep{wayback}. Only 70 of 168 validated resume URLs had existing records.}
    \Description{Bar graph by year of earliest URL timestamp from 2015 to 2023, peaking at 2019 with at least 25 samples.}
    \label{fig:resume_timestamp}
\end{figure}

\subsection{Children's information}
\label{sec:appx_childrens_info}

\begin{figure}[ht]
    \centering
    \begin{subfigure}[b]{0.48\textwidth}
        \includegraphics[width=\linewidth]{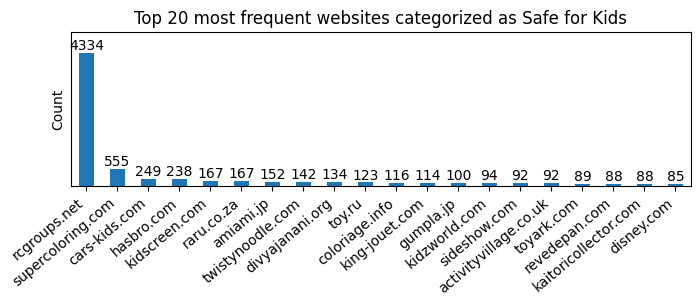}
        \caption{Sample counts of top 20 websites categorized by Cloudflare as \texttt{Safe for Kids}.}
        \Description{Bar graph with most common websites categorized as Safe for Kids with 4334 samples coming from rcgroups.net.}
        \label{fig:cloudflare_kids}
   \end{subfigure}
   \hfill
   \begin{subfigure}[b]{0.48\textwidth}
        \includegraphics[width=\linewidth]{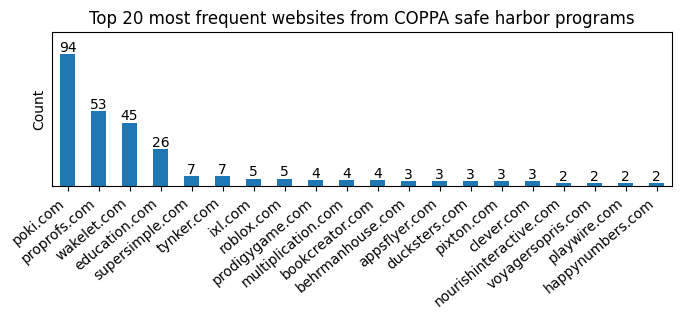}
        \caption{Sample counts of top 20 websites that are members of COPPA safe harbor programs \citep{coppa_safe_harbor}.}
        \Description{Bar graph with most common websites from COPPA safe harbor programs including poki.com, proprofs.com, wakelet.com, education.com with at least 25 samples.}
        \label{fig:coppa_sites}
    \end{subfigure}
    \caption{Website frequency of children-related information.}
    \label{fig:url_kids}
\end{figure}

\Cref{fig:url_kids} gives additional breakdown of popular websites relating to children based on our two approaches: Cloudflare URL categorization and COPPA safe harbor program membership described in \Cref{sec:audit_children}.

\subsection{Download errors of unavailable images}
\label{sec:appx_failed_imgs}

\begin{figure}[ht]
    \centering
    \includegraphics[width=0.8\linewidth]{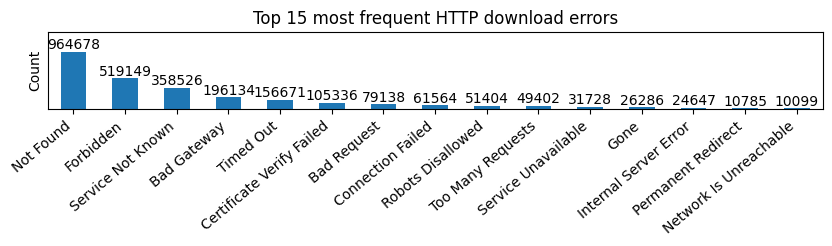}
    \caption{Sample counts of top 15 most common HTTP errors for images that failed to download during a download version run in April 2025.}
    \Description{Bar graph with most common HTTP download errors including Not Found, Forbidden, Service Not Known errors with at least 350 thousand samples.}
    \label{fig:http_error_freq}
\end{figure}

\begin{figure}[ht]
    \centering
    \begin{subfigure}[b]{0.4\textwidth}
        \includegraphics[width=\linewidth]{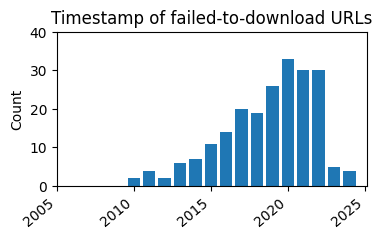}
        \caption{\textit{Failed-to-download}: 213 out of 1000 image URLs had existing records.}
        \Description{Bar graph by year of earliest URL timestamp from 2005 to 2024, peaking at around 30 samples per year from 2019-2022.}
        \label{fig:failed_url_timestamp}
   \end{subfigure}
   \hfill
   \begin{subfigure}[b]{0.42\textwidth}
        \includegraphics[width=\linewidth]{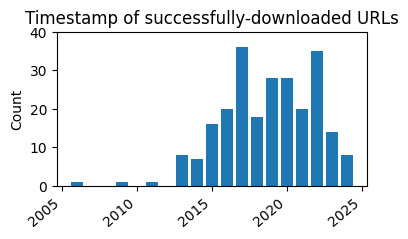}
        \caption{\textit{Successfully-downloaded}: 241 out of 1000 image URLs had existing records.}
        \Description{Bar graph by year of earliest URL timestamp from 2005 to 2024, peaking at around 30 samples from 2017-2022.}
        \label{fig:success_url_timestamp}
    \end{subfigure}
    \caption{Sample counts by year of earliest timestamps according to the Wayback Machine records for a random subsample of image URLs during a download version run in April 2025.}
    \label{fig:failed_vs_success_url_timestamp}
\end{figure}

\Cref{fig:http_error_freq} gives a breakdown of most common HTTP errors for images that failed to download, while \Cref{fig:failed_url_timestamp} shows the earliest recorded timestamp according to the Wayback Machine. \Cref{tab:failed_t_tests} refers to two-sample t-tests to measure statistical differences between samples that failed or succeeded in downloading.

\begin{figure}[ht]
    \centering
    \includegraphics[width=0.8\linewidth]{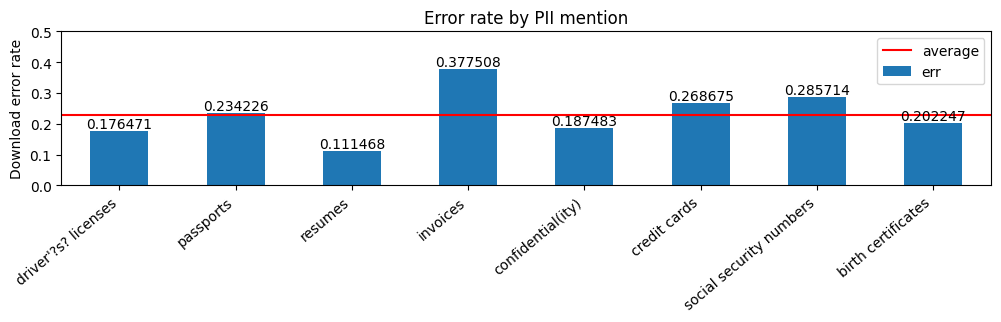}
    \caption{Download error rate for samples grouped by regular expression matches to instances of personal information. The average download error rate is plotted as a red line.}
    \Description{Bar graph of download error rates by type of PII mention. Invoices has the highest error rate at 0.38, while resumes have the lowest at 0.11, with the average at 0.21.}
    \label{fig:pii_mention_error_rate}
\end{figure}

\begin{table}
    \begin{subtable}{.48\linewidth}
      \centering
        \begin{tabular}{@{}llll@{}}
            \toprule
             \textbf{Image-related annotations} & $\mu_1$ & $\mu_2$ & \textbf{Adjusted p-value} \\
             \midrule
             \textbf{CLIP-similarity score} & 0.207 & 0.211 & $p < 0.001$ \\
             \textbf{Number of detected faces} & 0.558 & 0.587 & $p < 0.001$ \\
             \textbf{Image size} & 390.2 & 404.0 & $p < 0.001$ \\
             \bottomrule \\
        \end{tabular}
        \caption{For each row, $\mu_1$ refers to the sample mean of the variable for 1000 randomly-selected \textit{successfully-downloaded images}, while $\mu_2$ refers to the sample mean for 1000 randomly-selected \textit{failed-to-download images}. The alternative hypothesis is $\mu_1 < \mu_2$. The CLIP-similarity score refers to the cosine similarity in CLIP embeddings of the caption and image \citep{radford2021learning}. The number of detected faces is according to DataComp's SCRFD algorithm \citep{guo2021sample}.}
        \label{tab:failed_t_tests}
    \end{subtable}%
    \hfill
    \begin{subtable}{.48\linewidth}
      \centering
        \begin{tabular}{@{}llll@{}}
            \toprule
             \textbf{Image-related variable} & $\mu_1$ & $\mu_2$ & \textbf{Adjusted p-value} \\
             \midrule
             \textbf{Bounding box area} & 100.4 & 29.6 & $p < 0.001$ \\
             \textbf{Pixel brightness} & 119.6 & 98.2 & $p < 0.001$ \\
             \textbf{Predicted} \texttt{Female} & 0.42 & 0.33 & $p < 0.01$ \\
             \textbf{Predicted age} & 34.3 & 26.2 & $p < 0.001$ \\
             \bottomrule \\
        \end{tabular}
        \caption{For each row, $\mu_1$ refers to the sample mean of the variable for images with manually confirmed human faces \textit{detected (and therefore blurred)} by DataComp's face detection algorithm, while $\mu_2$ refers to the sample mean for images \textit{undetected (and therefore not blurred)}. The alternative hypothesis is $\mu_1 > \mu_2$. ``Predicted'' refers to the Rekognition gender and age annotations, where ``Predicted \texttt{Female}'' represents the the proportion among all images classified as \texttt{Female}, and ``Predicted age'' represents the average age prediction.}
        \label{tab:face_t_tests}
    \end{subtable} 
    \caption{Summary of two-sample one-tailed t-tests of various image-related annotations, adjusted by the Benjamini-Yekutieli procedure to control the false discovery rate for dependent tests \citep{benjamini2001control}.}
\label{tab:t_tests}
\end{table}

\subsection{Face detection}
\label{sec:appx_face_url}

\begin{figure}[ht]
    \centering
    \begin{subfigure}[b]{0.4\textwidth}
       \includegraphics[width=\textwidth]{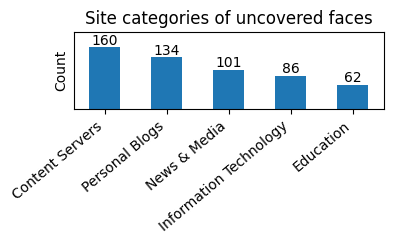}
       \caption{Cloudflare categorizations \citep{cloudflare} of URLs.}
       \Description{Bar graph of website categories with Content Servers, Personal Blogs, and News and Media as the most common with at least 100 samples.}
   \end{subfigure}
   \hfill
   \begin{subfigure}[b]{0.4\textwidth}
       \includegraphics[width=\textwidth]{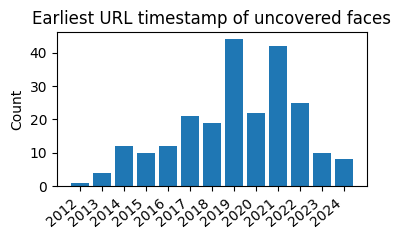}
       \caption{Earliest timestamp of URLs according to Internet Archive's Wayback Machine \citep{wayback} where 230 out of 854 image URLs had existing records.}
       \Description{Bar graph by year of earliest URL timestamp from 2012 to 2024, peaking at around 40 samples per year from 2019 to 2021.}
    \end{subfigure}
    \caption{Analysis of website URLs of manually confirmed images of faces not caught by SCRFD.}
    \label{fig:face_url_origin}
\end{figure}

\Cref{tab:face_t_tests} reports the differences in the sample means of the distributions of manually confirmed human faces that are blurred versus not blurred by DataComp. \Cref{fig:face_url_origin} includes analysis of the URL origins of the real faces uncovered by SCRFD.

